\documentclass{emulateapj}

\usepackage{natbib}
\usepackage{amsmath}
\shorttitle{IR extinction}
\shortauthors{FRITZ ET AL.}

\begin{document}

\title{Line derived infrared extinction towards the Galactic Center}

\author{ T.K.~Fritz \altaffilmark{1,$\star$}
, S.~Gillessen\altaffilmark{1}, K.~Dodds-Eden\altaffilmark{1}, D.~Lutz\altaffilmark{1},  R.~Genzel\altaffilmark{1,2}, W.~Raab\altaffilmark{1},   T.~Ott\altaffilmark{1},
O.~Pfuhl\altaffilmark{1}, F.~Eisenhauer\altaffilmark{1}, F.~Yusef-Zadeh\altaffilmark{3}
}

\altaffiltext{1}{Max Planck Institut f{\"u}r Extraterrestrische Physik, Postfach 1312, D-85741, Garching, Germany.}
\altaffiltext{2}{Department of Physics, University of California, Berkeley, 366 Le Comte Hall, Berkeley, CA 94720-7300}
\altaffiltext{3}{Department of Physics and Astronomy, Northwestern University, Evanston, IL 60208, USA}
\altaffiltext{$\star$}{E-mail: tfritz@mpe.mpg.de}

\keywords{  Galaxy: center - Galaxy: fundamental parameters- ISM extinction, dust}

\begin{abstract}
We derive the extinction curve towards the Galactic Center from 1 to 19 
$\mu$m. We use hydrogen emission lines of the 
minispiral observed by ISO-SWS and SINFONI. The extinction free flux reference is the 2 cm continuum
emission observed by the VLA.
Towards the inner $14'' \times 20''$ we find an extinction of A$_{2.166\,\mu\mathrm{m}}=2.62 \pm 0.11 $, with a power-law slope of
 $\alpha=-2.11 \pm 0.06$
shortward of $ 2.8$ $\mu$m, consistent with the average near infrared slope from the recent literature. At longer wavelengths, however, 
we find that the extinction is \emph{grayer} than shortward of 2.8 $\mu$m. We find it is not possible to fit the observed extinction 
curve with a dust model consisting of pure carbonaceous 
and silicate grains only, and the addition of composite particles, including ices, is needed to explain the observations.
Combining a distance dependent extinction with our distance independent extinction  we derive the distance to the GC to be R$_0=7.94
\pm 0.65$ kpc.
Towards Sgr~A* ($r<0.5''$) we obtain $A_{H}=4.21 \pm 0.10$, $A_{Ks}=2.42 \pm 0.10$ and $A_{L'}=1.09 \pm 0.13$.

% ApJ: citations are not allowed in the abstract!

\end{abstract}

\maketitle

\section{Introduction}
Knowledge of the extinction in the infrared \citep{Schultz_75, Cardelli_89, Mathis_90}  is important for obtaining the intrinsic luminosities of
 highly extincted objects. In addition, the extinction curve provides important constraints on the properties of interstellar dust (see e.g.,  \citealt{Compiegne_10}). 
Using the extinction curve together with the dust emission it is possible to  derive the composition and sizes of interstellar dust grains.
In this way it has been found that interstellar dust is mainly composed of silicate and carbonaceous dust grains \citep{Draine_03}. 
However, it is still uncertain if
composite particles are also required \citep{Mathis_96,Li_97,Weingartner_01,Zubko_04}, including, for example ices and voids in addition to the basic dust grains.  

\begin{figure}
\begin{center}

\includegraphics[width=0.707 \columnwidth,angle=-90]{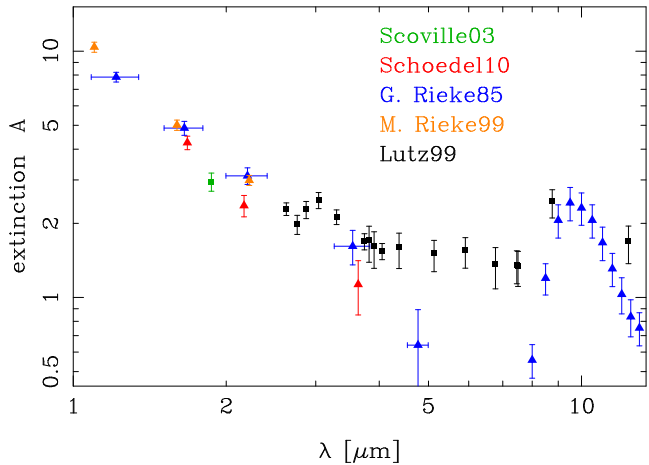}
\caption{Extinction towards Sgr~A* from the literature: triangles represent values derived from stars, boxes values derived from gas lines. Most values were obtained using slightly different regions around Sgr~A*. For the comparison, we convert the values to the direct sight line towards Sgr A* (i.e. the central $0.5''$, see Section \ref{sec:spat}) using the extinction map of \citet{Schoedel_09b}.
 In the case of \citet{Rieke_85} we use the extinction law of \citet{Rieke_85} and the absolute extinction of \citet{Rieke_89}.  
} 
\label{fig:_ex_now}
\end{center}
\end{figure}

The Galactic Center (GC), invisible in the optical, is one well-known example of a highly extincted region \citep{Becklin_68}. 
Because of this, it should be relatively easy to measure the infrared extinction, without the large relative errors implicated in using regions of small
 absolute extinction. \citet{Rieke_85}, for example, used infrared photometry of a few red supergiants present in the GC together with a visible to near infrared 
extinction law obtained outside of the Galactic Center, in order to derive a universal infrared extinction law (Figure~\ref{fig:_ex_now}).
This measurement was later slightly improved by adding NICMOS near infrared measurements by \citet{Rieke_99}.

Another method of measuring the IR extinction is to use a stellar population of known infrared luminosity \citep{Nishiyama_06}, 
such as the red clump  for which the absolute luminosity is measured locally \citep{Groenewegen_08}.
For this method, however, it is also necessary to know a precise distance to the object, which makes the method difficult for most objects. In the case of the GC the distance 
is however well known \citep{Reid_93,Genzel_10}. Using this method \citet{Schoedel_09b} obtained  A$_H=4.35 \pm 0.18$,   A$_{Ks}=2.46 \pm 0.12$ and 
 A$_{L'}=1.23 \pm 0.20$ towards Sgr~A*.
 
As an alternative to methods involving stars, one can also use nebular hydrogen lines to measure extinction. As a case in point, \citet{Lutz_96} 
and \citet{Lutz_99} used ISO-SWS spectra of the minispiral \citep[a bright HII region in the GC;][]{Lo_83}  to derive extinction from 2.6 to 19 $\mu$m. 
Typically this method is more precise, since the intrinsic uncertainty of the relative line strengths in this case \citep{Hummer_87} is 
smaller than the spectral uncertainty of stellar emission. An additional advantage is that the lines are much narrower than the bandpasses of the
 broadband filters, so that the effective wavelength  is known a priori, while for broadband filters it is extinction dependent. 
Furthermore, there are many more lines in the IR than the number of broadband filters in frequent use, such that it is easier to obtain a 
well-sampled extinction curve, that includes also extinction features.

Although it is only possible, using infrared lines alone, to obtain \emph{relative} extinction measurements within the infrared, stellar 
methods often have the same problem. 
Absolute extinction calibration is made possible by comparing with longer wavelengths where the extinction is negligible. 
The absolute extinction towards the GC  for the Paschen-$\alpha$ line at 1.87 $\mu$m was obtained in this way in \citet{Scoville_03} using 6~cm continuum 
data. 

The extinction towards Sgr A* from the various measurements mentioned above is shown in Figure~\ref{fig:_ex_now}. However, there is apparently 
some discrepancy between the different studies. The uncertainty in extinction creates, in turn, uncertainty in determining the intrinsic luminosities
 of young massive stars in the GC \citep{Martins_07,Fritz_10}, as well as the intrinsic luminosity of Sgr~A* 
\citep{Genzel_03}. In addition, there is no study that covers the full wavelength range, and only a few
studies of this kind are available outside of the Galactic Center.

In this paper we use the emission of the minispiral from 1.28 to 19 $\mu$m to derive relative extinction values from line emission, which are 
compared to 2 cm radio continuum data to compute the absolute extinction. 
In Section 2 we present our data, and discuss the extraction of line maps and flux calibration in Section 3.
In Section 4 we derive the extinction.
We discuss the results in Section 5. We compare the results with  other works for the GC and for other parts of the Galaxy
 and with theoretical dust models. We also use the data for estimating the distance to the Galactic Center, R$_0$. We summarize in Section 6.

\section{Dataset}

In this section we describe the observations used to derive the extinction curve towards the Galactic Center. We used a VLA radio (2cm continuum) map, ISO-SWS spectra (2.4 to 45 $\mu$m) and SINFONI imaging spectroscopy (1.2 to 2.4 $\mu$m), of which the latter is used to construct line maps at Brackett-$\gamma$ (2.166 $\mu$m), Brackett-$\zeta$ (1.736 $\mu$m), and Paschen-$\beta$ (1.283 $\mu$m). The fields of view for the different datasets are compared in Figure \ref{fig:_areas}.

\subsection{Radio}
% joined with \subsection{Radio data}
\label{sec:radio}

\begin{figure}
\begin{center}
\includegraphics[width=0.99 \columnwidth,angle=0]{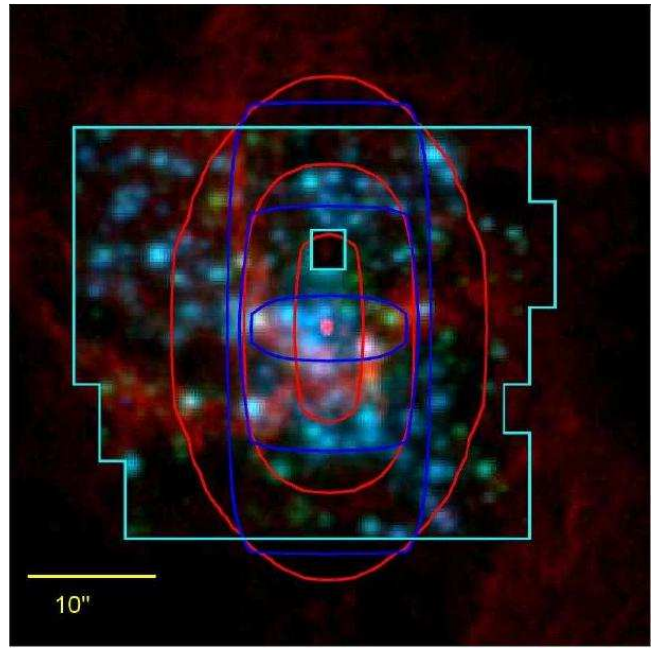} % astro-ph compressed
\caption{Field of view for different measurements. In red, the radio 2 cm continuum image; in blue and green, respectively, 
are the H and Ks broadband images constructed from SINFONI HK data. The radio image is larger than displayed here, but the outer areas are 
not used in this publication. Light green lines mark the borders of the SINFONI field which has a hole around IRS7. Blue lines indicate the contours 
of the ISO PSF around 2.6 $\mu$m, representing 90\% , 50\% and 10\% of the central intensity. The red lines indicate the same for the ISO PSF 
around 19 $\mu$m. % dieter spatial scale
} 
\label{fig:_areas}
\end{center}
\end{figure}

Figure~\ref{fig:_areas} presents the radio and near-infrared data used combined with the field of views of all measurements.
The radio data consists of combined 2 cm continuum observations in the A, B, C and D configurations of the VLA.
The phase center of the multi-configuration data set at 2cm is on Sgr A*.
Standard calibration was done on each data set before each data set was 
combined in the {\it uv} plane. Details of phase and amplitude 
calibration of each data set taken in different configurations of the VLA can be found 
in \citet{Yusef_93} and \citet{Yusef_98}.
 The combination of these configurations makes high resolution possible, while at the same time ensuring good coverage of extended structure.
 Due to the use of the smallest, D, configuration, we are able to detect structures of up to 50'' in size. The 
resolution of the final image is $0.42'' \times 0.3''$ with the longer axis of the Gaussian oriented approximately north (P.A.$\approx 4^{\circ}$). 
The image extends at least 48$''$ from Sgr~A* in all directions.
We use a radio continuum map because, for the data available to us, the SNR and the resolution are better than for line maps. 

\begin{figure}
\begin{center}
\includegraphics[width=0.99 \columnwidth]{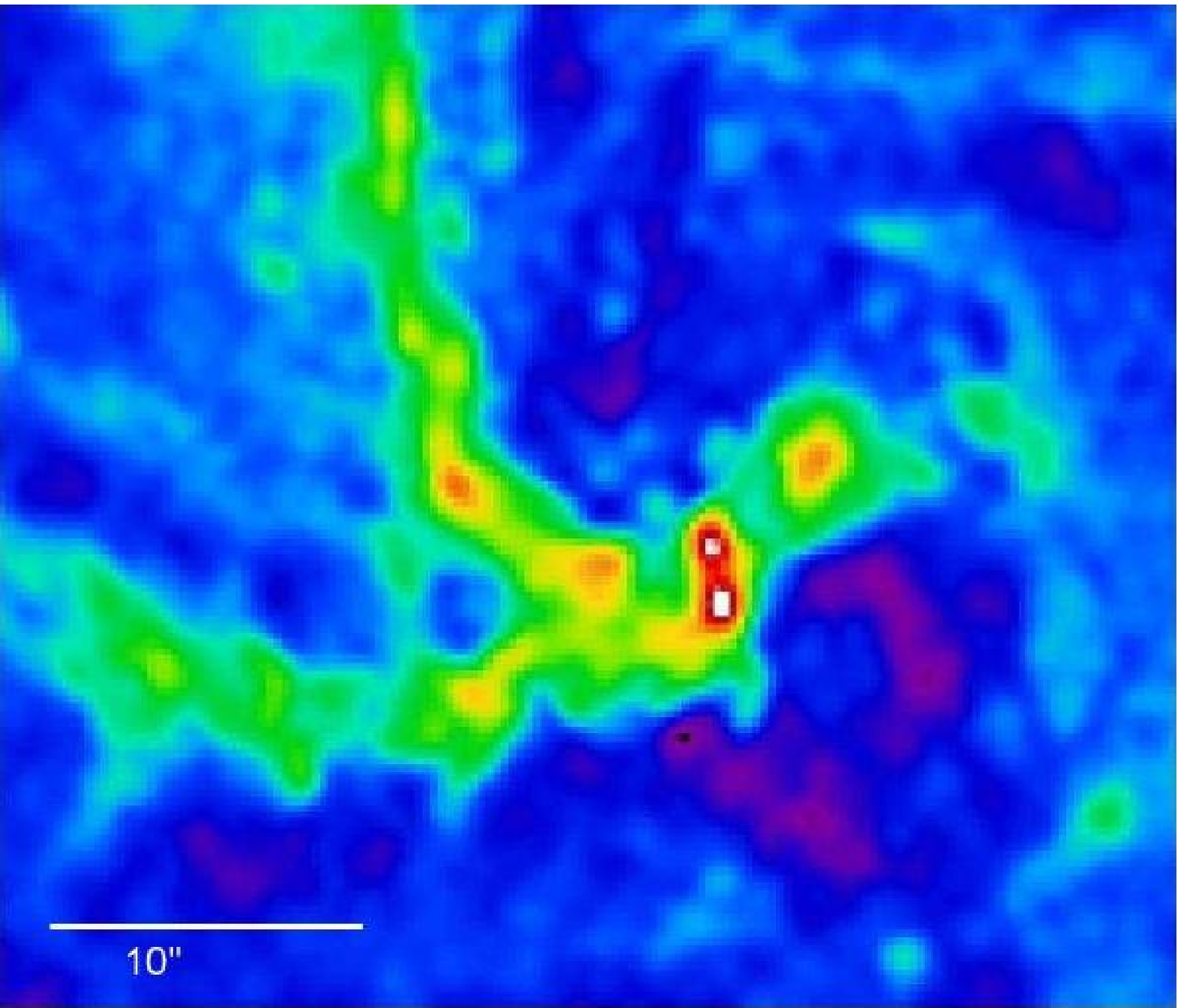}
\includegraphics[width=0.99 \columnwidth]{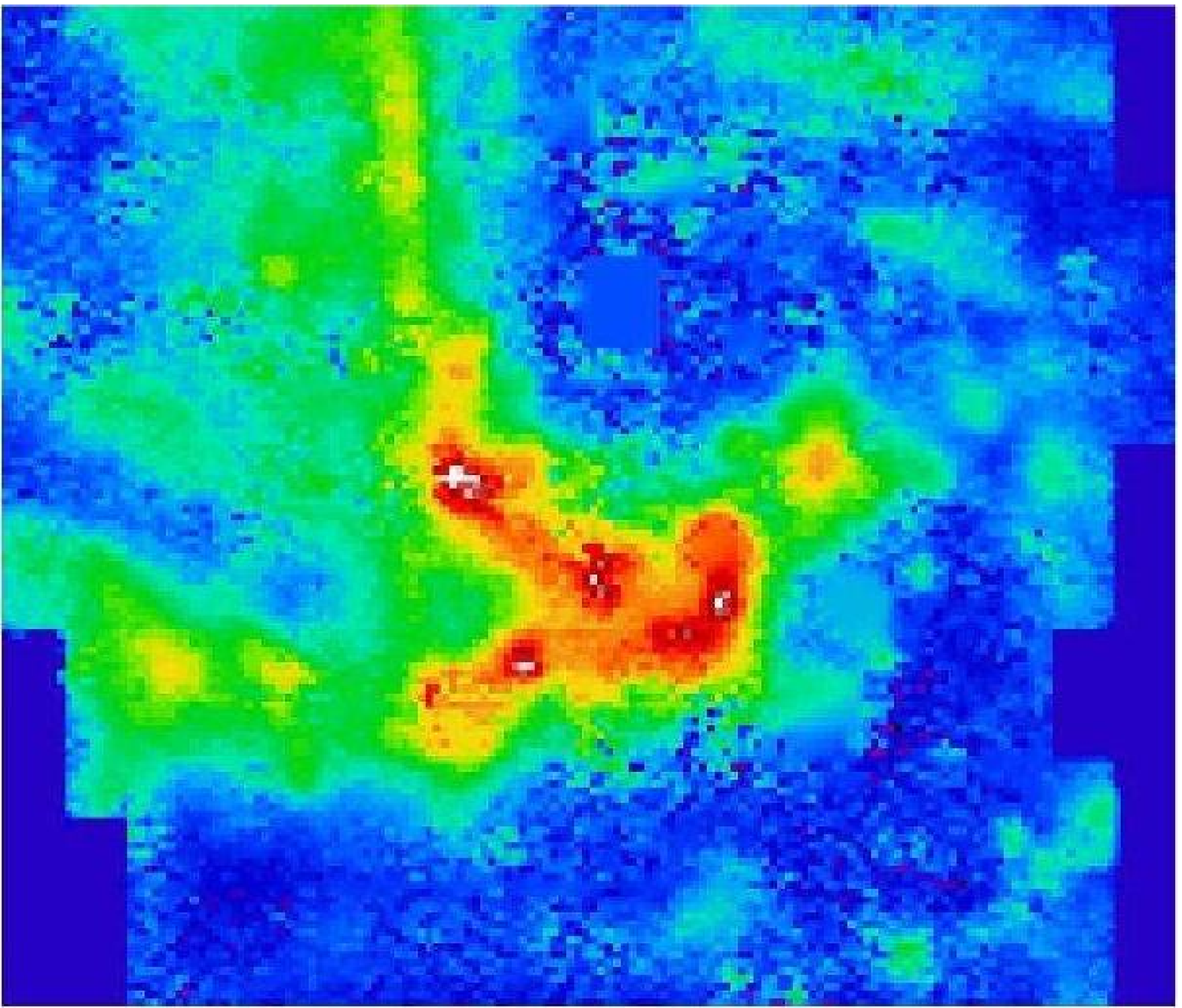}
\caption{Top: 2 cm continuum radio image from VLA. We subtracted Sgr~A* by fitting a Gaussian to it. The resolution is smoothed to $0.85''$.
Bottom: Brackett-$\gamma$ map derived from SINFONI. Areas with stellar lines are replaced by the neighboring gas emission.
 Some areas at the border were not mapped and are given zero flux. 
We fill the area around IRS7 that was not covered by the observations with the median flux of the surrounding.
Both maps show the same area, with 250 mas pixel$^{-1}$, the same resolution, and the same color scaling.
} 
\label{fig:_br_g_map}
\end{center}
\end{figure}

We chose to use a map at 2 cm because non-thermal emission, redder than the thermal emission of the minispiral, is less dominant at shorter wavelengths (\citet{Scoville_03}, on the other hand, used a 6 cm map). Naturally, some non-thermal contamination is expected at 2 cm as well. We control for this by testing whether there is a correlation between minispiral emission and 
extinction strength (see Section~\ref{sec:ext}). Sgr~A* is a clearly visible non-thermal source in the radio map, see Figure~\ref{fig:_areas}. However,
due to our high resolution it is well isolated from the minispiral. We are thus able to fit the source by a Gaussian and subtract it, smoothing the 
area around it. As a result, Sgr~A* is not visible in our final radio map (Figure~\ref{fig:_br_g_map}).

Two particularly problematic regions are IRS2 and IRS13, for which 
\citet{Roberts_96} and \citet{Shukla_04} derived significantly higher than average electron temperatures. 
In our analysis we simply ignore these regions, masking out IRS13 and IRS2. 

Difficulties can also arise, for interferometric observations, from the fact that only a finite range of spatial frequencies can be sampled. This has the consequence that structure larger than the fringe spacing of the shortest baseline is not measured, which can result in an underestimate of the flux, if the observed object is much larger than the largest angular scale sampled. 
For us, however, this is not a problem, since our map, due to the use of D configuration as mentioned above, 
we are able to detect structures of sizes up to 50'', while nearly all of the flux of the minispiral is concentrated within 50'' diameter (Figure~1 of \citet{Scoville_03}).

\subsection{ISO}
\begin{figure}
\begin{center}
\includegraphics[width=0.72 \columnwidth,angle=-90]{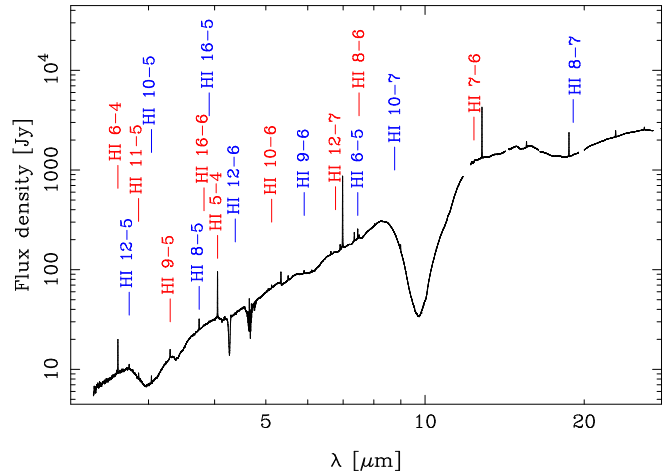}
\caption{ISO-SWS spectrum of the GC. The hydrogen lines used for the extinction measurements are marked by colored vertical lines.
} 
\label{fig:iso_spec}
\end{center}
\end{figure}

We use the ISO-SWS data of \citet{Lutz_99}. These data are reduced with the SWS interactive analysis system 
\citep{Wieprecht_98}. The data consist of a single aperture spectrum extending from 2.4 to 45 $\mu$m, shown in Figure \ref{fig:iso_spec}.

The field of view of SWS \citep{Gra_96} is approximately rectangular; it can be seen in Figure~\ref{fig:_areas}.
 We consider the deviations from rectangularity 
\citep{Beintema_03} in Section~\ref{sec:flux-cal}. 
In our observations the longer axis of the field of view is oriented north-south (P.A.$\,\approx 0^{\circ}$). 
The field of view is approximately $14'' \times 20''$, shortward of 12 $\mu$m, and   $14'' \times 27''$ between 12 and 28~$\mu$m.
The observations are centered on Sgr~A*.

\subsection{SINFONI}
For shorter wavelengths (1.2 to 2.4 $\mu$m) 
we use spectra obtained with the integral field spectrometer SINFONI \citep{Eisenhauer_etal2003, Bonnet_03} at UT4 of the
VLT\footnote{We use SINFONI data from ESO programs 70.A-0029(A), 183.B-100(P) and 183.B-100(R).
}. We use the following data:  H+K-band (spectral resolution 1500), seeing limited cubes (FWHM$=$0.85$''$)  covering an area of about 37$''$ $\times$ 30'' around Sgr~A* from 2003 April 9 with a spatial sampling of  $250\,$mas pixel$^{-1}$ $\times$ $250\,$mas pixel$^{-1}$. 
J-band (spectral resolution 2000), seeing limited cubes (FWHM$=$0.6$''$-1$''$) of most areas of the minispiral obtained 2010 May 21 and July 6, with a spatial sampling of $125\,$mas pixel$^{-1}$ $\times$ $250\,$mas pixel$^{-1}$.
We apply the standard data reduction SPRED \citep{Abuter_06, Schreiber_04} for SINFONI data, including detector calibrations (such as bad pixel correction, flat-fielding,
 and distortion correction) and cube reconstruction. The wavelength scale is calibrated with emission line lamps and finetuned with atmospheric OH lines. 
The HK data is atmosphere corrected and kept in ADU count units. No conversion in energy is applied. 
 We flux calibrate the J-band data with standards which were observed directly after the science observations, as is described in more detail in the next section. The Brackett-$\gamma$ map, one of three linemaps derived from the SINFONI observations (described in the next Section), can be seen in Figure~\ref{fig:_br_g_map}.

\section{Construction of Line Maps and Flux Calibration}

\subsection{Line maps}
\label{sec:line_maps}

In this section we describe the construction of Brackett-$\gamma$, Brackett-$\zeta$, and Paschen-$\beta$ line maps from the SINFONI data, see Table~\ref{tab:obs}.
In order to do this, we integrate over the channels which contain the nebular emission and subtract the average of the adjacent spectral channels on both sides
 as background. The line maps contain the following types of artifacts, unrelated to the minispiral emission:

\begin{itemize}
\item Bad pixels, in the sense of a large deviation of a few pixels from their neighbors. We identify them mostly manually.

\item Gaps between cubes, when not well aligned

\item Emission and absorption line stars
\end{itemize}

We interpolate these artifacts, treating them like bad pixels in the data reduction.
We also replace the hole around IRS7 in our map with the median flux of the surrounding for obtaining line maps, see the Brackett $\gamma$ map in
 Figure~\ref{fig:_br_g_map}.

\subsection{Flux calibration}
\label{sec:flux-cal}

We flux calibrate the Brackett-$\gamma$ map with a NACO Ks-image of the GC from April 29, 2006. On this day, a standard  star was observed with a zero-point uncertainty of 0.06 mag. 
In order to translate this calibration to the SINFONI data, we extract a mock Ks-image (F$_{\mathrm{SIN\,Ks}}$) from the SINFONI cube. This is done by multiplying the atmosphere corrected SINFONI data slice by slice with factors which represent the product of
atmospheric and NACO Ks-filter transmission\footnote{We use the NACO Ks-filter transmission from the NACO web site: http://www.eso.org/sci/facilities/paranal/instruments/naco
/inst/filters.html
} for every slice. The factors are scaled such that their integral is one. In this way we calibrate a single SINFONI slice 
at the isophotal wavelength of the Ks-band.

We measure the full flux in the NACO 
image and in  F$_{\mathrm{SIN\,Ks}}$ over the area covered by the SINFONI data. We estimate the uncertainty of the cross-calibration by dividing the two images in nine parts  and measuring in each part the count ratio. We obtain an  uncertainty of 0.03 mag from the standard
deviation of the nine count ratios. 

In total we use the following factor for calibration of the Brackett $\gamma$ line map:
\begin{eqnarray} 
f & = & \frac {F_{\lambda\,\mathrm{Ks}}\,\lambda_{\mathrm{iso}\,\mathrm{Ks}} }{ R \, \lambda_{\mathrm{Br}\,\gamma}} \, 10^{-0.4 \, \mathrm{ZP}} \\
& = & \frac {F_{\lambda\, \mathrm{Ks}}\,\lambda_{\mathrm{iso}\,\mathrm{Ks}} }{ R \, \lambda_{\mathrm{Br}\,\gamma}} \, 10^{-0.4 \, \left( \mathrm{ZP}_{\mathrm{NACO}}+2.5\, \log  \frac{F_{\mathrm{SIN\,Ks}}}{F_{\mathrm{NACO\,Ks}}} \right) }
\end{eqnarray}
Here, F$_{\lambda\,Ks}$ is the Ks calibration of Vega \citep{Tokunaga_05}, $R$ the number of slices per $\mu$m and ZP the zero point in magnitudes of the SINFONI data.
We assume an error of 0.03 mag for F$_{\lambda\,Ks}$.
Because we use ADU spectral data, the amount of energy per ADU depends on wavelength. Since the calibration is at the isophotal wavelength \citep{Tokunaga_05},  
we multiply  F$_{\lambda\,Ks}$ by $ \lambda_{\mathrm{iso}\,\mathrm{Ks}}/ \lambda_{\mathrm{Br}\,\gamma}$. 
Compared to using a standard star, our calibration has the advantage that we use the same NACO data as  \citet{Schoedel_09b} for calibration.
 In this way we reduce the uncertainty when comparing our results with theirs, see Section~\ref{sec:phot_dis}.

We calibrate Brackett-$\zeta$ in a similar way to Brackett-$\gamma$, using the integrated flux of an H-band NACO image from April 29, 2006, which is compared to a H-band image constructed from the SINFONI cube.
The errors of the calibration are 0.05 mag (NACO zero point), 0.04 mag (cross-calibration) and 0.03 mag uncertainty from the conversion to energy (F$_{\lambda\,H}$).

\begin{figure}
\begin{center}
 \includegraphics[width=0.99 \columnwidth]{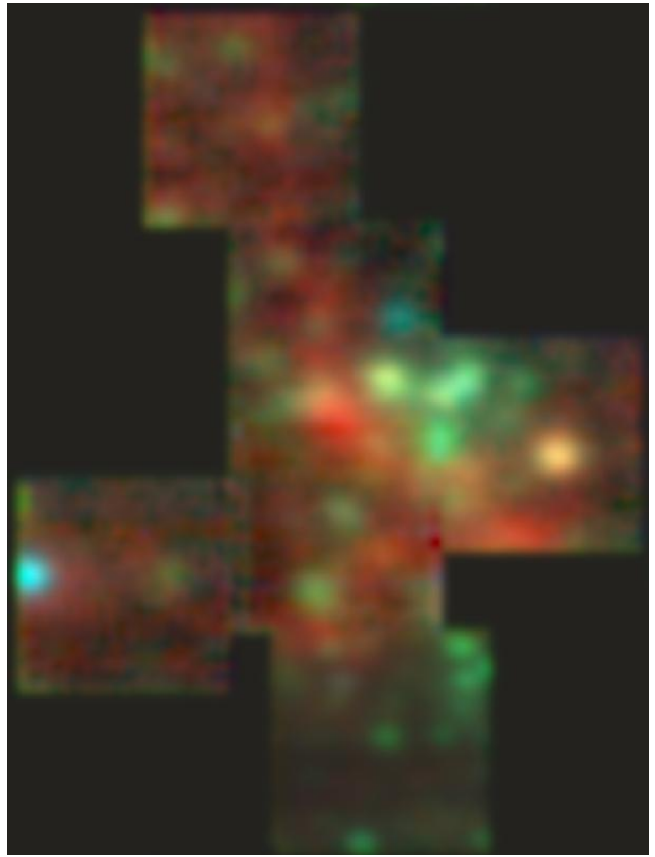}
\caption{Color image from J-band SINFONI data: blue 1.18 $\mu$m image, green 1.3 $\mu$m image, red Paschen-$\beta$ line map.
} 
\label{fig:_pa_b_map}
\end{center}
\end{figure}

We use data cubes flux calibrated with standard stars for Paschen-$\beta$ (see Figure~\ref{fig:_pa_b_map}), because a flux calibrated GC image in the J-band was not readily available. 
By comparing the fluxes of different standard stars from the two nights in April and July 2010 in which the data was obtained, we validate that the calibration is stable, with only 0.034 mag rms scatter and 0.005 mag bias over the J-band spectrum.
We test the calibration accuracy further by comparing adjacent areas in the data from April and July. 
In doing so, we do not find signs of significant discontinuities in the Brackett-$\gamma$ to Paschen-$\beta$ ratio at the border between the two areas. We estimate the systematic calibration error to be the same as for Brackett-$\gamma$: 0.075 mag.

For the ISO data we use the absolute calibration of \citet{Lutz_96}. The calibration error dominates the total error for bright lines. 
It is 10\% below 4.06 $\mu$m and 20\% above. 

Since the emission of the minispiral is not homogeneous, it is necessary to compute the NIR and radio comparison fluxes from the same area as the ISO-SWS beam profile.
Because of this we multiply the ISO-SWS beam profile \citep{Beintema_03} in every ISO band with the Brackett-$\gamma$ line map  and obtain the
 Brackett-$\gamma$ flux within the ISO-SWS beam profile. Since the Brackett-$\gamma$ map does not cover the full ISO beam, 
we also calculate the radio flux contained in the ISO-SWS beam profile in the same way. 
Over the Brackett-$\gamma$ field the difference in the beam correction is 3 \% between using radio and using Brackett-$\gamma$. 
This difference could be caused by non-thermal emission. Therefore, we use the radio data only for calculating how much flux is missed due to the smaller field of the Brackett-$\gamma$ data. In total, we find that the flux is about 7\% smaller below 12 $\mu$m and about 6\% higher above 12 $\mu$m compared to a sharp $14'' \times 20''$ field.
The sudden change at 12 $\mu$m is an artifact caused by the field size increase to  $14'' \times 27''$ at this wavelength. 
%Linesmaps/check.apertur2.dpuser

\section{Deriving extinction from hydrogen lines}
\label{sec:dev_ext}

The ratios between different hydrogen recombination lines in HII regions depend only weakly on the local physics \citep{Hummer_87}.
Hence, it is possible to derive the relative extinction ($A$) between two lines $a$ and $b$ by comparing the observed flux ratio 
($ F(a)_{\mathrm{obs}}/F(b)_{\mathrm{obs}}$) with the 
expected (extinction-free) flux ratio ($F(a)_{\mathrm{exp}}/F(b)_{\mathrm{exp}} $):

\begin{equation}
A_{a-b}=-2.5 \times \log\left(\frac{F(a)_{\mathrm{obs}}/F(b)_{\mathrm{obs}}}{F(a)_{\mathrm{exp}}/F(b)_{\mathrm{exp}}}\right)
\label{form:prin}
\end{equation}

Using infrared data only, and without making further assumptions about the shape of the extinction curve, it is only possible to obtain the 
relative extinction. In order to  obtain absolute extinction values it is necessary to use a wavelength for $b$ at which the extinction is 
known independently, or negligible.
Then it is possible to calculate the expected flux at $a$ from the observed flux at $b$ using the known flux ratios for emission from a 
gaseous nebula \citep{Baker_38}. We use their Case B, for which the nebula is opaque to Lyman radiation but transparent to all other radiation
(in Case A the nebula is also transparent to Lyman radiation.):
\begin{equation} F(a)_{\mathrm{exp}}=c\times F(b)_{\mathrm{obs}} \end{equation}
$c$ follows from the Case B calculation and depends on the radio frequency and 
infrared line used, as well as on the electron temperature, see also Section~\ref{sec:t_e_calc}.  

In our analysis, we use radio data for $b$ and infrared lines in place of $a$. The formula for absolute extinction is thus:

\begin{equation}
A_{\mathrm{IR}}=-2.5 \times \log \left( \frac {F(\mathrm{IR})_{\mathrm{obs}}} {c \times F(\mathrm{radio})_{\mathrm{obs}}}\right)
\label{form:prin2}
\end{equation}

We use the Case B line ratios in \citet{Hummer_87} for the $F_{\mathrm{obs}}$ ratios in Formula~\ref{form:prin}. The unextincted ratios of hydrogen emission originating
from different atomic levels depend on the radiation state (like Case B) \citep{Lutz_99}. 
As a first test of the validity of Case B we use the hydrogen line in the ISO-SWS spectrum at 
7.50 $\mu$m, which is a blend of the lines 6-8 and 8-11. After accounting for the blending the 7.50 $\mu$m extinction differs only by 0.05 mag
from the extinction for the 6-5 line at 7.46 $\mu$m, assuming Case B. 
This implies that Case B is indeed valid for the Galactic Center \citep{Lutz_99}.

\subsection{Electron Temperature}
\label{sec:t_e_calc}

The physical conditions of the plasma, in particular the electron temperature (T$_e$), have an important influence on the extinction free flux ratios of the IR line fluxes to 
the radio flux . For example the dependence of Paschen $\alpha$ to radio continuum (free-free) emission ($S_{\mathrm{ff}}$)  is
 $F_{\mathrm{Pa}\,\alpha}/S_{\mathrm{ff}}\propto T_e^{-0.52}$ \citep{Scoville_03}.
Accordingly, it is necessary to know T$_{e}$ for deriving the absolute extinction.

The electron temperature has been derived at  H92$\alpha$ (8.3 GHz) by \citet{Roberts_93} and at H41$\alpha$ (92 GHz) by \citet{Shukla_04}.
 Both obtain T$_e\approx7000$ K in most parts of the minispiral. The consistency of both measurements shows that all conditions, thus also Case~B,  \citep{Roberts_91} for
 deriving T$_e$ are fulfilled, even at the smaller and more problematic frequency of 8.3 GHz.
In this work, we take the measurement by \citet{Roberts_93} of T$_e=7000 \pm 500 $ K, since it has the highest SNR, however we correct it to account for the He$^+$ fraction (see below).

According to \citet{Roberts_93}, there is no significant spatial variation in the electron temperature.
Thus, we think it is justified to simply assume a constant electron temperature. Even if the electron temperature varies, the variation does not matter as long as the value used
 is equal to the flux weighted average of T$_e$ of the area over which we integrate.

The absolute value of T$_e$ also depends on the He$^+$ fraction: $Y^+=He^+/H^+$. \citet{Roberts_93} derived a 2$\sigma$ limit  for Y$^+$ of 3\%, and 
measured  $Y^+=5 \% \pm 2$\% in another area. \citet{Krabbe_91} derived  $Y^+ \approx 4$ \%  from the He I line at 2.06 $\mu$m. 
We obtain  $Y^+=2 \% \pm 0.7 $ \% using the same line. Considering all information, we assume $Y^+=3 \% \pm 1$ \% for the calculation of T$_e$. 

The electron temperature depends, according to \citet {Roberts_96} and \citet{Roelfsema_92}, in the following way on Y$^+$: 
\begin{equation}
T_e=\left(T_{e\, \mathrm{raw}} \frac {1} {1+Y^+}\right)^{0.87}
\label{form_t_eff}
\end{equation}

T$_{e\, \mathrm{raw}}$ is the T$_e$ obtained assuming $Y^+=0$.
Since \citet{Roberts_93} used $Y^+=0$, they calculated T$_{e\, \mathrm{raw}}$.
We therefore calculate the real T$_e$ from the T$_{e\, \mathrm{raw}}$  of \citet{Roberts_93} using our value of $Y^+=3$\%. We obtain T$_e= 6800 \pm 500$ K, and use this value hereafter. The error in the electron temperature results in an absolute extinction error of 0.043 mag.
 In principle, it is also necessary to know the electron density of the emitting plasma to calculate the absolute extinction from infrared lines and radio emission \citep{Hummer_87}. We use an electron density of 10$^4 \mathrm{cm}^{-3}$  \citep{Shukla_04},
but the results are not very sensitive to the actual value used \citep{Hummer_87}.

\subsection{Extinction calculation}
\label{sec:ext}

We first compute the Brackett-$\gamma$ extinction from direct comparison of the Brackett-$\gamma$ map with the radio data (Figure~\ref{fig:_br_g_map}), 
following Formula~\ref{form:prin2}. 
We then calculate all other extinction values with respect to the Brackett-$\gamma$ data.
This method  has the advantage that we can select the fiducial area by comparing with the high SNR Brackett-$\gamma$ data (Table~\ref{tab:obs}). 
This is done by calculating the extinction relative to Brackett-$\gamma$ extinction following Formula~\ref{form:prin}. 
The Brackett-$\gamma$ extinction is then added to this value to obtain the absolute extinction.

To calculate the expected flux ratio we convert Formula~3 of \citet{Scoville_03} to Brackett-$\gamma$, our continuum frequency of 15 GHz, and mJy, using 
the relative line strength in \citet{Hummer_87}:

\begin{equation} \left( \frac{F_{\mathrm{Br}\,\gamma}}{S_{ff} }\right)_{\mathrm{exp}}= \left( 1.327 \pm 0.052 \right) \, 10^{-11} \mathrm{erg}\,\mathrm{s}^{-1}\,\,\mathrm{cm}^{-2}\,\,\mathrm{mJy}^{-1}
\label{form:rad_brag} 
\end {equation}

Inserting this expected ratio and the measured Brackett-$\gamma$ to radio ratio in Formula~\ref{form:prin2}
 we obtain a Brackett-$\gamma$ extinction map.

\begin{figure}
\begin{center}
\includegraphics[width=0.72 \columnwidth,angle=-90]{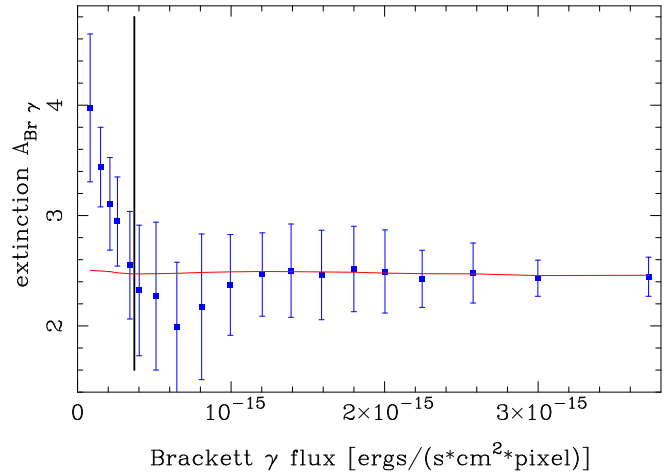}
\caption{Extinction in the central 14'' $\times$ 20'', binned in Brackett-$\gamma$ flux. The squares and error bars mark the median extinction per flux bin. 
(The error is based on the median deviation, scaled to a Gaussian $1\sigma$ error by multiplying it with 1.483.) While for small fluxes the extinction 
is anticorrelated with the flux, there is no such correlation at fluxes larger
 than the vertical line at $3.7 \times 10^{-16}\, \mathrm{ergs}\,\mathrm{s}^{-1}\,\mathrm{cm}^{-2} \,\mathrm{pixel}^{-1}$. The red line is the integrated extinction using 
all pixels with a flux larger than or equal than the flux in the given bin. There is nearly no variation. 
} 
\label{fig:_ex_flux}
\end{center}
\end{figure}

In order to estimate the impact of non-thermal emission, we measure the resulting extinction in Brackett-$\gamma$ flux bins, see Figure~\ref{fig:_ex_flux}.
For most line fluxes, the extinction is approximately constant.
However for line fluxes smaller than   $3.7 \times 10^{-16}\, \mathrm{ergs}\,\mathrm{s}^{-1}\,\mathrm{cm}^{-2} \,\mathrm{pixel}^{-1}$ the extinction is anticorrelated with the Brackett-$\gamma$ flux.
Most likely the anticorrelation is caused by non-thermal emission of Sgr~A~East which becomes more important in fainter regions of the minispiral.
Because of this, we mask out areas with fluxes smaller than $3.7 \times 10^{-16}\,\mathrm{ergs}\,\mathrm{s}^{-1}\,\mathrm{cm}^{-2} \,\mathrm{pixel}^{-1}$.
The choice of this flux cut is somewhat arbitrary. However, since we only use the integrated flux of the whole ISO beam, which is dominated by the brighter fluxes, the uncertainty of the flux cut
only introduces an extinction error of 0.03 mag. 
In addition, we mask out IRS2 and IRS13 (see Section~\ref{sec:radio}), as well as IRS16NE and IRS16C because of remaining stellar emission line flux there.

\begin{deluxetable*}{lll} 
\tabletypesize{\scriptsize}
\tablecolumns{3}
\tablewidth{0pc}
\tablecaption{Combinations of lines or continuum used for deriving extinctions \label{tab:obs}}
\tablehead{Line A & Line B & area used ('good' area) \\
}
\startdata
2 cm continuum & Brackett-$\gamma$  & 'ISO beam', excluding: low Br-$\gamma$ flux, IRS16C, IRS16NE, IRS13, IRS2 \\
Brackett-$\gamma$ & Brackett-$\zeta$ & 'ISO beam', excluding: low Br-$\gamma$ flux, Br-$\gamma$/ Br-$\zeta$  outliers, IRS16C, IRS16NE, IRS13, IRS2 \\
Brackett-$\gamma$ & Paschen-$\beta$ & 'ISO beam' inside the J-band data, excluding: low Br-$\gamma$ flux, IRS16C, IRS16NE, IRS13, IRS2 \\
Brackett-$\gamma$ & ISO lines  & in each case the adequate ISO beam \\
\enddata
\end{deluxetable*}

The extinction in the areas defined by the above constraints is similar to the extinction map of \citet{Schoedel_09b}. However, there is a scatter of 
about 0.4 mag in our map, when comparing  pixel by pixel, due to the low SNR per pixel in Brackett-$\gamma$ and radio data. Because of this, our 
extinction map is not of direct use.
Nevertheless, the small SNR per pixel does not affect our absolute measurement, because we use the flux from a total of 2854 pixels for deriving 
the Brackett-$\gamma$ extinction. 

\begin{figure}
\begin{center}
\includegraphics[width=0.99 \columnwidth,angle=-90]{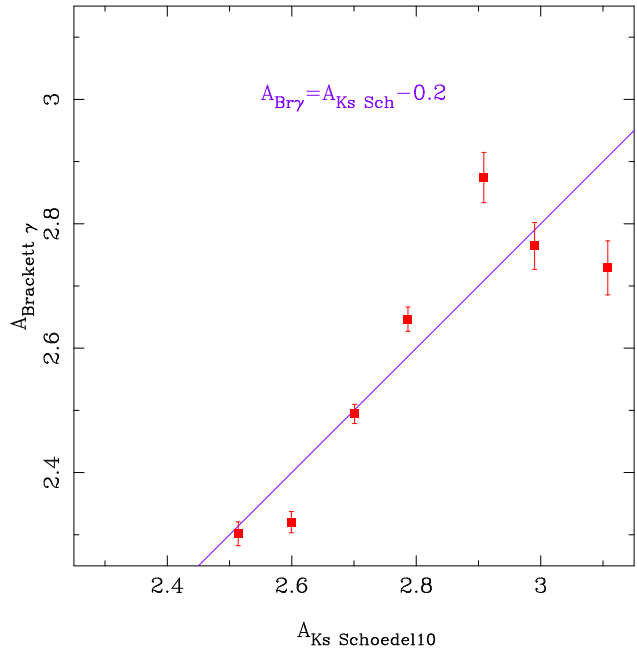}
\caption{Comparison of the extinction derived from our data and that of the Ks extinction map of \citet{Schoedel_09b}. We bin our data according to the map of \citet{Schoedel_09b}
and measure the extinction in every bin. The error bars show the expected error due to the single pixel SNR.  The rms scatter in the plot is 0.1 mag, exceeding the expectation.
} 
\label{fig:_ex_diff}
\end{center}
\end{figure}

For a comparison with \citet{Schoedel_09b} we add up the Brackett-$\gamma$ and radio fluxes within 0.1 mag bins. We use the extinction map of 
\citet{Schoedel_09b} to bin according to extinction. We then calculate the extinction in every bin from the total fluxes therein by means of Formula~\ref{form:rad_brag} (see Figure~\ref{fig:_ex_diff}).
 The differences between the extinction values from \citet{Schoedel_09b} and this work have an rms scatter of 0.10 mag.
 This scatter is bigger than would be expected from the SNR of both maps, but systematic problems, such as residual anisoplanatism in 
\citet{Schoedel_09b}, or T$_e$ variations for our 
map, could cause increased scatter. We assume that half of the scatter is caused by our data and add therefore 0.07 mag as additional extinction
 error. This estimate is conservative, because many problems
should average out over the full area.

Our aim is to measure the wavelength dependent extinction curve in one line of sight. Spatially variable extinction can however cause deviation of the line of sight extinction curve, if different measurement areas are used at different wavelengths. 
 In addition, for comparison with \citet{Schoedel_09b}, it is also necessary to correct for the difference between
the average extinction of this map and the Brackett-$\gamma$ flux weighted extinction in the same area.
In order to take both effects into account, we calculate the average extinction of the extinction map of \citet{Schoedel_09b} within the region of the ISO beam, and the flux weighted extinction within the ISO beam for the different line maps.  The difference in extinction is small for all lines. We find the biggest deviation is -0.03 mag for Paschen-$\beta$, for which line the smallest area was covered, see Figure~\ref{fig:_pa_b_map}. We apply these corrections to our raw extinction values.

There is an additional bias in integrated extinction measurements that can cause apparent flattening of the extinction law, and is especially important
 for integrated galaxy SEDs \citep{Calzetti_00}. This bias arises because the measurement of the extinction is flux-weighted: this flux being the 
observed (extincted) flux. This creates a problem, because if the extinction is spatially variable, the extincted flux will be brighter in 
low-extinction subregions of the image.
Thus more weight will be given to these bright, low-extinction sub-regions of the image. A similar effect of very inhomogeneously distributed extinction in the case 
of galaxy integrated measurements (mixed case) is visible in the top left panel of Figure 2 in \citet{Calzetti_01}.
The effect of this bias is negligible for negligible extinction (i.e. at 
long wavelengths), but increases towards shorter wavelength together with the extinction, producing an apparent flattening of the measured extinction
 law. 
We test and correct for this bias in our measurements, which is however small (up to 0.08 mag, see Appendix \ref{sec_ext_bias}).

In the H-band, the hydrogen lines are weaker, both intrinsically as well as being subject to higher extinction. The atmospheric OH-lines are also stronger in H-band.
Because of this, only Brackett-$\zeta$ at 1.736 $\mu$m is easily detectable, although even this line is polluted by an atmospheric OH-line,  
despite sky subtraction. The structure of this remaining OH-line emission/absorption follows the cubes which are combined to the mosaic and is caused by the time variability 
of the sky. The sky has to be observed offset from the GC in space and thus also in time due to the brightness of the GC. 
To correct the sky remnants, we derive OH-line strength maps from the next two strongest OH-lines at longer wavelengths, and subtract the average of the two maps from the
 raw Brackett-$\zeta$ map. We scale the subtraction such that any observable structure due to individual cubes in the combined cube vanishes. We use the uncertainty of the scaling 
factor and the difference between the two OH-line maps to estimate the error. 
In order to further exclude unphysical outliers we do not use pixels which deviate by more than 2.5~$\sigma$ from the median Brackett-$\gamma$/ Brackett-$\zeta$ flux ratio. 
Integrating over the useful area of both lines we obtain $F_{\mathrm{Br}\,\gamma}/F_{\mathrm{Br}\, \zeta}=16.1 \pm 1.3$. 
% Linesmaps/check.oh.problem.dpuser
The  error due to the OH line subtraction and the pixel selection uncertainty adds an error of 0.081 mag to the extinction for Brackett-$\zeta$. 

For Paschen-$\beta$ we used the smaller area (Figure~\ref{fig:_pa_b_map}) within its overlap with the central 14''$\times$20''.
We use the hydrogen lines detected in the ISO-SWS spectrum (Figure~\ref{fig:iso_spec}) to derive extinction values at the corresponding wavelengths. For the ISO data, which is not spatially resolved, stellar emission could be a problem. We test this by comparing the total Brackett $\gamma$ from our map with that computed upon declaring stars visible in the linemap to be ``bad pixels''. The flux difference between the two cases is only 0.04 mag.
This contribution, compared to the extinction error of the ISO data of at least 0.14 mag, is therefore not relevant. 
Since the intrinsic (non extincted) line 
ratios follow Case B (Section~\ref{sec:ext}), the intrinsic ratios are well known and do not cause an additional error apart of the T$_e$ uncertainty.

\subsection{Extinction derived}
\label{sec:ex_pham}

The extinction derived in this analysis is presented in Table~\ref{tab:lin_ext} and Figure~\ref{fig:feat_ext}. It decreases with wavelength, following a linear relation in this log-log plot
 between 1.2 and 2.8 $\mu$m.  At longer wavelengths, the extinction curve changes shape: it contains more bumpy features, and is higher than expected from linear extrapolation.
For the purpose of investigating the extinction, it is useful to separate the two regimes. Therefore, we define the near infrared (NIR) as the wavelength regime from 1.2 to 2.8 $\mu$m and the mid 
infrared (MIR) as the wavelength regime from 2.8 to 26 $\mu$m. 
The linear relation within the NIR in the log-log plot implies a power-law: 
\begin{equation} A_b=A_a \times(\lambda_a/\lambda_b)^{\alpha}
\label{form:pow}
\end{equation}

\begin{deluxetable}{ll} 
\tabletypesize{\scriptsize}
\tablecolumns{2}
\tablewidth{18pc}
\tablecaption{Average hydrogen line extinction values towards the central 14''$\times$ 20'' of the GC\label{tab:lin_ext}}
\tablehead{ $\lambda$ [$\mu$m] & extinction 
}
\startdata
1.282 & 7.91 $\pm$ 0.11 \\
1.736 & 4.30 $\pm$ 0.13 \\
2.166 & 2.49 $\pm$ 0.11 \\
2.625 & 1.83 $\pm$ 0.13 \\
2.758 & 1.51 $\pm$ 0.19 \\
2.873 & 1.84 $\pm$ 0.19 \\
3.039 & 2.07 $\pm$ 0.19 \\
3.297 & 1.66 $\pm$ 0.15 \\
3.74 & 1.19 $\pm$ 0.14 \\
3.819 & 1.19 $\pm$ 0.31 \\
3.907 & 1.09 $\pm$ 0.3 \\
4.052 & 1.01 $\pm$ 0.13 \\
4.376 & 1.09 $\pm$ 0.29 \\
5.128 & 0.99 $\pm$ 0.24 \\
5.908 & 1.04 $\pm$ 0.24 \\
6.772 & 0.84 $\pm$ 0.29 \\
7.459 & 0.81 $\pm$ 0.23 \\
7.502 & 0.79 $\pm$ 0.24 \\
8.76 & 2.04 $\pm$ 0.34 \\
12.371 & 1.34 $\pm$ 0.32 \\
19.062 & 1.34 $\pm$ 0.5 \\

\enddata
% \tablecomments{NIR extinction slope measurements of diffuse ISM extinction since 2005. 
 %}
\end{deluxetable}

\begin{deluxetable}{llll} 
\tabletypesize{\scriptsize}
\tablecolumns{4}
\tablewidth{0pc}
\tablecaption{NIR infrared power-law extinction fit parameters \label{tab:nir_law}}
\tablehead{Used range [$\mu$m] & $A_{Br \gamma}$ & slope $\alpha$  & $\chi^2$/d.o.f
}
\startdata
1.282 to 2.166 & $2.60 \pm 0.11$ & $-2.13 \pm 0.08$ & 2.05/1 \\
2.166 to 2.758 & $2.50 \pm 0.11$ & $-1.76 \pm 0.39 $ & 0.54/1 \\
\hline
1.282 to 2.758 & $2.62 \pm 0.11$ & $-2.11 \pm 0.06$ & 2.70/3 \\
\enddata
\end{deluxetable}

For the five NIR lines we obtain $\alpha=-2.11 \pm 0.06$ and $A_{\mathrm{Br}\, \gamma}=2.62\pm 0.11 $, see Table~\ref{tab:nir_law}.
The $\chi^2$ of the fit is 2.70 given 3 degrees of freedom. Hence, a power-law is a good description of the NIR extinction. 
Since the extinction is obviously grayer beyond 3.7 $\mu$m, we test if there is any indication for flattening
 within the NIR by fitting the red and blue part of the 5 NIR lines separately, see Table~\ref{tab:nir_law}. The slope of the red part, with $\alpha=-1.76\pm 0.39$, is only 0.9 $\sigma$ flatter than
 the blue part. Because this change in slope is not significant, we use the power-law obtained from all 5 NIR lines for deriving the broadband extinction values (Appendix~\ref{sec:filt_ext}). 
The extinction between 3.7 and 8 $\mu$m is fitted by the following power-law: 
$$A(\lambda)=\left(1.01 \pm 0.08\right) \times (\lambda/4.9\mu\mathrm{m})^{-0.47 \pm 0.29} $$
Thus, the MIR  extinction is grayer than the NIR extinction.

\section{Discussion}

\subsection {Comparison with literature for the GC}
\label{sec:oth_GC}
We now compare and discuss the extinction curve obtained in this work with the literature.

First of all, we compare with \citet{Rieke_85}, who used mainly data from GC stars to derive the infrared extinction law up to 13 $\mu$m. 
Their finding, however, a single power-law of slope $\approx-1.54 $ from J- to M-band, is in contrast with our results. The differences probably 
lie (i) in their assumption of an universal IR extinction law, whereby R$_V$ is determined only indirectly, and (ii) variability of the stars used. Firstly, for absolute calibration of 
the extinction law they used $o$ Sco, which is also detectable in the optical, together with the relation $E_{V-K}/E_{B-V}=$2.744 from 
\citet{Schultz_75} and \citet{Sneden_78}. However, this relation was obtained outside of the GC, making their extinction law vulnerable
 to line of sight variations, see Section~\ref{sec:nir-dis}. Secondly, they employ the $E_{V-M}/E_{B-V}$ of the GC and the bolometric
 luminosity of IRS7 to estimate R$_V=A_V/E_{B-V}=3.09$. In the process they use measurements of three stars in the M-band, which is dominated by 
atmospheric emission, to estimate their lower limit on R$_V$. 
Not only are all of the GC stars used by \citet{Rieke_85} supergiants, which are in general  variable, but variability has, in particular, been reported for 
IRS7 \citep{Blum_96}. 
Their indirectly determined R$_V$
could introduce a systematic error in their extinction measurement. For example, if we reduce R$_V$  to 2.98 in their calculation, 
we obtain from their data $\alpha=-2.04$ between J and K. This is compatible with our $\alpha$.

In a later work, \citet{Rieke_99} updated \citet{Rieke_85} using NICMOS data (F110M, F145M, F160W, F222M), again of stars in the GC region. 
They used spectra of the IRS16 stars \citep{Tamblyn_96} for deriving absolute extinctions. The extinction curve they 
found is flatter in 'HK' ($\alpha=-1.58$) than in 'JH' ($\alpha=-1.95$), in contrast to our extinction law (Section~\ref{sec:ex_pham})
 and to most other publications which derive a constant slope in the NIR \citep[e.g,][]{Draine_89}. In order to test whether
 the NICMOS data generally contradict our NIR extinction law (Section~\ref{sec:ex_pham}) we use stars which have NICMOS magnitudes published in \citet{Maillard_03}.
We use only stars which have been spectroscopically identified as early-type \citep{Paumard_06,Bartko_09}. Early-type stars have colors $\approx 0$ in the NIR. We exclude 
IRS13E2 and IRS13E4.0 from the analysis because they are dusty \citep{Fritz_10}.
Using the rescaled versions of the extinction map  and the Ks-band magnitudes of \citet{Schoedel_09b}, the NICMOS data are consistent with a single power-law with a slope 
of $\alpha =-2.16 \pm 0.08$. 
Thus, the NICMOS data do not contradict our result of $\alpha=-2.11 \pm 0.06$ (Section~\ref{sec:ex_pham}).
%We guess that the unusual NIR extinction law in \citet{Rieke_99} could be a consequence of trying to combine regions with different extinction laws.
% NICMOS /extlaw/schdata/plot.nic3.dpuser  

In another study, \citet{Viehmann_05} performed seeing-limited ISAAC L- and M-band photometry. Assuming stellar colors they derived a flat extinction slope from L to M: $A_M/A_L= 0.966 \pm 0.05 $. 
From our NACO  broadband extinction values  (Appendix~\ref{sec:filt_ext}) we obtain a consistent value: $A_M/A_L=0.88 \pm 0.23$. 
 However, these results are not directly comparable, because \citet{Viehmann_05} used a narrow M-band with a FWHM$=0.10\,\mu$m at 4.66 $\mu$m.
 At this wavelength, a CO absorption feature is visible in the 
ISO spectra \citep{Lutz_96, Moneti_01}, see Figure~\ref{fig:feat_ext}.

\citet{Scoville_03} obtained an absolute extinction map of the GC for Paschen-$\alpha$. To do this, they used  NICMOS narrowband imaging at Paschen-$\alpha$ (1.87 $\mu$m) and at 1.90 $\mu$m to construct the line map. They took a VLA continuum map at 6 cm to use as extinction-free data. 
Using their extinction map, we compute an average extinction of A$_{\mathrm{Pa}\,\alpha}=3.54$ integrated over the ISO beam. 
This value is consistent with our value of A$_{\mathrm{Pa}\,\alpha}=3.56 \pm 0.11$, obtained from the best-fit power-law. 

\citet{Schoedel_09b} measured the extinction towards the GC using NACO data in H-, Ks- and L'-band. 
By comparing the magnitude of the peak of the luminosity function (the red clump) with its expected magnitude, 
they obtained the total light modulus of the GC. They then used a distance to the GC of 8.03 $\pm$ 0.15 kpc, derived from different works,
 to obtain the extinction: A$_{H}~=~4.48 \pm 0.13$, A$_{Ks}~=~2.54 \pm 0.12$ and A$_{L'}~=~1.27 \pm 0.18$.

\citet{Schoedel_09b} also derived the selective extinction $E_{H-Ks}=(A_{H}-A_{Ks})/A_{Ks}$ from the red clump extinction values. 
They then used $E_{H-Ks}$ for obtaining a Ks-band extinction map from the observed stellar colors. Surprisingly, the average of the map is, with $A_{Ks}=2.70$, larger than the extinction obtained from the red clump.
It can be seen from  Table~A.2 in \citet{Schoedel_09b} that most individual stars have larger values of H-Ks than the stellar population of the red clump in the luminosity function, the reason for which is not clear.  

We do not know which of the two \citet{Schoedel_09b} extinction values should be preferred. Because of this, we use the average and enlarge the error by adding 
$\sqrt{1/2}\times(A_{Ks\,1}-A_{Ks\,2})=0.11$ mag. We obtain $A_{Ks}=2.62 \pm 0.16 $ for the full field of \citet{Schoedel_09b}. 
Their extinction map has
 an average of A$_{Ks}~=~2.70$ over the full map. However, the average is A$_{Ks}~=$~2.68 over the ISO field of view. 
Therefore, we correct the average by multiplying with $2.68/2.70$ and obtain $A_{Ks}~=~2.60~\pm~0.16$ as a final value
for comparison with our extinction. 

\begin{deluxetable}{lll} 
\tabletypesize{\scriptsize}
\tablecolumns{3}
\tablewidth{0pc}
\tablecaption{Error sources for the extinction \label{tab:err}}
\tablehead{Error source & $\Delta$mag  & $\Delta$mag \\
 & for Brackett $\gamma$ & for other lines 
}
\startdata
$\bullet$ IR calibration error & & \\
and line SNR & 0.073 & 0.073 to 0.50\\ 
$\bullet$ electron Temperature & 0.043 & 0.043 \\
$\bullet$ selection of good pixels & & \\
in Brackett-$\gamma$ & 0.03 & 0.03\\
$\bullet$ scatter comparing & & \\
with  \citet{Schoedel_09b}  & 0.07 & 0.07\\
$\bullet$ selection of good pixels & & \\
in other lines & 0 & 0 to 0.088\\
\hline
total & 0.114 & 0.114 to 0.507 

\enddata

\end{deluxetable}

The issue of two possible extinction values in the Ks-band also extends to the other bands. 
We assume a linear scaling of the extinction values based on stellar colors with the extinction values derived from the red clump in each band. We again use the average of red clump extinction and the selective extinction.
We obtain $A_{H}~=~4.58\pm 0.24 $ and  $A_{L'}~=~1.30 \pm 0.19$ as final values for comparison with our extinction values. 

If we now compare with our broadband extinction values (see Appendix~\ref{sec:filt_ext}),
we obtain the following differences between our results and \citet{Schoedel_09b}: $A_{H\,\mathrm{lines}}-A_{H\,\mathrm{Sch}}=0.07 \pm 0.25 $,
$A_{Ks\,\mathrm{lines}}-A_{Ks\,\mathrm{Sch}}= 0.07 \pm 0.17 $ and  $A_{L'\,\mathrm{lines}}-A_{L'\,\mathrm{Sch}}=-0.10 \pm 0.23 $. 
For the calculation of the errors here, we exclude the zero-point errors in the H and Ks-band because both results use the same calibration data, and we also exclude the error due to R$_0$ because we use the differences to measure R$_0$, see Section~\ref{sec:phot_dis}. 
To summarize, our values are all consistent with the values of \citet{Schoedel_09b}.

In Table~\ref{tab:err} we present the various sources of error for our extinction values.
Our data are well fitted with a power-law in the NIR ($\chi^2/d.o.f=0.71$), see Section~\ref{sec:ex_pham} as expected from the established literature \citep{Cardelli_89}. An error in the radio data (such as an error on T$_e$) affects all values in the same way, such that its 
influence on the $\chi^2$ of the power-law fit is much smaller than the influence of independent errors in the NIR data. 
Because the power-law does not continue to the MIR till to negligible extinction in our data, it is not possible to obtain the absolute extinction independent of the radio data, as was done in \citet{Landini_84}.
 However, the value for T$_e$ which we use is consistent with both \citet{Roberts_93} and \citet{Shukla_04}. Therefore, it is unlikely that there is a relevant systematic error in the
 extinction due to the radio data.

\subsection {A photometric distance to the GC}
\label{sec:phot_dis}
The distance to the Galactic Center R$_0$ is used to derive distances to all but the closest regions of the Galaxy and is one of the fundamental parameters for building Milky Way models. 
Knowing R$_0$ is hence of general relevance.
Prior to this publication there was no reliable measurement for the GC extinction that was independent of R$_0$. Works that did derive R$_0$ photometrically \citep{Nishiyama_06,Groenewegen_08b,Dambis_09,Matsunaga_09} used stars in the bulge, not directly in the GC.
On the other hand, stars directly in the GC have been 
 used to derive R$_0$ from the dynamics of their stellar orbits around the SMBH 
\citep{Eisenhauer_03, Ghez_09,Gillessen_09}, and by using the statistical parallax of the population of late-type stars in the GC \citep{Genzel_00,Trippe_08}.

With our extinction measurement we can obtain a photometric R$_0$ from stars in the GC.
We combine our extinction measurement with the extinction measurement of \citet{Schoedel_09b}. These authors obtained the total luminosity modulus of the red clump stars in the GC. 
Because the luminosity modulus involves both the extinction and the distance modulus it was necessary for
\citet{Schoedel_09b} to assume a distance for calculating extinction values, for which they used R$_0=8.03 \pm 0.15 $ kpc. 
 Here, we can use the extinction differences between the two works, see Section~\ref{sec:oth_GC} in order to estimate R$_0$.
For the Ks-band, we obtain R$_0=7.78 \pm 0.63 $ kpc. From the H-band we obtain R$_0=7.78 \pm 0.95$ kpc, and from the L'-band  we obtain  R$_0=8.41 \pm 0.94$ kpc. The errors follow from the errors of the extinction differences 
between our extinction values and those of \citet{Schoedel_09b}, see Section~\ref{sec:oth_GC}.
We use the weighted average of all values as our final value:
 R$_0 = 7.94 \pm 0.65$ kpc. For the error we use the smallest single error, 
the relative error of Ks-band, because the errors are correlated between the filters.
% weighted distance: extlaw/schdata get.random.dpuser 

\citet{Nishiyama_06} used red clump stars in the inner bulge to derive a photometric distance and obtained R$_0=7.52 \pm 0.36$ kpc.
 Our value is about 0.6~$\sigma$ larger. Since in both works the same 
absolute red clump magnitude is used, the consistency indicates only that other uncertainties (like the extinction) are not larger than assumed. However, it is still possible that the 
absolute magnitude has a bigger error than assumed. 
Our result is also consistent with the review of \citet{Reid_93}, who determined R$_0=8.0 \pm 0.5$ kpc, and the recent review of \citet{Genzel_10}, 
who determined R$_0=8.15 \pm 0.14 \pm 0.35$ kpc from direct and indirect measurements, and R$_0=8.23 \pm 0.2 \pm 0.19$ kpc from only direct measurements. 

Conversely, assuming that the direct estimate of R$_0$ by \citet{Genzel_10} is correct we can test the red clump magnitude:
The red clump  in the GC has then a magnitude of M$_{Ks}=-1.59 \pm 0.13$. The value used by us and \citet{Schoedel_09b} is M$_{Ks}=-1.47$, and the
 two values are consistent, although the GC star formation history (\citet{Blum_02}, Pfuhl et al. in prep) and metallicity \citep{Cunha_07} was not 
modeled to obtain M$_{Ks}$.
Therefore, as expected according to \citet{Salaris_02}, M$_{Ks}$ of the red clump is therefore relatively independent of the star formation history 
and metallicity and thus is a reliable distance indicator.

\subsection{Spatial distribution of the extinction and the extinction towards Sgr~A*}
\label{sec:spat}

Our data do not have enough SNR to obtain a good extinction map. 
However, because our absolute values for the extinction are more accurate than the extinction of \citet{Schoedel_09b},
combining our absolute value and the extinction map of \citet{Schoedel_09b} is useful. 
For this, we adjust the extinction map of \citet{Schoedel_09b} such that in it the extinction 
 is the same as our interpolated A$_{\mathrm{Br}\,\gamma}$ of the same area. 
This means that we multiply the map of \citet{Schoedel_09b} by 0.976.

From the adjusted map we use  A$_{Br\,\gamma}$ towards Sgr~A* to derive broadband extinctions (Appendix~\ref{sec:filt_ext}) of: 
$A_{H}=4.21 \pm 0.10 $, $ A_{Ks}=2.42 \pm 0.10$ and $A_{L'}=1.09 \pm 0.13$. 
Effectively, this is the extinction to the stars with r$\leq 0.5$'' around Sgr~A* due to the procedure used by \citet{Schoedel_09b}.
However, it is adjusted for the SED difference between stars and Sgr~A*  (Appendix~\ref{sec:filt_ext}).

The light that reaches Earth from the GC crosses many regions of the Galaxy. 
 It is possible that dust associated with the minispiral extincts the light in the GC itself. 
 However, because the extinction derived from the minispiral is consistent with the extinction derived from stars of \citet{Schoedel_09b},
 the extinction must occur mainly in front of the GC.
For testing this further, we smooth the Brackett-$\gamma$ flux to the resolution of the extinction map of \citet{Schoedel_09b}. 
This we compare with the extinction map of \citet{Schoedel_09b} in bins defined by the smoothed minispiral flux. 
We do not find a correlation between the minispiral flux and the extinction derived from stars. Furthermore, the scatter of the median extinction is 
only 0.04 mag over the different flux means. 
In addition, the small far infrared  flux \citep{Becklin_82,Guesten_87} of the central few parsecs shows that hardly any UV radiation is absorbed there \citep{Brown_84}. All in all, the NIR extinction must be very small (at least A$_{Ks}<0.1$) inside the central parsec.

In order to further constrain the location of the extinction in the line of sight, we use the $H-Ks$ values of the stars in Table~A.2 of \citet{Schoedel_09b}. Since the intrinsic color $|H-Ks|<0.2$ for nearly all stars, the $H-Ks$ of each star depends nearly only on the extinction. We then measure how many stars in the table have a $H-Ks$ compatible with
zero extinction. Thereby, we exclude stars which are so blue and faint in Ks that they
would be too faint for detection if they would have the extinction of the GC. We exclude them in order to avoid a bias
 towards foreground stars. After this, we obtain that only five of 6324 stars have a  $H-Ks$ color compatible with zero extinction. 
According to \citet{Philipp_99}, the Galactic disk and bulge (outside of 300 pc)
 have 2.3 \% of the flux of the GC at r$=10''$\footnote{We exclude the central 10'' because there the total flux is dominated by
young stars. Since there are only a few young stars due to the top heavy IMF \citep{Bartko_10} they are irrelevant for the star number ratio.}. 
Because the ratio of extinction free stars to all stars of 0.1\% is much smaller than the flux contribution of Galactic bulge and disk to the flux in the center 
of 2.3 \%, there must be extinction within the Galactic disk.
We then measure up to which extinction it is necessary to include stars in order to account for the 2.3 \% star contribution of Galactic disk and bulge. 
We find that it is necessary to include stars with extinction up 
A$_{Ks}=2.0$.  This is about  3/4 of the total extinction towards the GC.

Therefore, the measured extinction is mainly not associated with the Giant Molecular Clouds in the Nuclear bulge \citep{Mezger_96} and as such is not related to special processes in the central 50~pc. Because most bulges do not contain a lot of dust, it is likely that most of the extinction is caused by dust in the Galactic disk. 
As a result, the measured extinction curve is likely a typical extinction curve of
dust in the Galactic disk.

The extinction towards the GC is higher than the average extinction of the bulge behind the Galactic plane \citep{Marshall_06}. 
However there are also regions with much higher extinction close to the GC \citep{Ramirez_08}. Therefore, the extinction of the Galactic Center is not exceptional for an 8 kpc view through the 
Galactic disk.

\subsection{The NIR extinction}
\label{sec:nir-dis}

Our NIR data can be well fitted with a power-law of $\alpha=-2.11\pm 0.06$ ($\chi^2/d.o.f.=2.70/3$). This strengthens the case for the use of a power law
as model for the extinction in the NIR \citep[e.g.,][]{Cardelli_89}.
Our slope is steeper than the slope of $\alpha\approx-1.75$ of most reviews, see e.g. 
\citet{Savage_79, Mathis_90} and \citet{Draine_03}. 
However, most of the measurements \citep[e.g.,][]{Schultz_75,Landini_84,Whittet_88,He_95} combined in these reviews used relatively few stars; 
these stars are detectable in the optical and partly also in the UV, have  A$_V\leq5$ and are mostly closer than 3 kpc, see e.g. \citet{He_95}. 
The extinction  measurement of \citet{Rieke_85} towards the GC was also tied partly to measurements of stars which are visible also in the optical.

\begin{deluxetable}{lll} 
\tabletypesize{\scriptsize}
\tablecolumns{3}
\tablewidth{0pc}
\tablecaption{NIR infrared extinction law from literature \label{tab:nir_law_lit}}
\tablehead{Publication & $E_{J-H}/E_{H-K}$ & slope $\alpha$  
}
\startdata
\citet{Indebetouw_05} &  & $ -1.65 \pm 0.12 $  \\
\citet{Messineo_05}& & $-1.9 \pm 0.1 $\\
\citet{Nishiyama_06b} & & $-1.99 \pm 0.08 $ \\
\citet{Straizys_08b} & $2\pm 0.13$ & $-2.07\pm 0.23$ \\
\citet{Gosling_09} & & $-2.64 \pm 0.52 $\\
\citet{Nishiyama_09} & $2.09 \pm 0.13 $ & $-2.23 \pm 0.23 $\\
\citet{Stead_09}& & $-2.14 \pm 0.045 $ \\
\citet{Zasowski_09} & $2.11 \pm 0.1 $ & $-2.26 \pm 0.17 $\\
\citet{Schoedel_09b} & & $-2.21 \pm 0.24 $\\
Our work &  & $-2.11 \pm 0.06$ \\
\hline
weighted average & & $-2.07 \pm 0.16 $\\
\enddata
\tablecomments{NIR extinction slope measurements of diffuse ISM extinction since 2005. 
When $\alpha$ is not given in the publication, or the effective wavelength are unusual we calculate $\alpha$ from $E_{J-H}/E_{H-K}$. 
Thereby we use for $\lambda'_{\mathrm{eff}}$  the wavelength 1.24, 1.664 and 2.164 $\mu$m \citep{Nishiyama_09}. 
The error is either the measurement error in the publication or the scatter of different sight lines.
From \citet{Nishiyama_09} we only use the 2MASS data, since the SIRIUS data are identical to \citet{Nishiyama_06b}. 
 We use all values for the calculation of the weighted average.
 }
\end{deluxetable}

Since about 2005, large infrared surveys have become available and 
are now used by most publications about extinction. Since now no detection in the optical is necessary, and because it is easier to characterize high extinction, many publications (see Table~\ref{tab:nir_law_lit})  measured the extinction towards the highly extincted inner Galactic disk and bulge, at about 8 kpc distance. 
Most of these publications measure $\alpha\approx-2.1$.

It is possible that the change in the measured $\alpha$ around 2005 is due to systematic errors.
\citet{Stead_09}, for example, suggested that the use of the isophotal wavelength instead of the effective wavelength caused the flatter slope in the 
 measurement prior to 2005. We think, however, that it is unlikely that errors in the effective wavelengths are the reason for the discrepancy: 
while it is 
correct that using the isophotal filter wavelengths can lead to errors in $\alpha$, using the effective wavelength as presented in 
\citet{Stead_09} overestimates 
$\alpha$ slightly, see Appendix~\ref{sec:alpha_oth}. Furthermore, even if the isophotal wavelength is used, the systematic error on $\alpha$ 
is at maximum 0.07 
for the hot stars (see Appendix~\ref{sec:alpha_oth} for a 9480~K star) used in most works before 2005.
Additionally, \citet{Fitzpatrick_04} used 2MASS data of solar neighborhood stars and, using the effective wavelength, obtained 
 $\alpha\approx-1.84$.

Since the dust probed by studies which obtain $\alpha\approx-1.75$ and $\alpha\approx-2.1$, respectively, is not identical  
it seems likely that the extinction law varies between these regions.
A strong piece of evidence for truly variable extinction is the correlation of R$_V$ with $\alpha$ in \citet{Fitzpatrick_09},
 whereas for the standard R$_V=3.1$ $\alpha=-1.77\pm 0.05$. This work used  14 stars with observations from 120 nm to Ks-band.
All in all, we think it is likely that there is a transition of a mostly flatter NIR extinction in the solar neighborhood to a
steeper one in most parts of the Galactic disk. Naturally, further tests of the NIR extinction slope variation via
measurement of the NIR extinction slope in the local low extinction sight lines,
and in the Galactic disk, using the best methods available today would be very valuable.

For most molecular clouds, the extinction is, with $\alpha \approx -1.8$ \citep{Roman_07,Naoi_06,Kenyon_98,Flaherty_07, Lombardi_06}, 
flatter than for the Galactic
 disk sight lines. 
However, there are also clouds with deviating $\alpha$, with an $\alpha$ range from -1 to -2.4 \citep{Froebrich_06,Whittet_88,Racca_02,Naoi_07}.

There are some features typical of molecular clouds that are visible in the ISO-SWS spectrum. 
 \citet{Whittet_97} determined that about a third of the extinction towards the GC is caused by molecular clouds with substructure and the rest by 
diffuse extinction. A relatively small contribution of the extinction by molecular clouds is also supported by the fact, that the extinction 
variation towards the central parsec is less than one third of the maximum extinction, as can be seen in the extinction map of \citet{Schoedel_09b}. 
This means also that the molecular cloud contribution is not visible in terms of star counts as it is in the case of e.g. the coalsack.
Accordingly, the line of sight towards the GC is not dominated by molecular cloud extinction. 
Thus, we exclude molecular cloud extinction in our quantitative comparison with the GC.

For comparison we use publications about the extinction towards the Galactic disk and bulge. In practice this means
only publications since 2005, see  Table~\ref{tab:nir_law_lit}. We use these data for calculating the weighted average:  $\alpha=-2.07 \pm 0.16$. 
Of these 9 publications used, only  \citet{Indebetouw_05} is inconsistent with the others.
Hence, the extinction law is likely constant towards the inner Galactic disk and bulge and can probably
be used also for other fields in that region.

Due to the high absolute extinction in the inner Galactic disk, this region
contributes more to the global extinction of the Galaxy than the solar neighborhood, or the halo  of the Galaxy \citep{Schlafly_10}. 
Therefore, the steeper extinction law in the NIR of the inner Galactic disk is probably more important when integrated over the volume of the 
Milky Way. Similarly, for other galaxies for which our Galaxy is typical, a steeper extinction law could also be more important.

\subsection{The optical extinction}
\label{sec:opt_ext}

\citet{Rieke_89} estimated an optical extinction of A$_V=31$ towards the GC using the extinction law of \citet{Rieke_85}. 
The GC is, due to this high extinction,  undetectable in the optical. As a result, any estimate of A$_V$ for the GC is indirect.
 Since both this paper and \citet{Schoedel_09b} obtain a different infrared 
extinction law to \citet{Rieke_85}, it seems possible that the optical extinction
towards the GC is also different from the one assumed in \citet{Rieke_85}.

In order to test this, we use data which has a more direct connection to the GC than $o$ Sco which was used by \citet{Rieke_85}. % start new
The shortest wavelength at which the GC has been observed is the  
 z-band \citep{Henry_84,Rosa_92,Liu_93}. In order to estimate the z-band extinction, we use the magnitudes of the IRS16 stars from these works, because 
these stars and their intrinsic colors are well known. 
We neglect the data for IRS16C in \citet{Liu_93} because the star is brighter in this work compared to all other works. We use
the extinction map and the IRS16 magnitudes of \citet{Schoedel_09b} in order to calculate the dereddened Ks-band magnitudes of the IRS16 stars. 
We then subtract the dereddened Ks-band magnitudes
and an intrinsic color of $z-Ks=-0.25$ from the measured z-band magnitudes of IRS16 to obtain z-band extinctions.

We convert the z-band extinctions into an extinction power-law between Paschen-$\beta$ and the true effective wavelength for extinction 
measurements ($\lambda_{\mathrm{true}}$) of the GC (see Appendix~\ref{sec:alpha_oth}).
In the case of \citet{Henry_84} and \citet{Liu_93} the wavelengths given in the publications are probably $\lambda_{\mathrm{true}}$. We assign to
them an error of 0.01 $\mu$m. \citet{Rosa_92} give the central wavelength for stars extincted by GC extinction and unextincted stars. 
Since both are incorrect, see Appendix~\ref{sec:alpha_oth}, we calculate from the given central wavelength and FWHM of the optical system
$\lambda_{\mathrm{true}}=0.97$ $\mu$m.
Due to the additional uncertainties in calculating the effective wavelength we assume an error of 0.015 $\mu$m.
The effective wavelength uncertainty and the scatter of the IRS16 stars are used for the calculation of the errors. 

\begin{figure*}
\begin{center}
\includegraphics[width=1.48 \columnwidth,angle=-90]{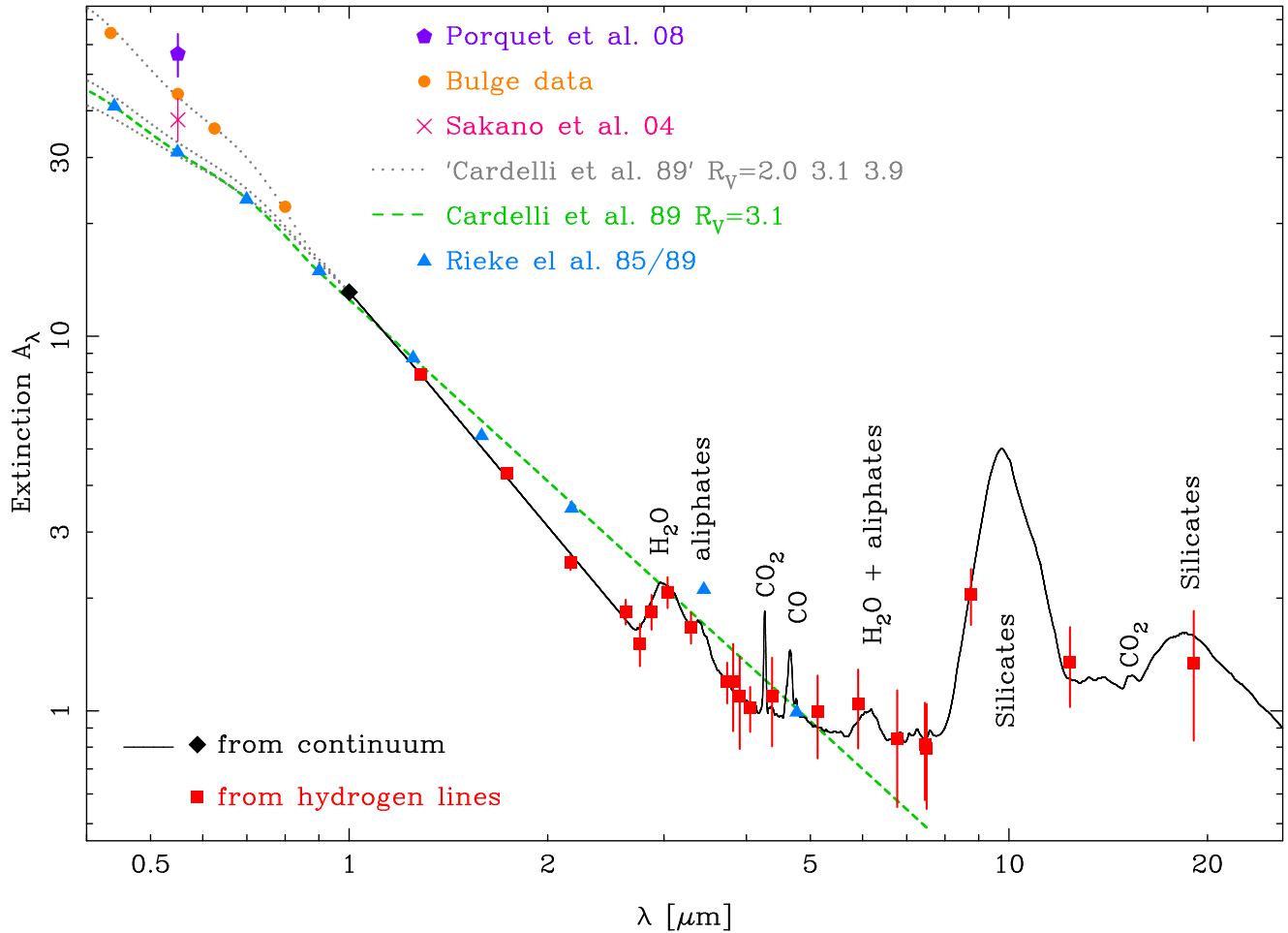}
\caption{Extinction towards the central 14''$\times$20'' of the GC. We use hydrogen lines for obtaining the extinction between 1.28 and 18 $\mu$m (red boxes) and stellar colors for 1 $\mu$m (black diamond). We interpolate the data by use of the continuum emission (black line). 
The spectral resolution of the interpolation is not high enough to resolve all features like the CO feature at $4.7\, \mu$m fully.
 We mark the larger 
 extinction features \citep{Lutz_96,Chiar_00}. 
We use the central bulge data of \citet{Sumi_04,Nishiyama_08} and \citet{Revnivtsev_10} for extending the extinction curve to the visible (orange dots). 
For comparison we add a value for A$_V$ derived from the X-ray spectrum of Sgr~A* \citep{Porquet_08} (violet pentagon).
In the same way we use the gas in front of Sgr~A~East \citep{Sakano_04} (pink cross).
The extinction 
curve differs from \citet{Rieke_85,Rieke_89} (blue triangles), the results of which are only partly based on GC data.
The GC data differs also from the \citet{Cardelli_89} curve (fitted to R$_V=$3.1 observations; green dashed line). 
The optical extinction towards the GC is uncertain. We use three curves
of the type used by \citet{Cardelli_89} with R$_V=2.3$, 3.1, 3.9 (dotted gray lines from top down) for showing the 
possible range.
} 
\label{fig:feat_ext}
\end{center}
\end{figure*}

In this way we obtain $\alpha=1.99\pm0.09$, $\alpha=1.91 \pm 0.14$ and $\alpha=2.165\pm 0.13$ 
from \citet{Henry_84,Rosa_92} and \citet{Liu_93}. The average slope is, with $\alpha=-2.02\pm0.07$, consistent with our determined NIR-slope of
 $\alpha=2.11\pm0.06$ (Section~\ref{sec:ex_pham}). 
Using the average of the three Paschen-$\beta$ z slopes we obtain  A$_{1\,\mu m}=13.11 \pm 0.30$. This extinction is slightly higher than interpolating
 \citet{Rieke_89} although our K-band extinction is smaller, see Figure~\ref{fig:feat_ext}.

We now consider how the z-band extinction should be extrapolated into the optical.
Molecular cloud features are probably responsible for a third of the extinction towards the GC, see Section~\ref{sec:spat}. 
An above average contribution of molecular cloud extinction to the GC is also supported by the fact, that
the extinction towards the GC is higher than the average bulge extinction (Section~\ref{sec:spat}).
If we assume a high R$_V=5.5$ for the molecular cloud extinction and 3.1 for the other two thirds, the average R$_V\approx3.9$ for the GC.
Use a \citet{Cardelli_89} curve with this R$_V$ (but with $\alpha=-1.85$ between 0.91 $\mu$m and 1 $\mu$m for reducing the jumps in the slope)
 to extrapolate to the optical, we obtain A$_{\lambda=0.55 \mu\mathrm{m}}=30.3$.

However, most molecular clouds do not have an R$_V$ of 5.5. Thus, even assuming that one third is caused by molecular cloud extinction and
two thirds by normal extinction, the real R$_V$ is probably closer to 3.1.
In addition, the high strength of aliphate (3.4 $\mu$m) and silicate  (9.7 $\mu$m) features
are different from the extinction in molecular clouds (Section~\ref{sec:mir_dis}). These features are even stronger than in most diffuse extinction
sight lines. Furthermore the steep NIR extinction slope of $\alpha=-2.11$
is not typical of molecular cloud extinction (Section~\ref{sec:nir-dis}). Therefore, we obtain another estimate, extrapolating to the visible also with the normal R$_V=3.1$
 (using $\alpha=-1.95$ between 0.91 $\mu$m and 1 $\mu$m). In this case we obtain  A$_{\lambda=0.55 \mu\mathrm{m}}=33$.

A much steeper extinction law than the standard R$_V=3.1$ from B-band to J-band is measured \citep{Udalski_03,Sumi_04,Revnivtsev_10,Nishiyama_08} 
towards parts of the bulge which have a much 
smaller extinction than the GC (A$_V<7$ in case of \citet{Nishiyama_08}).
In particular, \citet{Nishiyama_08} measured $A_V/A_J=5.32\pm0.14$, while according to \citet{Rieke_85} $A_V/A_J=3.55 \pm 0.16$ for the
 standard R$_V=3.1$. We fit these bulge data points with a \citet{Cardelli_89} curve with R$_V=2.0$ 
(using $\alpha=-2.02$ between 0.91 $\mu$m and 1 $\mu$m).
Extrapolating the z-band extinction of the GC with this curve we obtain A$_{\lambda=0.55 \mu\mathrm{m}}=44$.
The aliphate and silicate  features towards the GC are stronger than in diffuse extinction, while they are even weaker in molecular cloud extinction, 
see Section~\ref{sec:mir_dis}. This means, that the strength of these features is anticorrelated with R$_V$ for $3.1\leq R_V\leq5.5$. 
Extrapolating this anticorrelation to the stronger feature towards the GC implies, that R$_V<3.1$ towards the GC, which in turn implies an A$_V>33$.

Figure~4 in \citet{Fitzpatrick_09}  shows a correlation between the infrared power-law slope $\alpha$ and R$_V$. The correlation implies that the slope in the 
infrared approximately continues into the red.
Using a second order polynomial to fit their 13 measurements (neglecting one outlier) we obtain  R$_V=2.48\pm 0.06$ for the GC, using our
 measurement of $\alpha=-2.11\pm 0.06$ in the GC. Thus, because a small R$_V$ implies a high A$_V$, this again implies the optical 
extinction towards the GC is probably large.

X-rays can shed another light on $A_V$. X-ray photons, for example, are attenuated by scattering and absorption which are related to the extinction \citep{Morrison_83,Predehl_95}.
At the energy at which the GC is observed ($\approx3$ to 8 keV), absorption by astronomical metals is much stronger than scattering by dust grains \citep{Predehl_95,Porquet_08}. For the absorption the chemical state of the absorbing material is irrelevant and
 depends only on the integrated column density of astronomical metals. Thus the X-ray absorption can be used to constrain dust models and the extinction.
  
Here we investigate the implications only for the extinction. Observationally, the X-ray absorption is well correlated with A$_{V}$  \citep{Predehl_95} in most 
Galactic sight lines.
However, there are AGNs with different X-ray absorption to A$_{V}$ ratios \citep{Maiolino_01,Li_07}. 
Towards Sgr~A* we obtain a column density of  N$_H=10.5 \pm 1.4 \times 10^{22} \mathrm{cm}^{-2}$, averaging over the different states \citep{Porquet_08}.
 Thereby we assume an intrinsic X-ray 
power-law for the flares of Sgr~A* \citep{Katie_09}. We thus  obtain an X-ray derived A$_V=56.7 \pm 7.4$ for the GC (Figure~\ref{fig:feat_ext}) using the N$_H$/A$_V$ relation of \citet{Predehl_95}.
Using UV derived N$_H$/A$_V$ relations \citep{Bohlin_78,Draine_89,Zubko_04}  we obtain A$_V$ from 53 to 59. 
Our best fitting dust model  (Section~\ref{sec:dust-models}) gives A$_V\approx48$ for the measured N$_H$. 

There is the possibility, that part of the column density  towards Sgr~A* arises in ionized gas in the halo of the SNR Sgr~A~East \citep{Maeda_02,Porquet_08}, in which case not all of the $N_H$ (measured by X-rays) towards Sgr A* is caused by dust. For example, \citet{Sakano_04} measured an N$_H=15 \times 10^{22} \mathrm{cm}^{-2}$ for hot
plasma, and  N$_H=7 \times 10^{22} \mathrm{cm}^{-2}$ for somewhat colder plasma of Sgr~A~East. The second value is perhaps typical of plasma lying in the line of sight to Sgr~A*. Accordingly, it is possible that the column towards the colder plasma corresponds to the column to Sgr~A*. If this is the case we obtain a lower A$_V=37.8$, using the relation of \citet{Predehl_95}. However, probably the foreground
extinction towards Sgr~A~East and Sgr~A* is not fully homogeneous.  In principle it
 is possible to obtain a better N$_H$ towards the Sgr A region by
measuring N$_H$ towards  point sources outside of Sgr~A~East. By comparing such a map with an IR excess map, it should be possible to obtain a better estimate for N$_H$ for the GC.

Overall, the evidence for an R$_V<3.1$ is slightly larger than for  R$_V>3.1$, which means that, A$_{\lambda=0.55 \mu\mathrm{m}}>33$.
However, it is clear that only a direct measurement in the visible can clarify the value of R$_V$. The GC extinction attenuates 16NW, the brightest blue star in the GC with unextincted $m_{V}\approx 5.5$, to a magnitude of m$_V=34$ to 43 
depending on R$_V$ (since this simulated measurement integrates over the V-band the extinctions
are smaller than the corresponding 0.55 $\mu$m extinctions).

\subsection{The MIR extinction}
\label{sec:mir_dis}

At wavelengths $>2.8\,\mu$m extinction features are visible in the ISO-SWS spectrum, even shortward of the deep silicate feature at 9.7 
$\mu$m (Figure~\ref{fig:iso_spec}). These features are known from \citet{Butchert_86,Willner_79,Lutz_96} and \citet{Chiar_00}.

Most of these features are caused by ices like the H$_2$O feature at 3.1 $\mu$m  \citep{Butchert_86, Lutz_96, Chiar_00} and the CO$_2$ feature at 4.3 $\mu$m 
\citep{Lutz_96, Degraauw_96}. Among others, \citet{Whittet_88b,Rosenthal_00} and \citet{Knez_05} detected these features in many other sight lines with different 
strengths, both compared to one other as well as to the continuum extinction.
\citet{Rawlings_03} and \citet{Whittet_97} did not detect
H$_2$O ice in sight lines consisting of diffuse interstellar material with A$_V\approx10$.
 Thus, these features are likely only visible in sight lines through molecular clouds and not in purely diffuse extinction regions \citep{Whittet_97,Chiar_00}. 
This view is supported by the CO clouds in the Galactic arms in front of the GC \citep{Sutton_90}. This CO is also visible in the ISO-SWS spectrum 
\citep{Lutz_96,Moneti_01}.
The detection of molecular cloud features towards the GC is not unique, ice features are also
detected towards the Quintuplet cluster at a distance of about 12' to the GC, where they are slightly weaker compared to the continuum extinction \citep{Chiar_00}.

Other extinction features, like the strong aliphatic hydrocarbon feature at 3.4 $\mu$m \citep{Willner_79}, are caused by diffuse dust and are thus a 
general visible feature of extinction \citep{Chiar_00,Rawlings_03}.
The feature at 3.4 $\mu$m is about a factor of two stronger towards the GC 
than in the local diffuse extinction \citep{Rawlings_03,Gao_10}. 
The 3.4 $\mu$m feature is typical for diffuse dust. In molecular clouds it is not detected \citep{Pendleton_02}.

In order to better constrain the shape of the extinction curve, we use the continuum of the ISO-SWS spectrum, see Appendix~\ref{sec:int_ext} and 
Figure~\ref{fig:feat_ext}. 

For the silicate feature at 9.7 $\mu$m  we obtain an optical depth of $\Delta \tau_{\mathrm{Si}\,9.7}=3.84 \pm 0.52$ relative to the continuum at 
7 $\mu$m 
from our interpolated extinction curve. 
The depth is similar to that obtained by \citet{Chiar_00} of $ \Delta \tau_{\mathrm{Si}\,9.7}=3.46$ and by \citet{Roche_85} of $ \tau_{\mathrm{Si}\,9.7}=3.6$. 
According to \citet{Vanbreemen_10} the shape of 9.7 $\mu$m silicate feature of the GC is identical to sight lines with diffuse extinction 
and slightly different to sight lines with molecular
 clouds.
Using our broadband extinction values (Appendix~\ref{sec:filt_ext}) we obtain $\Delta \tau_{\mathrm{Si}\,9.7}/E(J-K)=0.70 \pm 0.10$,
 consistent with \citep{Roche_85}. 
This is more than the value of
$\Delta \tau_{\mathrm{Si}\,9.7}/E(J-K)=0.34$ in nearly all other diffuse sight lines \citep{Roche_84,Vanbreemen_10}. 
Sight lines with molecular clouds also have $\tau_{\mathrm{Si}\,9.7}/E(J-K)=0.34$ or even smaller values \citep{Vanbreemen_10}.
The large $\Delta \tau_{\mathrm{Si}\,9.7}/E(J-K)$ towards the GC is probably caused by an abnormally high $\Delta \tau_{\mathrm{Si}\,9.7}$. 
A larger silicate dust to carbon dust ratio in the inner Galaxy compared to the local diffuse medium \citep{Roche_85,Vanbreemen_10}
could explain the high $\Delta \tau_{\mathrm{Si}\,9.7}/E(J-K)$ towards the GC. With a higher silicate abundance $\Delta \tau_{3.4\mu\mathrm{m}}/E(J-K)$
 towards the GC should be identical to the local value. However, it is twice the local value. 
Porous dust grains cause both a strong 3.4 $\mu$m and a strong 9.7 $\mu$m feature \citep{Gao_10}. Porosity is also one element of the
dust model of \citet{Zubko_04}, which best fits our continuum extinction data, see Section~\ref{sec:dust-models}.

In the continuum, CO$_2$ is  visible at 15 $\mu$m \citep{Gerakines_99}. It is, however, much weaker than in the observations of extinction 
in molecular clouds observed by \citet{Knez_05}. In contrast, the silicate feature at 18 $\mu$m towards the GC is much stronger than the CO$_2$ feature,
 see Figure~\ref{fig:feat_ext}.

\citet{Rieke_85} and \citet{Rosenthal_00} can fit all their extinction values up to 7.5 $\mu$m with a single power-law. This single power-law extinction 
is inconsistent with the flattening of the extinction curve in our data at around 4~$\mu$m, see Section~\ref{sec:ex_pham}.
In the continuum, there  are no strong extinction features apparent between 3.7 and 8 $\mu$m, see Figure~\ref{fig:feat_ext}.
Some weak features due to H$_2$O, NH$_3$, CO$_2$, HCOOH and aliphatic hydrocarbon are tentatively identified by \citet{Lutz_96} and \citet{Chiar_00}.
 All the features are weaker than $\tau=0.15$. Thus, these features are  too weak to explain  a significant part of
the extinction difference between the measured extinction and a single power-law.
There is some indication for a steeper slope around 3.9 $\mu$m and a flatter one at longer wavelengths.
Also, the data of \citet{Nishiyama_09} and most theoretical models (Section~\ref{sec:dust-models}) exhibit a smooth flattening in the transition region.

\citet{Zasowski_09,Jiang_03,Jiang_06,Gao_09,Flaherty_07,Roman_07,Nishiyama_09} and \citet{Indebetouw_05} also detected a flattening of
 the extinction at MIR wavelengths 
towards different regions of the Galaxy. Some of these observations, such as, \citet{Nishiyama_09} and \citet{Zasowski_09}, targeted 
 diffuse extinction in the
 bulge and the Galactic disk. Therefore, the flattening of the curve in the MIR is not caused by the molecular cloud in front of the GC.
 
The extinction curves of other galaxies in the MIR are more difficult to interpret, because the extinction in other galaxies is not likely to be caused by an uniform foreground screen. However, 
since a mixed model \citep{Foerster_01} produces a flattening of the extinction curve at higher extinction, while in the GC a flattening at smaller extinction is observed, 
it is still possible to find qualitative signs for the GC extinction curve in the spectra of other galaxies.
The fact that a single power-law extinction is slightly preferred when fitting the continuum in 
some ULIRGS \citep{Tran_01} could also partially be caused by a  GC-like extinction and the mixed model. 
 On the other hand,  \citet{Thuan_99} found an extinction curve similar to our results for the extremely metal poor galaxy
 SBS 0335-052, also using continuum emission. In the case of the central region of M82, \citet{Foerster_01} found, using hydrogen lines, that an extinction curve similar to the one in the GC
 provides a better fit to the data than a single power-law.
All in all, it is likely that the  GC extinction \citep{Lutz_96}, considered to be unusual at the time of discovery,
is in general the more widely spread type of extinction.

\subsection{Broadband extinction curve}
\label{sec:curve}

Since the GC extinction is consistent with many extinction measurements outside of the GC (Section \ref{sec:nir-dis} and \ref{sec:mir_dis}),
the extinction curve towards the GC (Appendix~\ref{sec:int_ext}) is also useful for other sight lines. We derive from this extinction curve 
broadband extinction values (Appendix~\ref{sec:filt_ext}). 
In principle it is necessary to know the extinction amplitude and the intrinsic object spectrum for deriving precise broadband extinctions. 
The error caused by neglecting these aspects is below 2 \%. On the other hand, the error due to transmission differences between slightly different, 
but equally named broadband filters can be 
up to 2.5 \% (Appendix~\ref{sec:filt_ext}).
We give in Table~\ref{tab:sim_broad} broadband extinctions ratios relative to the NACO Ks-band.
For extinction values of many additional filters and for different source spectra, see Appendix~\ref{sec:filt_ext}.

\begin{deluxetable}{ll} 
\tabletypesize{\scriptsize}
\tablecolumns{2}
\tablewidth{20pc}
\tablecaption{Relative broadband extinction values derived from the GC extinction curve
\label{tab:sim_broad}}
\tablehead{ Broadband & A$_{\mathrm{band}}$/A$_{Ks}$  
}
\startdata
VIRCAM Y & 4.634 $\pm$ 0.103 \\
NACO J & 3.051 $\pm$ 0.069 \\
NACO H & 1.737 $\pm$ 0.027 \\
NACO Ks &     1.000  $\pm$  0.000 \\
NACO L' & 0.450 $\pm$  0.053 \\
NACO M' & 0.391  $\pm$  0.094 \\
IRAC 1 & 0.547 $\pm$ 0.052 \\
IRAC 2 & 0.396 $\pm$ 0.082 \\
IRAC 3 & 0.340 $\pm$ 0.094 \\
IRAC 4 & 0.383 $\pm$ 0.116 \\

\enddata
 \tablecomments{The extinction ratios are obtained from the average of two calculations for the effective filter extinction
 as described in Appendix \ref{sec:filt_ext} using Vega-like stars with A$_{Br\,\gamma}=1$ and A$_{Br\,\gamma}=5$. The errors 
are made up of half of the extinction ratio difference between the two different Brackett-$\gamma$ extinction and the extinction curve errors. We assume that the NIR extinction follows a power-law of 
$\alpha=-2.11\pm 0.06$. Broadband filters of other instruments can have another transmission curve. This can result in an additional error of up 3.5\%. 
For further filters, see Appendix \ref{sec:filt_ext}.
}
\end{deluxetable}

\subsection{Dust models}
\label{sec:dust-models}

Extinction curves, together with the emission spectrum of dust, as well as the elemental abundances and depletions, constrain the properties of 
interstellar dust. Ideally, the dust model should also be plausible with regards to the formation and destruction of the dust \citep{Compiegne_10}. 
We use published extinction models which fulfilled most constraints at the time of their publication and compare them with the  extinction curve 
towards the GC.

The classical grain model \citep{Mathis_77} is composed of silicate and graphite grains, where both follow a power-law size distribution with a 
lower and an upper cutoff for the exclusion of very large and very small grains.
\citet{Li_01} improved this simple model to to size distributions which include
even smaller grains in order to account for the PAH emissions by interstellar clouds.  We use a slight variant of this dust model considered by
 \citet{Weingartner_01} (Wg-model), who also consider different R$_V$.
This model consists of a trimodal carbonaceous grain size distribution and a simple silicate grain size distribution. 
With increasing R$_V$, the number of small 
grains decreases while the maximum grain size increases.

\begin{deluxetable*}{lllll} 
\tabletypesize{\scriptsize}
\tablecolumns{5}
\tablewidth{0pc}
\tablecaption{ Fit goodness of different dust models \label{tab:dust_mod}}
\tablehead{Author & model & total $\chi^2/\mathrm{d.o.f.}$ & best fit  $\chi^2/\mathrm{points}$ of NIR   & best fit  $\chi^2/\mathrm{points}$ of MIR  \\
                              }
\startdata
Wg01 & R$_V=$3.1  & 114.3/14 & 106.6.3/6 & 7.7/9   \\
Wg01 & case B R$_V=$4.0  & 228/14 & 178/6 & 51/9   \\
Zu04 & BARE-GR-S  & 18.9/14 & 4.4/6 & 14.5/9   \\
Zu04 & COMP-AC-S  & 6.4/12 & 4.7/4 & 1.7/9   \\
Dw04 &   & 30.9/14 & 21.7/6 & 9.2/9  \\
Vo06 & p$=0.18$ &  111/14 & 66/6 & 45/9  \\
Vo06 & p$=0.6$ &  117/14 & 111/6 & 5.2/9  \\
\enddata
\tablecomments{We fit the different dust models to our data by scaling the full extinction curve.
The NIR data are between 1 and 2.8 $\mu$m, the MIR lines are between 3.7 and 7.5 $\mu$m. 
In $\chi^2/\mathrm{points}$ we give the number of data points used for the fitting which are within the NIR and MIR.
The models are from \citet{Weingartner_01,Dwek_04,Zubko_04,Voshchinnikov_06}.
}
\end{deluxetable*}

Similarly to most later models the Wg-models do not have the H$_2$O features at 3 $\mu$m and 6 $\mu$m. For constraining the dust properties, 
independent of the features, we exclude 
the four lines concerned in our quantitative comparisons. We also exclude the lines beyond 8 $\mu$m because the
silicate feature is not modeled in all works
and in no case matches our silicate extinction towards the GC. 
In addition, excluding these features  we are also more likely to match other extinction measurements in the infrared, such as, for  
example \citet{Indebetouw_05,Nishiyama_09}, where the features could 
not be measured due to too low spectral resolution. 
We fit all models to the data by adjusting the extinction curve with a global scaling factor. 

\begin{figure}
\begin{center}
\includegraphics[width=0.99 \columnwidth,angle=-90]{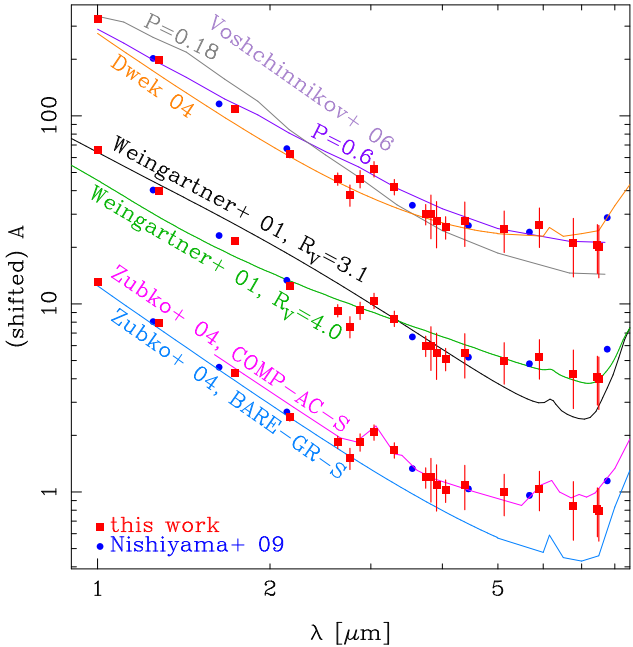}
\caption{Comparison of our data with other data and models. 
To our extinction measurements derived from hydrogen lines (red boxes) we add 
 the inner bulge observations of \citet{Nishiyama_09} (blue circles)
 scaled to our A$_{\mathrm{Br} \gamma}$. 
The models (lines) are from \citet{Weingartner_01,Zubko_04,Dwek_04} and \citet{Voshchinnikov_06}.
Everything apart from the lowest data is shifted for better visualization.
} 
\label{fig:_draine}
\end{center}
\end{figure}

All Wg-models  have a relatively flat NIR slope $\alpha \geq -1.7$ and hence fit the data badly in the NIR 
 ($\lambda<2.8\,\mu$m), see Table~\ref{tab:dust_mod}.
They differ in the MIR from each other. There the models with higher R$_V$ are flatter. On the one hand, when fitting all lines the models with
 R$_V\geq4.0$ have a worse $\chi^2$ than the R$_V=3.1$ model because of the flatter slope of the R$_V\geq4.0$ models in the K-band, which is due to the
 smaller extinction errors 
there more important than the MIR.  On the other hand, the R$_V=3.1$ model  is incompatible with the data of \citet{Nishiyama_09} in the MIR, 
see Figure~\ref{fig:_draine}. Compared with the \citet{Nishiyama_09} data, the models with R$_V \geq 4.0$ are a better fit, although 
none of the Wg-models show such a strong and sharp flattening from NIR to MIR as observed.

One possibility for more complex dust models is the inclusion of voids in the dust grains \citep{Voshchinnikov_06}.  
In this model, mainly two size distributions of porous silicate grains and one of small graphite grains contribute to the extinction.
Still, the extinction features shortward of 7~$\mu$m are unexplained. This model has the advantage that the NIR slope between J and K depends on the 
porosity.
None of their three models with different porosity and also no linear combination of these models can fit
 our data, see Figure~\ref{fig:_draine} and Table~\ref{tab:dust_mod}.

\citet{Zubko_04} obtained dust models by fitting grain size distribution to an R$_V=3.1$ extinction curve, dust emission and elemental abundances. 
 They use the extinction dispersion between different sight lines as their error estimate. Thus, the extinction has a relatively small weight which 
could be 
part of the reason that various dust models can fit their data. \citet{Zubko_04} used different element abundances which have some impact on the result 
in the 
case of carbon abundances. We only use their solar abundances. More likely, the real  solar abundances \citep{Asplund_09} are smaller than those used
 by  
\citep{Zubko_04}. However the abundances in HII regions are higher than the solar abundances \citep{Asplund_09} and at least the B-star carbon
abundance is too small for causing the observed magnitude of extinction \citep{Li_05}. 

The simplest model of \citet{Zubko_04}, BARE-GR, consists of only graphite, PAHs and silicates grains and as such is similar to the Wg-models. 
BARE-GR only uses graphite grains  smaller than 300 nm while the Wg-models use grains up to 0.8 $\mu$m and 5 $\mu$m respectively. 
Since grain sizes around 300 nm cause the steepest infrared slopes in contrast to grains of other sizes \citep{Moore_05,Gao_09},
the NIR extinction slope of the BARE-GR model is steeper than that of Wg-models. 
 Overall, the BARE-GR model can fit the data with $\chi^2/d.o.f.=18.9/14$ relatively well. 
But this model does not have the H$_2$O features in contrast to the ISO-SWS spectrum. Because this model nearly does not flatten at all in the MIR,
 it cannot fit the inner bulge data of \citet{Nishiyama_09}, see Figure~\ref{fig:_draine}. 

The COMP models  of \citet{Zubko_04} use, in addition to silicate and carbonaceous grains, composite grains consisting of silicates, organic refractory material, water ice and voids. These models seem to be more promising for the GC, because they contain water ice which is visible in extinction features towards the GC. The best fitting subtype is AC in which the carbon is mainly amorphous. 
The composite particles are, with sizes up to 0.8 $\mu$m, bigger than the other particles.
The other subtypes of composite models have a steeper slope, at least in the MIR, and hence make a worse fit to the data.
The COMP-AC-S model has a $\chi^2/d.o.f.=6.4/12$, slightly better than the BARE-GR-S model. In contrast to the
other models, it also matches the \citet{Nishiyama_09} data, see Figure~\ref{fig:_draine}.

\citet{Dwek_04} added metallic needles to the BARE-GR model of \citet{Zubko_04}. The extinction caused by the needles rises with wavelength up 
to 8 $\mu$m, thus flattening the Wg-model between 3 and 8 $\mu$m, see Figure~\ref{fig:_draine}. The extinction due to needles drops
 fast longward of 8~$\mu$m for avoiding a weakening of the silicate feature. 
This model does not fit the data with $\chi^2/d.o.f.=30.9/14$. The inconsistency is caused by an overly 
gradual transition to a flatter slope  around 3 $\mu$m, while the observed slopes changes faster, see Figure~\ref{fig:_draine}.
While it might be possible to fit our data better with some dust model variant which consists of metallic needles as well as simple carbonaceous and silicate grains, finding a fit to the data will not be easy, since the extinction due to both components varies 
so slowly with wavelength that it is difficult to obtain the observed sharp flattening around 3 $\mu$m.

All together, the COMP-AC-S model of \citet{Zubko_04} seems to be the best model for extinction towards the GC because of its best
$\chi^2/d.o.f.$ and the presence of the H$_2$O features. However, the lack of H$_2$O ice features in most sight lines \citep{Rawlings_03,Whittet_97}, while MIR-flattening is observed in many regions  
\citep{Zasowski_09,Jiang_03,Jiang_06,Gao_09,Flaherty_07,Roman_07,Nishiyama_09,Indebetouw_05}
renders the idea that composite particles glued by ices are responsible the flat MIR extinction questionable, also for the GC. 
This is because if composite particles can exist only together with ices, the other regions should show a steeper MIR extinction than the GC. 
 However, because the sight lines with extinction probed in the MIR do not overlap with the sight lines which are tested for H$_2$O ice,  
it is still possible that the ice feature could exist in most sight lines with flatter MIR extinction.

An alternative model could be that something else aside from ices produces the flat extinction in the MIR towards the GC and elsewhere, 
and towards the GC additional pure ice grains produce the 
 extinction features. However, all other models tested here have the problem that they do not have at once a steep NIR extinction and a flat
 MIR extinction. It is perhaps possible to change the shape of the extinction curves within the models, without changing the type of
 particles in the model. However, this seems difficult for the pure carbonaceous and silicate grain model, because a smooth particle
 size distribution of carbonaceous and silicate grains should also produce a smooth extinction curve. 
It might be possible, on the other hand, to change the extinction shape only around 3 $\mu$m by omitting only H$_2$O ice from 
the composite particles in 
the COMP-AC-S model.  If this does not change the other extinction properties of the dust such model could fit the 
extinction also in other sight lines.

For solving the issue, further modeling seems to be necessary, maybe concentrating first on the detailed extinction curve towards the 
Galactic Center. In the future, similarly detailed extinction curves in other sight lines would also be very useful. Especially promising
 could be searching for the H$_2$O ice feature in sight lines which are already probed by broadband measurements.
Also, adding more extinction measurements in between for testing the commonness of the sharp transition to a flatter extinction in the MIR  
might be very interesting for further study.

\section{Summary}

The simplest way to derive an extinction curve is to observe a well understood object at different wavelengths, including at least one extinction-free wavelength. The minispiral in the GC fulfills these conditions. It is a HII region and Case B is valid.
As extinction-free emission we use the 2 cm radio continuum observed with the VLA. In the infrared, we obtain line fluxes between 1.28 $\mu$m and 2.17 $\mu$m from SINFONI. For lines at longer wavelengths of up to 19 $\mu$m we use ISO-SWS observations. We obtain the following results:

\begin{itemize}
 \item By interpolation we obtain  A$_{2.166\,\mu\mathrm{m}}=2.62 \pm 0.11$ as the average 2.166 $\mu$m extinction of the ISO field of 14'' $\times 20$'' about Sgr~A*.

\item  Using the extinction map of \citet{Schoedel_09b} for the relative spatial extinction we obtain, for the direction towards Sgr~A*: 
$A_{H}=4.21 \pm 0.10$, $A_{Ks}=2.42 \pm 0.10$ and $A_{L'}=1.09 \pm 0.13$.

 \item \citet{Schoedel_09b} measured the total luminosity modulus of the red clump consisting of the extinction and distance modulus towards the
Galactic Center. Since we measure the extinction independent of the distance, the combination of \citet{Schoedel_09b} with our extinction 
yields the distance  R$_0$, the distance to the Galactic Center: we obtain R$_0=7.94 \pm 0.65$ kpc in agreement
 with current measurements.

 \item  The extinction in the NIR (1.2 to 2.8 $\mu$m) is well fitted by a power-law of slope $\alpha=-2.11 \pm 0.06$.  
 This law is  steeper than the value of about $\alpha \approx-1.75$, which was mostly reported in the literature before 2004, see for example \citep{Draine_89}. However, 
since 2005 most publications about diffuse 
extinction towards the bulge and Galactic disk (for example \citet{Stead_09}) yield a steeper law. 
We obtain $\alpha=-2.07 \pm 0.16$ as weighted average of all publications about diffuse extinction since 2005.

\item At longer wavelengths, several extinction features are visible, such as H$_2$O at 3.1 $\mu$m and silicates at 9.7~$\mu$m. 
Even aside from such features, the extinction is higher than expected from extrapolating the NIR power-law. 
 Our data agree well with several IRAC publications for the inner bulge, such as \citet{Nishiyama_09} for example. 
Because we are able to use many lines, it is apparent that the change in slope is sharper and stronger than in any of the extinction curves 
produced by pure carbonaceous and silicate grains \citep{Weingartner_01}. 
The best fitting model adds composite particles which contain also H$_2$O ice \citep{Zubko_04}.

\end{itemize}

\bibliography{mspap}

\begin{thebibliography}{147}
\expandafter\ifx\csname natexlab\endcsname\relax\def\natexlab#1{#1}\fi

\bibitem[{{Abuter} {et~al.}(2006){Abuter}, {Schreiber}, {Eisenhauer}, {Ott},
  {Horrobin}, \& {Gillesen}}]{Abuter_06}
{Abuter}, R., {Schreiber}, J., {Eisenhauer}, F., {Ott}, T., {Horrobin}, M., \&
  {Gillesen}, S. 2006, 50, 398

\bibitem[{{Asplund} {et~al.}(2009){Asplund}, {Grevesse}, {Sauval}, \&
  {Scott}}]{Asplund_09}
{Asplund}, M., {Grevesse}, N., {Sauval}, A.~J., \& {Scott}, P. 2009, \araa, 47,
  481

\bibitem[{{Baker} \& {Menzel}(1938)}]{Baker_38}
{Baker}, J.~G. \& {Menzel}, D.~H. 1938, \apj, 88, 52

\bibitem[{{Bartko} {et~al.}(2009){Bartko}, {Martins}, {Fritz}, {Genzel},
  {Levin}, {Perets}, {Paumard}, {Nayakshin}, {Gerhard}, {Alexander},
  {Dodds-Eden}, {Eisenhauer}, {Gillessen}, {Mascetti}, {Ott}, {Perrin},
  {Pfuhl}, {Reid}, {Rouan}, {Sternberg}, \& {Trippe}}]{Bartko_09}
{Bartko}, H., {Martins}, F., {Fritz}, T.~K., {Genzel}, R., {Levin}, Y.,
  {Perets}, H.~B., {Paumard}, T., {Nayakshin}, S., {Gerhard}, O., {Alexander},
  T., {Dodds-Eden}, K., {Eisenhauer}, F., {Gillessen}, S., {Mascetti}, L.,
  {Ott}, T., {Perrin}, G., {Pfuhl}, O., {Reid}, M.~J., {Rouan}, D.,
  {Sternberg}, A., \& {Trippe}, S. 2009, \apj, 697, 1741

\bibitem[{{Bartko} {et~al.}(2010){Bartko}, {Martins}, {Trippe}, {Fritz},
  {Genzel}, {Ott}, {Eisenhauer}, {Gillessen}, {Paumard}, {Alexander},
  {Dodds-Eden}, {Gerhard}, {Levin}, {Mascetti}, {Nayakshin}, {Perets},
  {Perrin}, {Pfuhl}, {Reid}, {Rouan}, {Zilka}, \& {Sternberg}}]{Bartko_10}
{Bartko}, H., {Martins}, F., {Trippe}, S., {Fritz}, T.~K., {Genzel}, R., {Ott},
  T., {Eisenhauer}, F., {Gillessen}, S., {Paumard}, T., {Alexander}, T.,
  {Dodds-Eden}, K., {Gerhard}, O., {Levin}, Y., {Mascetti}, L., {Nayakshin},
  S., {Perets}, H.~B., {Perrin}, G., {Pfuhl}, O., {Reid}, M.~J., {Rouan}, D.,
  {Zilka}, M., \& {Sternberg}, A. 2010, \apj, 708, 834

\bibitem[{{Becklin} {et~al.}(1982){Becklin}, {Gatley}, \&
  {Werner}}]{Becklin_82}
{Becklin}, E.~E., {Gatley}, I., \& {Werner}, M.~W. 1982, \apj, 258, 135

\bibitem[{{Becklin} \& {Neugebauer}(1968)}]{Becklin_68}
{Becklin}, E.~E. \& {Neugebauer}, G. 1968, \apj, 151, 145

\bibitem[{{Beintema} {et~al.}(2003){Beintema}, {Salama}, \&
  {Lorente}}]{Beintema_03}
{Beintema}, D.~A., {Salama}, A., \& {Lorente}, R. 2003, in ESA Special
  Publication, Vol. 481, The Calibration Legacy of the ISO Mission, ed.
  {L.~Metcalfe, A.~Salama, S.~B.~Peschke, \& M.~F.~Kessler}, 215--+

\bibitem[{{Blum} {et~al.}(2003){Blum}, {Ram{\'{\i}}rez}, {Sellgren}, \&
  {Olsen}}]{Blum_02}
{Blum}, R.~D., {Ram{\'{\i}}rez}, S.~V., {Sellgren}, K., \& {Olsen}, K. 2003,
  \apj, 597, 323

\bibitem[{{Blum} {et~al.}(1996){Blum}, {Sellgren}, \& {Depoy}}]{Blum_96}
{Blum}, R.~D., {Sellgren}, K., \& {Depoy}, D.~L. 1996, \apj, 470, 864

\bibitem[{{Bohlin} {et~al.}(1978){Bohlin}, {Savage}, \& {Drake}}]{Bohlin_78}
{Bohlin}, R.~C., {Savage}, B.~D., \& {Drake}, J.~F. 1978, \apj, 224, 132

\bibitem[{{Bonnet} {et~al.}(2003){Bonnet}, {Str{\"o}bele}, {Biancat-Marchet},
  {Brynnel}, {Conzelmann}, {Delabre}, {Donaldson}, {Farinato}, {Fedrigo},
  {Hubin}, {Kasper}, \& {Kissler-Patig}}]{Bonnet_03}
{Bonnet}, H., {Str{\"o}bele}, S., {Biancat-Marchet}, F., {Brynnel}, J.,
  {Conzelmann}, R.~D., {Delabre}, B., {Donaldson}, R., {Farinato}, J.,
  {Fedrigo}, E., {Hubin}, N.~N., {Kasper}, M.~E., \& {Kissler-Patig}, M. 2003,
  in Society of Photo-Optical Instrumentation Engineers (SPIE) Conference
  Series, Vol. 4839, Society of Photo-Optical Instrumentation Engineers (SPIE)
  Conference Series, ed. {P.~L.~Wizinowich \& D.~Bonaccini}, 329--343

\bibitem[{{Brown} \& {Liszt}(1984)}]{Brown_84}
{Brown}, R.~L. \& {Liszt}, H.~S. 1984, \araa, 22, 223

\bibitem[{{Butchart} {et~al.}(1986){Butchart}, {McFadzean}, {Whittet},
  {Geballe}, \& {Greenberg}}]{Butchert_86}
{Butchart}, I., {McFadzean}, A.~D., {Whittet}, D.~C.~B., {Geballe}, T.~R., \&
  {Greenberg}, J.~M. 1986, \aap, 154, L5

\bibitem[{{Calzetti}(2001)}]{Calzetti_01}
{Calzetti}, D. 2001, \pasp, 113, 1449

\bibitem[{{Calzetti} {et~al.}(2000){Calzetti}, {Armus}, {Bohlin}, {Kinney},
  {Koornneef}, \& {Storchi-Bergmann}}]{Calzetti_00}
{Calzetti}, D., {Armus}, L., {Bohlin}, R.~C., {Kinney}, A.~L., {Koornneef}, J.,
  \& {Storchi-Bergmann}, T. 2000, \apj, 533, 682

\bibitem[{{Cardelli} {et~al.}(1989){Cardelli}, {Clayton}, \&
  {Mathis}}]{Cardelli_89}
{Cardelli}, J.~A., {Clayton}, G.~C., \& {Mathis}, J.~S. 1989, \apj, 345, 245

\bibitem[{{Chiar} {et~al.}(2000){Chiar}, {Tielens}, {Whittet}, {Schutte},
  {Boogert}, {Lutz}, {van Dishoeck}, \& {Bernstein}}]{Chiar_00}
{Chiar}, J.~E., {Tielens}, A.~G.~G.~M., {Whittet}, D.~C.~B., {Schutte}, W.~A.,
  {Boogert}, A.~C.~A., {Lutz}, D., {van Dishoeck}, E.~F., \& {Bernstein}, M.~P.
  2000, \apj, 537, 749

\bibitem[{{Compi{\`e}gne} {et~al.}(2011){Compi{\`e}gne}, {Verstraete}, {Jones},
  {Bernard}, {Boulanger}, {Flagey}, {Le Bourlot}, {Paradis}, \&
  {Ysard}}]{Compiegne_10}
{Compi{\`e}gne}, M., {Verstraete}, L., {Jones}, A., {Bernard}, J., {Boulanger},
  F., {Flagey}, N., {Le Bourlot}, J., {Paradis}, D., \& {Ysard}, N. 2011, \aap,
  525, A103+

\bibitem[{{Cunha} {et~al.}(2007){Cunha}, {Sellgren}, {Smith}, {Ramirez},
  {Blum}, \& {Terndrup}}]{Cunha_07}
{Cunha}, K., {Sellgren}, K., {Smith}, V.~V., {Ramirez}, S.~V., {Blum}, R.~D.,
  \& {Terndrup}, D.~M. 2007, \apj, 669, 1011

\bibitem[{{Dambis}(2009)}]{Dambis_09}
{Dambis}, A.~K. 2009, \mnras, 396, 553

\bibitem[{{de Graauw} {et~al.}(1996{\natexlab{a}}){de Graauw}, {Haser},
  {Beintema}, {Roelfsema}, {van Agthoven}, {Barl}, {Bauer}, {Bekenkamp},
  {Boonstra}, {Boxhoorn}, {Cote}, {de Groene}, {van Dijkhuizen}, {Drapatz},
  {Evers}, {Feuchtgruber}, {Frericks}, {Genzel}, {Haerendel}, {Heras}, {van der
  Hucht}, {van der Hulst}, {Huygen}, {Jacobs}, {Jakob}, {Kamperman},
  {Katterloher}, {Kester}, {Kunze}, {Kussendrager}, {Lahuis}, {Lamers},
  {Leech}, {van der Lei}, {van der Linden}, {Luinge}, {Lutz}, {Melzner},
  {Morris}, {van Nguyen}, {Ploeger}, {Price}, {Salama}, {Schaeidt}, {Sijm},
  {Smoorenburg}, {Spakman}, {Spoon}, {Steinmayer}, {Stoecker}, {Valentijn},
  {Vandenbussche}, {Visser}, {Waelkens}, {Waters}, {Wensink}, {Wesselius},
  {Wiezorrek}, {Wieprecht}, {Wijnbergen}, {Wildeman}, \& {Young}}]{Gra_96}
{de Graauw}, T., {Haser}, L.~N., {Beintema}, D.~A., {Roelfsema}, P.~R., {van
  Agthoven}, H., {Barl}, L., {Bauer}, O.~H., {Bekenkamp}, H.~E.~G., {Boonstra},
  A., {Boxhoorn}, D.~R., {Cote}, J., {de Groene}, P., {van Dijkhuizen}, C.,
  {Drapatz}, S., {Evers}, J., {Feuchtgruber}, H., {Frericks}, M., {Genzel}, R.,
  {Haerendel}, G., {Heras}, A.~M., {van der Hucht}, K.~A., {van der Hulst}, T.,
  {Huygen}, R., {Jacobs}, H., {Jakob}, G., {Kamperman}, T., {Katterloher},
  R.~O., {Kester}, D.~J.~M., {Kunze}, D., {Kussendrager}, D., {Lahuis}, F.,
  {Lamers}, H.~J.~G.~L.~M., {Leech}, K., {van der Lei}, S., {van der Linden},
  R., {Luinge}, W., {Lutz}, D., {Melzner}, F., {Morris}, P.~W., {van Nguyen},
  D., {Ploeger}, G., {Price}, S., {Salama}, A., {Schaeidt}, S.~G., {Sijm}, N.,
  {Smoorenburg}, C., {Spakman}, J., {Spoon}, H., {Steinmayer}, M., {Stoecker},
  J., {Valentijn}, E.~A., {Vandenbussche}, B., {Visser}, H., {Waelkens}, C.,
  {Waters}, L.~B.~F.~M., {Wensink}, J., {Wesselius}, P.~R., {Wiezorrek}, E.,
  {Wieprecht}, E., {Wijnbergen}, J.~J., {Wildeman}, K.~J., \& {Young}, E.
  1996{\natexlab{a}}, \aap, 315, L49

\bibitem[{{de Graauw} {et~al.}(1996{\natexlab{b}}){de Graauw}, {Whittet},
  {Gerakines}, {Bauer}, {Beintema}, {Boogert}, {Boxhoorn}, {Chiar},
  {Ehrenfreund}, {Feuchtgruber}, {Helmich}, {Heras}, {Huygen}, {Kester},
  {Kunze}, {Lahuis}, {Leech}, {Lutz}, {Morris}, {Prusti}, {Roelfsema},
  {Salama}, {Schaeidt}, {Schutte}, {Spoon}, {Tielens}, {Valentijn},
  {Vandenbusshe}, {van Dishoeck}, {Wesselius}, {Wieprecht}, \&
  {Wright}}]{Degraauw_96}
{de Graauw}, T., {Whittet}, D.~C.~B., {Gerakines}, P.~A., {Bauer}, O.~H.,
  {Beintema}, D.~A., {Boogert}, A.~C.~A., {Boxhoorn}, D.~R., {Chiar}, J.~E.,
  {Ehrenfreund}, P., {Feuchtgruber}, H., {Helmich}, F.~P., {Heras}, A.~M.,
  {Huygen}, R., {Kester}, D.~J.~M., {Kunze}, D., {Lahuis}, F., {Leech}, K.~J.,
  {Lutz}, D., {Morris}, P.~W., {Prusti}, T., {Roelfsema}, P.~R., {Salama}, A.,
  {Schaeidt}, S.~G., {Schutte}, W.~A., {Spoon}, H.~W.~W., {Tielens},
  A.~G.~G.~M., {Valentijn}, E.~A., {Vandenbusshe}, B., {van Dishoeck}, E.~F.,
  {Wesselius}, P.~R., {Wieprecht}, E., \& {Wright}, C.~M. 1996{\natexlab{b}},
  \aap, 315, L345

\bibitem[{{Dodds-Eden} {et~al.}(2009){Dodds-Eden}, {Porquet}, {Trap},
  {Quataert}, {Haubois}, {Gillessen}, {Grosso}, {Pantin}, {Falcke}, {Rouan},
  {Genzel}, {Hasinger}, {Goldwurm}, {Yusef-Zadeh}, {Clenet}, {Trippe},
  {Lagage}, {Bartko}, {Eisenhauer}, {Ott}, {Paumard}, {Perrin}, {Yuan},
  {Fritz}, \& {Mascetti}}]{Katie_09}
{Dodds-Eden}, K., {Porquet}, D., {Trap}, G., {Quataert}, E., {Haubois}, X.,
  {Gillessen}, S., {Grosso}, N., {Pantin}, E., {Falcke}, H., {Rouan}, D.,
  {Genzel}, R., {Hasinger}, G., {Goldwurm}, A., {Yusef-Zadeh}, F., {Clenet},
  Y., {Trippe}, S., {Lagage}, P., {Bartko}, H., {Eisenhauer}, F., {Ott}, T.,
  {Paumard}, T., {Perrin}, G., {Yuan}, F., {Fritz}, T.~K., \& {Mascetti}, L.
  2009, \apj, 698, 676

\bibitem[{{Draine}(1989)}]{Draine_89}
{Draine}, B.~T. 1989, in ESA Special Publication, Vol. 290, Infrared
  Spectroscopy in Astronomy, ed. {E.~B{\"o}hm-Vitense}, 93--98

\bibitem[{{Draine}(2003)}]{Draine_03}
{Draine}, B.~T. 2003, \araa, 41, 241

\bibitem[{{Dwek}(2004)}]{Dwek_04}
{Dwek}, E. 2004, \apjl, 611, L109

\bibitem[{{Eisenhauer} {et~al.}(2003{\natexlab{a}}){Eisenhauer}, {Abuter},
  {Bickert}, {Bianchet-Marchet}, {Bonnet}, {Brynnel}, {Conzelmann}, {Delabre},
  {Donaldson}, {Farinato}, {Fedrigo}, {Genzel}, {Hubin}, {Iserlohe}, {Kasper},
  {Kissler-Patig}, {Monnet}, {Roehrle}, {Scheiber}, {Stroebele}, {Tecza},
  {Thatte}, \& {Weisz}}]{Eisenhauer_etal2003}
{Eisenhauer}, F., {Abuter}, R., {Bickert}, K., {Bianchet-Marchet}, F.,
  {Bonnet}, H., {Brynnel}, J., {Conzelmann}, R.~D., {Delabre}, B., {Donaldson},
  R., {Farinato}, J., {Fedrigo}, E., {Genzel}, R., {Hubin}, N.~N., {Iserlohe},
  C., {Kasper}, M.~E., {Kissler-Patig}, M., {Monnet}, G.~J., {Roehrle}, C.,
  {Scheiber}, J., {Stroebele}, S., {Tecza}, M., {Thatte}, N.~A., \& {Weisz}, H.
  2003{\natexlab{a}}, SPIE, 4841, 1548

\bibitem[{{Eisenhauer} {et~al.}(2005){Eisenhauer}, {Genzel}, {Alexander},
  {Abuter}, {Paumard}, {Ott}, {Gilbert}, {Gillessen}, {Horrobin}, {Trippe},
  {Bonnet}, {Dumas}, {Hubin}, {Kaufer}, {Kissler-Patig}, {Monnet},
  {Str{\"o}bele}, {Szeifert}, {Eckart}, {Sch{\"o}del}, \&
  {Zucker}}]{Eisenhauer_05}
{Eisenhauer}, F., {Genzel}, R., {Alexander}, T., {Abuter}, R., {Paumard}, T.,
  {Ott}, T., {Gilbert}, A., {Gillessen}, S., {Horrobin}, M., {Trippe}, S.,
  {Bonnet}, H., {Dumas}, C., {Hubin}, N., {Kaufer}, A., {Kissler-Patig}, M.,
  {Monnet}, G., {Str{\"o}bele}, S., {Szeifert}, T., {Eckart}, A.,
  {Sch{\"o}del}, R., \& {Zucker}, S. 2005, \apj, 628, 246

\bibitem[{{Eisenhauer} {et~al.}(2003{\natexlab{b}}){Eisenhauer}, {Sch{\"o}del},
  {Genzel}, {Ott}, {Tecza}, {Abuter}, {Eckart}, \& {Alexander}}]{Eisenhauer_03}
{Eisenhauer}, F., {Sch{\"o}del}, R., {Genzel}, R., {Ott}, T., {Tecza}, M.,
  {Abuter}, R., {Eckart}, A., \& {Alexander}, T. 2003{\natexlab{b}}, \apjl,
  597, L121

\bibitem[{{Espinoza} {et~al.}(2009){Espinoza}, {Selman}, \&
  {Melnick}}]{Espinoza_09}
{Espinoza}, P., {Selman}, F.~J., \& {Melnick}, J. 2009, \aap, 501, 563

\bibitem[{{Fitzpatrick}(2004)}]{Fitzpatrick_04}
{Fitzpatrick}, E.~L. 2004, in Astronomical Society of the Pacific Conference
  Series, Vol. 309, Astrophysics of Dust, ed. {A.~N.~Witt, G.~C.~Clayton, \&
  B.~T.~Draine}, 33--+

\bibitem[{{Fitzpatrick} \& {Massa}(2009)}]{Fitzpatrick_09}
{Fitzpatrick}, E.~L. \& {Massa}, D. 2009, \apj, 699, 1209

\bibitem[{{Flaherty} {et~al.}(2007){Flaherty}, {Pipher}, {Megeath}, {Winston},
  {Gutermuth}, {Muzerolle}, {Allen}, \& {Fazio}}]{Flaherty_07}
{Flaherty}, K.~M., {Pipher}, J.~L., {Megeath}, S.~T., {Winston}, E.~M.,
  {Gutermuth}, R.~A., {Muzerolle}, J., {Allen}, L.~E., \& {Fazio}, G.~G. 2007,
  \apj, 663, 1069

\bibitem[{{F{\"o}rster Schreiber} {et~al.}(2001){F{\"o}rster Schreiber},
  {Genzel}, {Lutz}, {Kunze}, \& {Sternberg}}]{Foerster_01}
{F{\"o}rster Schreiber}, N.~M., {Genzel}, R., {Lutz}, D., {Kunze}, D., \&
  {Sternberg}, A. 2001, \apj, 552, 544

\bibitem[{{Fritz} {et~al.}(2010){Fritz}, {Gillessen}, {Dodds-Eden}, {Martins},
  {Bartko}, {Genzel}, {Paumard}, {Ott}, {Pfuhl}, {Trippe}, {Eisenhauer}, \&
  {Gratadour}}]{Fritz_10}
{Fritz}, T.~K., {Gillessen}, S., {Dodds-Eden}, K., {Martins}, F., {Bartko}, H.,
  {Genzel}, R., {Paumard}, T., {Ott}, T., {Pfuhl}, O., {Trippe}, S.,
  {Eisenhauer}, F., \& {Gratadour}, D. 2010, \apj, 721, 395

\bibitem[{{Froebrich} \& {del Burgo}(2006)}]{Froebrich_06}
{Froebrich}, D. \& {del Burgo}, C. 2006, \mnras, 369, 1901

\bibitem[{{Gao} {et~al.}(2009){Gao}, {Jiang}, \& {Li}}]{Gao_09}
{Gao}, J., {Jiang}, B.~W., \& {Li}, A. 2009, \apj, 707, 89

\bibitem[{{Gao} {et~al.}(2010){Gao}, {Jiang}, \& {Li}}]{Gao_10}
---. 2010, Earth, Planets, and Space, 62, 63

\bibitem[{{Genzel} {et~al.}(2010){Genzel}, {Eisenhauer}, \&
  {Gillessen}}]{Genzel_10}
{Genzel}, R., {Eisenhauer}, F., \& {Gillessen}, S. 2010, Reviews of Modern
  Physics, 82, 3121

\bibitem[{{Genzel} {et~al.}(2000){Genzel}, {Pichon}, {Eckart}, {Gerhard}, \&
  {Ott}}]{Genzel_00}
{Genzel}, R., {Pichon}, C., {Eckart}, A., {Gerhard}, O.~E., \& {Ott}, T. 2000,
  \mnras, 317, 348

\bibitem[{{Genzel} {et~al.}(2003){Genzel}, {Sch{\"o}del}, {Ott}, {Eckart},
  {Alexander}, {Lacombe}, {Rouan}, \& {Aschenbach}}]{Genzel_03}
{Genzel}, R., {Sch{\"o}del}, R., {Ott}, T., {Eckart}, A., {Alexander}, T.,
  {Lacombe}, F., {Rouan}, D., \& {Aschenbach}, B. 2003, \nat, 425, 934

\bibitem[{{Gerakines} {et~al.}(1999){Gerakines}, {Whittet}, {Ehrenfreund},
  {Boogert}, {Tielens}, {Schutte}, {Chiar}, {van Dishoeck}, {Prusti},
  {Helmich}, \& {de Graauw}}]{Gerakines_99}
{Gerakines}, P.~A., {Whittet}, D.~C.~B., {Ehrenfreund}, P., {Boogert},
  A.~C.~A., {Tielens}, A.~G.~G.~M., {Schutte}, W.~A., {Chiar}, J.~E., {van
  Dishoeck}, E.~F., {Prusti}, T., {Helmich}, F.~P., \& {de Graauw}, T. 1999,
  \apj, 522, 357

\bibitem[{{Ghez} {et~al.}(2008){Ghez}, {Salim}, {Weinberg}, {Lu}, {Do}, {Dunn},
  {Matthews}, {Morris}, {Yelda}, {Becklin}, {Kremenek}, {Milosavljevic}, \&
  {Naiman}}]{Ghez_09}
{Ghez}, A.~M., {Salim}, S., {Weinberg}, N.~N., {Lu}, J.~R., {Do}, T., {Dunn},
  J.~K., {Matthews}, K., {Morris}, M.~R., {Yelda}, S., {Becklin}, E.~E.,
  {Kremenek}, T., {Milosavljevic}, M., \& {Naiman}, J. 2008, \apj, 689, 1044

\bibitem[{{Gillessen} {et~al.}(2006){Gillessen}, {Eisenhauer}, {Quataert},
  {Genzel}, {Paumard}, {Trippe}, {Ott}, {Abuter}, {Eckart}, {Lagage},
  {Lehnert}, {Tacconi}, \& {Martins}}]{Gillessen_06}
{Gillessen}, S., {Eisenhauer}, F., {Quataert}, E., {Genzel}, R., {Paumard}, T.,
  {Trippe}, S., {Ott}, T., {Abuter}, R., {Eckart}, A., {Lagage}, P.~O.,
  {Lehnert}, M.~D., {Tacconi}, L.~J., \& {Martins}, F. 2006, \apjl, 640, L163

\bibitem[{{Gillessen} {et~al.}(2009){Gillessen}, {Eisenhauer}, {Trippe},
  {Alexander}, {Genzel}, {Martins}, \& {Ott}}]{Gillessen_09}
{Gillessen}, S., {Eisenhauer}, F., {Trippe}, S., {Alexander}, T., {Genzel}, R.,
  {Martins}, F., \& {Ott}, T. 2009, \apj, 692, 1075

\bibitem[{{Gosling} {et~al.}(2009){Gosling}, {Bandyopadhyay}, \&
  {Blundell}}]{Gosling_09}
{Gosling}, A.~J., {Bandyopadhyay}, R.~M., \& {Blundell}, K.~M. 2009, \mnras,
  394, 2247

\bibitem[{{Groenewegen}(2008)}]{Groenewegen_08}
{Groenewegen}, M.~A.~T. 2008, \aap, 488, 935

\bibitem[{{Groenewegen} {et~al.}(2008){Groenewegen}, {Udalski}, \&
  {Bono}}]{Groenewegen_08b}
{Groenewegen}, M.~A.~T., {Udalski}, A., \& {Bono}, G. 2008, \aap, 481, 441

\bibitem[{{Guesten} {et~al.}(1987){Guesten}, {Genzel}, {Wright}, {Jaffe},
  {Stutzki}, \& {Harris}}]{Guesten_87}
{Guesten}, R., {Genzel}, R., {Wright}, M.~C.~H., {Jaffe}, D.~T., {Stutzki}, J.,
  \& {Harris}, A.~I. 1987, \apj, 318, 124

\bibitem[{{He} {et~al.}(1995){He}, {Whittet}, {Kilkenny}, \& {Spencer
  Jones}}]{He_95}
{He}, L., {Whittet}, D.~C.~B., {Kilkenny}, D., \& {Spencer Jones}, J.~H. 1995,
  \apjs, 101, 335

\bibitem[{{Henry} {et~al.}(1984){Henry}, {Depoy}, \& {Becklin}}]{Henry_84}
{Henry}, J.~P., {Depoy}, D.~L., \& {Becklin}, E.~E. 1984, \apjl, 285, L27

\bibitem[{{Hornstein} {et~al.}(2007){Hornstein}, {Matthews}, {Ghez}, {Lu},
  {Morris}, {Becklin}, {Rafelski}, \& {Baganoff}}]{Hornstein_07}
{Hornstein}, S.~D., {Matthews}, K., {Ghez}, A.~M., {Lu}, J.~R., {Morris}, M.,
  {Becklin}, E.~E., {Rafelski}, M., \& {Baganoff}, F.~K. 2007, \apj, 667, 900

\bibitem[{{Hummer} \& {Storey}(1987)}]{Hummer_87}
{Hummer}, D.~G. \& {Storey}, P.~J. 1987, \mnras, 224, 801

\bibitem[{{Indebetouw} {et~al.}(2005){Indebetouw}, {Mathis}, {Babler}, {Meade},
  {Watson}, {Whitney}, {Wolff}, {Wolfire}, {Cohen}, {Bania}, {Benjamin},
  {Clemens}, {Dickey}, {Jackson}, {Kobulnicky}, {Marston}, {Mercer},
  {Stauffer}, {Stolovy}, \& {Churchwell}}]{Indebetouw_05}
{Indebetouw}, R., {Mathis}, J.~S., {Babler}, B.~L., {Meade}, M.~R., {Watson},
  C., {Whitney}, B.~A., {Wolff}, M.~J., {Wolfire}, M.~G., {Cohen}, M., {Bania},
  T.~M., {Benjamin}, R.~A., {Clemens}, D.~P., {Dickey}, J.~M., {Jackson},
  J.~M., {Kobulnicky}, H.~A., {Marston}, A.~P., {Mercer}, E.~P., {Stauffer},
  J.~R., {Stolovy}, S.~R., \& {Churchwell}, E. 2005, \apj, 619, 931

\bibitem[{{Jiang} {et~al.}(2006){Jiang}, {Gao}, {Omont}, {Schuller}, \&
  {Simon}}]{Jiang_06}
{Jiang}, B.~W., {Gao}, J., {Omont}, A., {Schuller}, F., \& {Simon}, G. 2006,
  \aap, 446, 551

\bibitem[{{Jiang} {et~al.}(2003){Jiang}, {Omont}, {Ganesh}, {Simon}, \&
  {Schuller}}]{Jiang_03}
{Jiang}, B.~W., {Omont}, A., {Ganesh}, S., {Simon}, G., \& {Schuller}, F. 2003,
  \aap, 400, 903

\bibitem[{{Kemper} {et~al.}(2004){Kemper}, {Vriend}, \& {Tielens}}]{Kemper_04}
{Kemper}, F., {Vriend}, W.~J., \& {Tielens}, A.~G.~G.~M. 2004, \apj, 609, 826

\bibitem[{{Kenyon} {et~al.}(1998){Kenyon}, {Lada}, \& {Barsony}}]{Kenyon_98}
{Kenyon}, S.~J., {Lada}, E.~A., \& {Barsony}, M. 1998, \aj, 115, 252

\bibitem[{{Knez} {et~al.}(2005){Knez}, {Boogert}, {Pontoppidan},
  {Kessler-Silacci}, {van Dishoeck}, {Evans}, {Augereau}, {Blake}, \&
  {Lahuis}}]{Knez_05}
{Knez}, C., {Boogert}, A.~C.~A., {Pontoppidan}, K.~M., {Kessler-Silacci}, J.,
  {van Dishoeck}, E.~F., {Evans}, II, N.~J., {Augereau}, J., {Blake}, G.~A., \&
  {Lahuis}, F. 2005, \apjl, 635, L145

\bibitem[{{Krabbe} {et~al.}(1991){Krabbe}, {Genzel}, {Drapatz}, \&
  {Rotaciuc}}]{Krabbe_91}
{Krabbe}, A., {Genzel}, R., {Drapatz}, S., \& {Rotaciuc}, V. 1991, \apjl, 382,
  L19

\bibitem[{{Landini} {et~al.}(1984){Landini}, {Natta}, {Salinari}, {Oliva}, \&
  {Moorwood}}]{Landini_84}
{Landini}, M., {Natta}, A., {Salinari}, P., {Oliva}, E., \& {Moorwood},
  A.~F.~M. 1984, \aap, 134, 284

\bibitem[{{Li}(2005)}]{Li_05}
{Li}, A. 2005, \apj, 622, 965

\bibitem[{{Li}(2007)}]{Li_07}
{Li}, A. 2007, in Astronomical Society of the Pacific Conference Series, Vol.
  373, The Central Engine of Active Galactic Nuclei, ed. {L.~C.~Ho \&
  J.-W.~Wang}, 561--+

\bibitem[{{Li} \& {Draine}(2001)}]{Li_01}
{Li}, A. \& {Draine}, B.~T. 2001, \apj, 554, 778

\bibitem[{{Li} \& {Greenberg}(1997)}]{Li_97}
{Li}, A. \& {Greenberg}, J.~M. 1997, \aap, 323, 566

\bibitem[{{Liu} {et~al.}(1993){Liu}, {Becklin}, {Henry}, \& {Simons}}]{Liu_93}
{Liu}, T., {Becklin}, E.~E., {Henry}, J.~P., \& {Simons}, D. 1993, \aj, 106,
  1484

\bibitem[{{Lo} \& {Claussen}(1983)}]{Lo_83}
{Lo}, K.~Y. \& {Claussen}, M.~J. 1983, \nat, 306, 647

\bibitem[{{Lombardi} {et~al.}(2006){Lombardi}, {Alves}, \&
  {Lada}}]{Lombardi_06}
{Lombardi}, M., {Alves}, J., \& {Lada}, C.~J. 2006, \aap, 454, 781

\bibitem[{{Lord}(1992)}]{Lord_92}
{Lord}, S.~D. 1992, NASA Technical Memorandum

\bibitem[{{Lutz}(1999)}]{Lutz_99}
{Lutz}, D. 1999, in ESA Special Publication, Vol. 427, The Universe as Seen by
  ISO, ed. {P.~Cox \& M.~Kessler}, 623--+

\bibitem[{{Lutz} {et~al.}(1996){Lutz}, {Feuchtgruber}, {Genzel}, {Kunze},
  {Rigopoulou}, {Spoon}, {Wright}, {Egami}, {Katterloher}, {Sturm},
  {Wieprecht}, {Sternberg}, {Moorwood}, \& {de Graauw}}]{Lutz_96}
{Lutz}, D., {Feuchtgruber}, H., {Genzel}, R., {Kunze}, D., {Rigopoulou}, D.,
  {Spoon}, H.~W.~W., {Wright}, C.~M., {Egami}, E., {Katterloher}, R., {Sturm},
  E., {Wieprecht}, E., {Sternberg}, A., {Moorwood}, A.~F.~M., \& {de Graauw},
  T. 1996, \aap, 315, L269

\bibitem[{{Maeda} {et~al.}(2002){Maeda}, {Baganoff}, {Feigelson}, {Morris},
  {Bautz}, {Brandt}, {Burrows}, {Doty}, {Garmire}, {Pravdo}, {Ricker}, \&
  {Townsley}}]{Maeda_02}
{Maeda}, Y., {Baganoff}, F.~K., {Feigelson}, E.~D., {Morris}, M., {Bautz},
  M.~W., {Brandt}, W.~N., {Burrows}, D.~N., {Doty}, J.~P., {Garmire}, G.~P.,
  {Pravdo}, S.~H., {Ricker}, G.~R., \& {Townsley}, L.~K. 2002, \apj, 570, 671

\bibitem[{{Maillard} {et~al.}(2004){Maillard}, {Paumard}, {Stolovy}, \&
  {Rigaut}}]{Maillard_03}
{Maillard}, J.~P., {Paumard}, T., {Stolovy}, S.~R., \& {Rigaut}, F. 2004, \aap,
  423, 155

\bibitem[{{Maiolino} {et~al.}(2001){Maiolino}, {Marconi}, {Salvati},
  {Risaliti}, {Severgnini}, {Oliva}, {La Franca}, \& {Vanzi}}]{Maiolino_01}
{Maiolino}, R., {Marconi}, A., {Salvati}, M., {Risaliti}, G., {Severgnini}, P.,
  {Oliva}, E., {La Franca}, F., \& {Vanzi}, L. 2001, \aap, 365, 28

\bibitem[{{Marshall} {et~al.}(2006){Marshall}, {Robin}, {Reyl{\'e}},
  {Schultheis}, \& {Picaud}}]{Marshall_06}
{Marshall}, D.~J., {Robin}, A.~C., {Reyl{\'e}}, C., {Schultheis}, M., \&
  {Picaud}, S. 2006, \aap, 453, 635

\bibitem[{{Martins} {et~al.}(2007){Martins}, {Genzel}, {Hillier}, {Eisenhauer},
  {Paumard}, {Gillessen}, {Ott}, \& {Trippe}}]{Martins_07}
{Martins}, F., {Genzel}, R., {Hillier}, D.~J., {Eisenhauer}, F., {Paumard}, T.,
  {Gillessen}, S., {Ott}, T., \& {Trippe}, S. 2007, \aap, 468, 233

\bibitem[{{Mathis}(1990)}]{Mathis_90}
{Mathis}, J.~S. 1990, \araa, 28, 37

\bibitem[{{Mathis}(1996)}]{Mathis_96}
---. 1996, \apj, 472, 643

\bibitem[{{Mathis} {et~al.}(1977){Mathis}, {Rumpl}, \& {Nordsieck}}]{Mathis_77}
{Mathis}, J.~S., {Rumpl}, W., \& {Nordsieck}, K.~H. 1977, \apj, 217, 425

\bibitem[{{Matsunaga} {et~al.}(2009){Matsunaga}, {Kawadu}, {Nishiyama},
  {Nagayama}, {Hatano}, {Tamura}, {Glass}, \& {Nagata}}]{Matsunaga_09}
{Matsunaga}, N., {Kawadu}, T., {Nishiyama}, S., {Nagayama}, T., {Hatano}, H.,
  {Tamura}, M., {Glass}, I.~S., \& {Nagata}, T. 2009, \mnras, 399, 1709

\bibitem[{{Messineo} {et~al.}(2005){Messineo}, {Habing}, {Menten}, {Omont},
  {Sjouwerman}, \& {Bertoldi}}]{Messineo_05}
{Messineo}, M., {Habing}, H.~J., {Menten}, K.~M., {Omont}, A., {Sjouwerman},
  L.~O., \& {Bertoldi}, F. 2005, \aap, 435, 575

\bibitem[{{Mezger} {et~al.}(1996){Mezger}, {Duschl}, \& {Zylka}}]{Mezger_96}
{Mezger}, P.~G., {Duschl}, W.~J., \& {Zylka}, R. 1996, \aapr, 7, 289

\bibitem[{{Moneti} {et~al.}(2001){Moneti}, {Cernicharo}, \&
  {Pardo}}]{Moneti_01}
{Moneti}, A., {Cernicharo}, J., \& {Pardo}, J.~R. 2001, \apjl, 549, L203

\bibitem[{{Moore} {et~al.}(2005){Moore}, {Lumsden}, {Ridge}, \&
  {Puxley}}]{Moore_05}
{Moore}, T.~J.~T., {Lumsden}, S.~L., {Ridge}, N.~A., \& {Puxley}, P.~J. 2005,
  \mnras, 359, 589

\bibitem[{{Morrison} \& {McCammon}(1983)}]{Morrison_83}
{Morrison}, R. \& {McCammon}, D. 1983, \apj, 270, 119

\bibitem[{{Naoi} {et~al.}(2007){Naoi}, {Tamura}, {Nagata}, {Nakajima}, {Suto},
  {Murakawa}, {Kandori}, {Sasaki}, {Nishiyama}, {Oasa}, \&
  {Sugitani}}]{Naoi_07}
{Naoi}, T., {Tamura}, M., {Nagata}, T., {Nakajima}, Y., {Suto}, H., {Murakawa},
  K., {Kandori}, R., {Sasaki}, S., {Nishiyama}, S., {Oasa}, Y., \& {Sugitani},
  K. 2007, \apj, 658, 1114

\bibitem[{{Naoi} {et~al.}(2006){Naoi}, {Tamura}, {Nakajima}, {Nagata}, {Suto},
  {Murakawa}, {Kandori}, {Sasaki}, {Baba}, {Kato}, {Kurita}, {Nagashima},
  {Nagayama}, {Nakaya}, {Nishiyama}, {Oasa}, {Sato}, \& {Sugitani}}]{Naoi_06}
{Naoi}, T., {Tamura}, M., {Nakajima}, Y., {Nagata}, T., {Suto}, H., {Murakawa},
  K., {Kandori}, R., {Sasaki}, S., {Baba}, D., {Kato}, D., {Kurita}, M.,
  {Nagashima}, C., {Nagayama}, T., {Nakaya}, H., {Nishiyama}, S., {Oasa}, Y.,
  {Sato}, S., \& {Sugitani}, K. 2006, \apj, 640, 373

\bibitem[{{Nishiyama} {et~al.}(2006{\natexlab{a}}){Nishiyama}, {Nagata},
  {Kusakabe}, {Matsunaga}, {Naoi}, {Kato}, {Nagashima}, {Sugitani}, {Tamura},
  {Tanab{\'e}}, \& {Sato}}]{Nishiyama_06b}
{Nishiyama}, S., {Nagata}, T., {Kusakabe}, N., {Matsunaga}, N., {Naoi}, T.,
  {Kato}, D., {Nagashima}, C., {Sugitani}, K., {Tamura}, M., {Tanab{\'e}}, T.,
  \& {Sato}, S. 2006{\natexlab{a}}, \apj, 638, 839

\bibitem[{{Nishiyama} {et~al.}(2006{\natexlab{b}}){Nishiyama}, {Nagata},
  {Sato}, {Kato}, {Nagayama}, {Kusakabe}, {Matsunaga}, {Naoi}, {Sugitani}, \&
  {Tamura}}]{Nishiyama_06}
{Nishiyama}, S., {Nagata}, T., {Sato}, S., {Kato}, D., {Nagayama}, T.,
  {Kusakabe}, N., {Matsunaga}, N., {Naoi}, T., {Sugitani}, K., \& {Tamura}, M.
  2006{\natexlab{b}}, \apj, 647, 1093

\bibitem[{{Nishiyama} {et~al.}(2008){Nishiyama}, {Nagata}, {Tamura}, {Kandori},
  {Hatano}, {Sato}, \& {Sugitani}}]{Nishiyama_08}
{Nishiyama}, S., {Nagata}, T., {Tamura}, M., {Kandori}, R., {Hatano}, H.,
  {Sato}, S., \& {Sugitani}, K. 2008, \apj, 680, 1174

\bibitem[{{Nishiyama} {et~al.}(2009){Nishiyama}, {Tamura}, {Hatano}, {Kato},
  {Tanab{\'e}}, {Sugitani}, \& {Nagata}}]{Nishiyama_09}
{Nishiyama}, S., {Tamura}, M., {Hatano}, H., {Kato}, D., {Tanab{\'e}}, T.,
  {Sugitani}, K., \& {Nagata}, T. 2009, \apj, 696, 1407

\bibitem[{{Paumard} {et~al.}(2006){Paumard}, {Genzel}, {Martins}, {Nayakshin},
  {Beloborodov}, {Levin}, {Trippe}, {Eisenhauer}, {Ott}, {Gillessen}, {Abuter},
  {Cuadra}, {Alexander}, \& {Sternberg}}]{Paumard_06}
{Paumard}, T., {Genzel}, R., {Martins}, F., {Nayakshin}, S., {Beloborodov},
  A.~M., {Levin}, Y., {Trippe}, S., {Eisenhauer}, F., {Ott}, T., {Gillessen},
  S., {Abuter}, R., {Cuadra}, J., {Alexander}, T., \& {Sternberg}, A. 2006,
  \apj, 643, 1011

\bibitem[{{Pendleton} \& {Allamandola}(2002)}]{Pendleton_02}
{Pendleton}, Y.~J. \& {Allamandola}, L.~J. 2002, \apjs, 138, 75

\bibitem[{{Philipp} {et~al.}(1999){Philipp}, {Zylka}, {Mezger}, {Duschl},
  {Herbst}, \& {Tuffs}}]{Philipp_99}
{Philipp}, S., {Zylka}, R., {Mezger}, P.~G., {Duschl}, W.~J., {Herbst}, T., \&
  {Tuffs}, R.~J. 1999, \aap, 348, 768

\bibitem[{{Porquet} {et~al.}(2008){Porquet}, {Grosso}, {Predehl}, {Hasinger},
  {Yusef-Zadeh}, {Aschenbach}, {Trap}, {Melia}, {Warwick}, {Goldwurm},
  {B{\'e}langer}, {Tanaka}, {Genzel}, {Dodds-Eden}, {Sakano}, \&
  {Ferrando}}]{Porquet_08}
{Porquet}, D., {Grosso}, N., {Predehl}, P., {Hasinger}, G., {Yusef-Zadeh}, F.,
  {Aschenbach}, B., {Trap}, G., {Melia}, F., {Warwick}, R.~S., {Goldwurm}, A.,
  {B{\'e}langer}, G., {Tanaka}, Y., {Genzel}, R., {Dodds-Eden}, K., {Sakano},
  M., \& {Ferrando}, P. 2008, \aap, 488, 549

\bibitem[{{Predehl} \& {Schmitt}(1995)}]{Predehl_95}
{Predehl}, P. \& {Schmitt}, J.~H.~M.~M. 1995, \aap, 293, 889

\bibitem[{{Racca} {et~al.}(2002){Racca}, {G{\'o}mez}, \& {Kenyon}}]{Racca_02}
{Racca}, G., {G{\'o}mez}, M., \& {Kenyon}, S.~J. 2002, \aj, 124, 2178

\bibitem[{{Ram{\'{\i}}rez} {et~al.}(2008){Ram{\'{\i}}rez}, {Arendt},
  {Sellgren}, {Stolovy}, {Cotera}, {Smith}, \& {Yusef-Zadeh}}]{Ramirez_08}
{Ram{\'{\i}}rez}, S.~V., {Arendt}, R.~G., {Sellgren}, K., {Stolovy}, S.~R.,
  {Cotera}, A., {Smith}, H.~A., \& {Yusef-Zadeh}, F. 2008, \apjs, 175, 147

\bibitem[{{Rawlings} {et~al.}(2003){Rawlings}, {Adamson}, \&
  {Whittet}}]{Rawlings_03}
{Rawlings}, M.~G., {Adamson}, A.~J., \& {Whittet}, D.~C.~B. 2003, \mnras, 341,
  1121

\bibitem[{{Reid}(1993)}]{Reid_93}
{Reid}, M.~J. 1993, \araa, 31, 345

\bibitem[{{Revnivtsev} {et~al.}(2010){Revnivtsev}, {van den Berg}, {Burenin},
  {Grindlay}, {Karasev}, \& {Forman}}]{Revnivtsev_10}
{Revnivtsev}, M., {van den Berg}, M., {Burenin}, R., {Grindlay}, J.~E.,
  {Karasev}, D., \& {Forman}, W. 2010, \aap, 515, A49+

\bibitem[{{Rieke} \& {Lebofsky}(1985)}]{Rieke_85}
{Rieke}, G.~H. \& {Lebofsky}, M.~J. 1985, \apj, 288, 618

\bibitem[{{Rieke} {et~al.}(1989){Rieke}, {Rieke}, \& {Paul}}]{Rieke_89}
{Rieke}, G.~H., {Rieke}, M.~J., \& {Paul}, A.~E. 1989, \apj, 336, 752

\bibitem[{{Rieke}(1999)}]{Rieke_99}
{Rieke}, M.~J. 1999, in Astronomical Society of the Pacific Conference Series,
  Vol. 186, The Central Parsecs of the Galaxy, ed. {H.~Falcke, A.~Cotera,
  W.~J.~Duschl, F.~Melia, \& M.~J.~Rieke}, 32--+

\bibitem[{{Roberts} \& {Goss}(1993)}]{Roberts_93}
{Roberts}, D.~A. \& {Goss}, W.~M. 1993, \apjs, 86, 133

\bibitem[{{Roberts} {et~al.}(1991){Roberts}, {Goss}, {van Gorkom}, \&
  {Leahy}}]{Roberts_91}
{Roberts}, D.~A., {Goss}, W.~M., {van Gorkom}, J.~H., \& {Leahy}, J.~P. 1991,
  \apjl, 366, L15

\bibitem[{{Roberts} {et~al.}(1996){Roberts}, {Yusef-Zadeh}, \&
  {Goss}}]{Roberts_96}
{Roberts}, D.~A., {Yusef-Zadeh}, F., \& {Goss}, W.~M. 1996, \apj, 459, 627

\bibitem[{{Roche} \& {Aitken}(1984)}]{Roche_84}
{Roche}, P.~F. \& {Aitken}, D.~K. 1984, \mnras, 208, 481

\bibitem[{{Roche} \& {Aitken}(1985)}]{Roche_85}
---. 1985, \mnras, 215, 425

\bibitem[{{Roelfsema} {et~al.}(1992){Roelfsema}, {Goss}, \&
  {Mallik}}]{Roelfsema_92}
{Roelfsema}, P.~R., {Goss}, W.~M., \& {Mallik}, D.~C.~V. 1992, \apj, 394, 188

\bibitem[{{Rom{\'a}n-Z{\'u}{\~n}iga} {et~al.}(2007){Rom{\'a}n-Z{\'u}{\~n}iga},
  {Lada}, {Muench}, \& {Alves}}]{Roman_07}
{Rom{\'a}n-Z{\'u}{\~n}iga}, C.~G., {Lada}, C.~J., {Muench}, A., \& {Alves},
  J.~F. 2007, \apj, 664, 357

\bibitem[{{Rosa} {et~al.}(1992){Rosa}, {Zinnecker}, {Moneti}, \&
  {Melnick}}]{Rosa_92}
{Rosa}, M.~R., {Zinnecker}, H., {Moneti}, A., \& {Melnick}, J. 1992, \aap, 257,
  515

\bibitem[{{Rosenthal} {et~al.}(2000){Rosenthal}, {Bertoldi}, \&
  {Drapatz}}]{Rosenthal_00}
{Rosenthal}, D., {Bertoldi}, F., \& {Drapatz}, S. 2000, \aap, 356, 705

\bibitem[{{Sakano} {et~al.}(2004){Sakano}, {Warwick}, {Decourchelle}, \&
  {Predehl}}]{Sakano_04}
{Sakano}, M., {Warwick}, R.~S., {Decourchelle}, A., \& {Predehl}, P. 2004,
  \mnras, 350, 129

\bibitem[{{Salaris} \& {Girardi}(2002)}]{Salaris_02}
{Salaris}, M. \& {Girardi}, L. 2002, \mnras, 337, 332

\bibitem[{{Savage} \& {Mathis}(1979)}]{Savage_79}
{Savage}, B.~D. \& {Mathis}, J.~S. 1979, \araa, 17, 73

\bibitem[{{Schlafly} {et~al.}(2010){Schlafly}, {Finkbeiner}, {Schlegel},
  {Juri{\'c}}, {Ivezi{\'c}}, {Gibson}, {Knapp}, \& {Weaver}}]{Schlafly_10}
{Schlafly}, E.~F., {Finkbeiner}, D.~P., {Schlegel}, D.~J., {Juri{\'c}}, M.,
  {Ivezi{\'c}}, {\v Z}., {Gibson}, R.~R., {Knapp}, G.~R., \& {Weaver}, B.~A.
  2010, \apj, 725, 1175

\bibitem[{{Sch{\"o}del} {et~al.}(2010){Sch{\"o}del}, {Najarro}, {Muzic}, \&
  {Eckart}}]{Schoedel_09b}
{Sch{\"o}del}, R., {Najarro}, F., {Muzic}, K., \& {Eckart}, A. 2010, \aap, 511,
  A18+

\bibitem[{{Schreiber} {et~al.}(2004){Schreiber}, {Thatte}, {Eisenhauer},
  {Tecza}, {Abuter}, \& {Horrobin}}]{Schreiber_04}
{Schreiber}, J., {Thatte}, N., {Eisenhauer}, F., {Tecza}, M., {Abuter}, R., \&
  {Horrobin}, M. 2004, in Astronomical Society of the Pacific Conference
  Series, Vol. 314, Astronomical Data Analysis Software and Systems (ADASS)
  XIII, ed. {F.~Ochsenbein, M.~G.~Allen, \& D.~Egret}, 380--+

\bibitem[{{Schultz} \& {Wiemer}(1975)}]{Schultz_75}
{Schultz}, G.~V. \& {Wiemer}, W. 1975, \aap, 43, 133

\bibitem[{{Scoville} {et~al.}(2003){Scoville}, {Stolovy}, {Rieke},
  {Christopher}, \& {Yusef-Zadeh}}]{Scoville_03}
{Scoville}, N.~Z., {Stolovy}, S.~R., {Rieke}, M., {Christopher}, M., \&
  {Yusef-Zadeh}, F. 2003, \apj, 594, 294

\bibitem[{{Shukla} {et~al.}(2004){Shukla}, {Yun}, \& {Scoville}}]{Shukla_04}
{Shukla}, H., {Yun}, M.~S., \& {Scoville}, N.~Z. 2004, \apj, 616, 231

\bibitem[{{Sneden} {et~al.}(1978){Sneden}, {Gehrz}, {Hackwell}, {York}, \&
  {Snow}}]{Sneden_78}
{Sneden}, C., {Gehrz}, R.~D., {Hackwell}, J.~A., {York}, D.~G., \& {Snow},
  T.~P. 1978, \apj, 223, 168

\bibitem[{{Stead} \& {Hoare}(2009)}]{Stead_09}
{Stead}, J.~J. \& {Hoare}, M.~G. 2009, \mnras, 400, 731

\bibitem[{{Strai{\v z}ys} \& {Laugalys}(2008)}]{Straizys_08b}
{Strai{\v z}ys}, V. \& {Laugalys}, V. 2008, Baltic Astronomy, 17, 253

\bibitem[{{Sumi}(2004)}]{Sumi_04}
{Sumi}, T. 2004, \mnras, 349, 193

\bibitem[{{Sutton} {et~al.}(1990){Sutton}, {Danchi}, {Jaminet}, \&
  {Masson}}]{Sutton_90}
{Sutton}, E.~C., {Danchi}, W.~C., {Jaminet}, P.~A., \& {Masson}, C.~R. 1990,
  \apj, 348, 503

\bibitem[{{Tamblyn} {et~al.}(1996){Tamblyn}, {Rieke}, {Hanson}, {Close},
  {McCarthy}, \& {Rieke}}]{Tamblyn_96}
{Tamblyn}, P., {Rieke}, G.~H., {Hanson}, M.~M., {Close}, L.~M., {McCarthy},
  Jr., D.~W., \& {Rieke}, M.~J. 1996, \apj, 456, 206

\bibitem[{{Thuan} {et~al.}(1999){Thuan}, {Sauvage}, \& {Madden}}]{Thuan_99}
{Thuan}, T.~X., {Sauvage}, M., \& {Madden}, S. 1999, \apj, 516, 783

\bibitem[{{Tokunaga} \& {Vacca}(2005)}]{Tokunaga_05}
{Tokunaga}, A.~T. \& {Vacca}, W.~D. 2005, \pasp, 117, 421

\bibitem[{{Tran} {et~al.}(2001){Tran}, {Lutz}, {Genzel}, {Rigopoulou}, {Spoon},
  {Sturm}, {Gerin}, {Hines}, {Moorwood}, {Sanders}, {Scoville}, {Taniguchi}, \&
  {Ward}}]{Tran_01}
{Tran}, Q.~D., {Lutz}, D., {Genzel}, R., {Rigopoulou}, D., {Spoon}, H.~W.~W.,
  {Sturm}, E., {Gerin}, M., {Hines}, D.~C., {Moorwood}, A.~F.~M., {Sanders},
  D.~B., {Scoville}, N., {Taniguchi}, Y., \& {Ward}, M. 2001, \apj, 552, 527

\bibitem[{{Trippe} {et~al.}(2008){Trippe}, {Gillessen}, {Gerhard}, {Bartko},
  {Fritz}, {Maness}, {Eisenhauer}, {Martins}, {Ott}, {Dodds-Eden}, \&
  {Genzel}}]{Trippe_08}
{Trippe}, S., {Gillessen}, S., {Gerhard}, O.~E., {Bartko}, H., {Fritz}, T.~K.,
  {Maness}, H.~L., {Eisenhauer}, F., {Martins}, F., {Ott}, T., {Dodds-Eden},
  K., \& {Genzel}, R. 2008, \aap, 492, 419

\bibitem[{{Udalski}(2003)}]{Udalski_03}
{Udalski}, A. 2003, \apj, 590, 284

\bibitem[{{van Breemen} {et~al.}(2010){van Breemen}, {Min}, {Chiar}, {Waters},
  {Kemper}, {Boogert}, {Cami}, {Decin}, {Knez}, {Sloan}, \&
  {Tielens}}]{Vanbreemen_10}
{van Breemen}, J.~M., {Min}, M., {Chiar}, J.~E., {Waters}, L.~B.~F.~M.,
  {Kemper}, F., {Boogert}, A.~C.~A., {Cami}, J., {Decin}, L., {Knez}, C.,
  {Sloan}, G.~C., \& {Tielens}, A.~G.~G.~M. 2010, ArXiv e-prints

\bibitem[{{Viehmann} {et~al.}(2005){Viehmann}, {Eckart}, {Schoedel},
  {Moultaka}, {Straubmeier}, \& {Pott}}]{Viehmann_05}
{Viehmann}, T., {Eckart}, A., {Schoedel}, R., {Moultaka}, J., {Straubmeier},
  C., \& {Pott}, J. 2005, VizieR Online Data Catalog, 343, 30117

\bibitem[{{Voshchinnikov} {et~al.}(2006){Voshchinnikov}, {Il'in}, {Henning}, \&
  {Dubkova}}]{Voshchinnikov_06}
{Voshchinnikov}, N.~V., {Il'in}, V.~B., {Henning}, T., \& {Dubkova}, D.~N.
  2006, \aap, 445, 167

\bibitem[{{Weingartner} \& {Draine}(2001)}]{Weingartner_01}
{Weingartner}, J.~C. \& {Draine}, B.~T. 2001, \apj, 548, 296

\bibitem[{{Whittet}(1988)}]{Whittet_88}
{Whittet}, D.~C.~B. 1988, \mnras, 230, 473

\bibitem[{{Whittet} {et~al.}(1988){Whittet}, {Bode}, {Longmore}, {Adamson},
  {McFadzean}, {Aitken}, \& {Roche}}]{Whittet_88b}
{Whittet}, D.~C.~B., {Bode}, M.~F., {Longmore}, A.~J., {Adamson}, A.~J.,
  {McFadzean}, A.~D., {Aitken}, D.~K., \& {Roche}, P.~F. 1988, \mnras, 233, 321

\bibitem[{{Whittet} {et~al.}(1997){Whittet}, {Boogert}, {Gerakines}, {Schutte},
  {Tielens}, {de Graauw}, {Prusti}, {van Dishoeck}, {Wesselius}, \&
  {Wright}}]{Whittet_97}
{Whittet}, D.~C.~B., {Boogert}, A.~C.~A., {Gerakines}, P.~A., {Schutte}, W.,
  {Tielens}, A.~G.~G.~M., {de Graauw}, T., {Prusti}, T., {van Dishoeck}, E.~F.,
  {Wesselius}, P.~R., \& {Wright}, C.~M. 1997, \apj, 490, 729

\bibitem[{{Wieprecht} {et~al.}(1998){Wieprecht}, {Lahuis}, {Bauer}, {Boxhoorn},
  {Huygen}, {Kester}, {Leech}, {Roelfsema}, {Sturm}, {Sym}, \&
  {Vandenbussche}}]{Wieprecht_98}
{Wieprecht}, E., {Lahuis}, F., {Bauer}, O.~H., {Boxhoorn}, D., {Huygen}, R.,
  {Kester}, D., {Leech}, K.~J., {Roelfsema}, P., {Sturm}, E., {Sym}, N.~J., \&
  {Vandenbussche}, B. 1998, in Astronomical Society of the Pacific Conference
  Series, Vol. 145, Astronomical Data Analysis Software and Systems VII, ed.
  {R.~Albrecht, R.~N.~Hook, \& H.~A.~Bushouse}, 279--+

\bibitem[{{Willner} {et~al.}(1979){Willner}, {Russell}, {Puetter}, {Soifer}, \&
  {Harvey}}]{Willner_79}
{Willner}, S.~P., {Russell}, R.~W., {Puetter}, R.~C., {Soifer}, B.~T., \&
  {Harvey}, P.~M. 1979, \apjl, 229, L65

\bibitem[{{Yusef-Zadeh} {et~al.}(1998){Yusef-Zadeh}, {Roberts}, \&
  {Biretta}}]{Yusef_98}
{Yusef-Zadeh}, F., {Roberts}, D.~A., \& {Biretta}, J. 1998, \apjl, 499, L159+

\bibitem[{{Yusef-Zadeh} \& {Wardle}(1993)}]{Yusef_93}
{Yusef-Zadeh}, F. \& {Wardle}, M. 1993, \apj, 405, 584

\bibitem[{{Zasowski} {et~al.}(2009){Zasowski}, {Majewski}, {Indebetouw},
  {Meade}, {Nidever}, {Patterson}, {Babler}, {Skrutskie}, {Watson}, {Whitney},
  \& {Churchwell}}]{Zasowski_09}
{Zasowski}, G., {Majewski}, S.~R., {Indebetouw}, R., {Meade}, M.~R., {Nidever},
  D.~L., {Patterson}, R.~J., {Babler}, B., {Skrutskie}, M.~F., {Watson}, C.,
  {Whitney}, B.~A., \& {Churchwell}, E. 2009, \apj, 707, 510

\bibitem[{{Zubko} {et~al.}(2004){Zubko}, {Dwek}, \& {Arendt}}]{Zubko_04}
{Zubko}, V., {Dwek}, E., \& {Arendt}, R.~G. 2004, \apjs, 152, 211

\end{thebibliography}

\appendix
\section{A: Correcting for the extinction law flattening bias due to inhomogeneous extinction}
\label{sec_ext_bias}

Here, we estimate the extinction law flattening bias  due to inhomogeneous extinction. 

We  make the assumption that the true extinction map $A_i(\lambda)$, for resolution element $i$ and wavelength $\lambda$,
 has the same relative spatial distribution of extinction as the \citet{Schoedel_09b} map, such that we can obtain the extinction map at a given 
wavelength $\lambda$ via a simple scaling of an (as yet unknown) factor $x(\lambda)$: 

\begin{equation} 
A_{i}(\lambda)= x(\lambda) \times A_{i\,\mathrm{Sch}} 
\end{equation} 

We then derive an unextincted Brackett-$\gamma$ flux map from our observed Brackett-$\gamma$ flux map, using the extinction map of 
\citet{Schoedel_09b} (scaled to our measured Brackett-$\gamma$ extinction) as model for the spatial inhomogeneity in the extinction at this wavelength. For resolution element $i$, the unextincted flux map can be written as:
\begin{equation} 
F_i(\,\lambda=\mathrm{Br}\,\gamma\,)_{\mathrm{unext}} =
F_i(\,\lambda=\mathrm{Br}\,\gamma\,)_{\mathrm{obs}}*10^{0.4\,x_0\,A\left(\mathrm{map}\right)_{\mathrm{Sch}}}.
\end{equation} 
Here we use as scaling factor, $x_0$, our integrated measurement of the Brackett gamma extinction (Section \ref{sec:ext}), divided by the (observed) flux-weighted extinction of the \citet{Schoedel_09b} map:
\begin{equation} 
x_0= A(\,\lambda=\mathrm{Br}\,\gamma)_{\mathrm{integrated,measured}}  \left(
 \frac { \sum_{i\in \mathrm{ISO}} F_i(\,\lambda=\mathrm{Br}\,\gamma\,)_{\mathrm{ext}}\, A_{i\,{\mathrm{Sch}}}  }
{\sum_{i \in \mathrm{ISO}} F_i(\,\lambda=\mathrm{Br}\,\gamma\,)_\mathrm{ext}} \right)^{-1}.
\end{equation}
We use the unextincted Brackett-$\gamma$ flux map as a model for the spatial distribution of the intrinsic (unextincted) flux at all wavelengths (i.e. the correct distribution of relative flux weights).

The true (extinction-law-conserving) integrated extinction should be flux-weighted by the unextincted flux, not the observed flux: \begin{equation} 
A\left(\lambda\right)_{\mathrm{integrated,true}}=x(\lambda) \times \frac { \sum_{i\in \mathrm{ISO}} F_i(\,\lambda=\mathrm{Br}
\,\gamma\,)_{\mathrm{unext}}\, A_{i\,\mathrm{Sch}}  }
{\sum_{i\in \mathrm{ISO}} F_i(\,\lambda=\mathrm{Br}\,\gamma\,)_\mathrm{unext}}
\label{eq:trueextinction}\end{equation} 
Our extinction measurement, on the other hand, measures rather the ratio of integrated fluxes: %(provided a wavelength at which the extinction is negligible) 
\begin{equation} 
A(\lambda)_{\mathrm{integrated,measured}}=-2.5\times \log\left( \frac { \sum_{i\in \mathrm{ISO}} F_i(\,\lambda=\mathrm{Br}\,\gamma\,)_\mathrm{ext}  } 
 { \sum_{i\in \mathrm{ISO}} F_i(\,\lambda=\mathrm{Br}\,\gamma\,)_\mathrm{unext}  } 
 \right) 
\end{equation} 
In this equation, areas with smaller extinction have higher observed fluxes relative to the unextincted flux and
 thus are given a higher weight. Upon integration this then leads to a smaller measured extinction than the true integrated
 extinction of Equation \ref{eq:trueextinction}.

To estimate this bias, we simulate our extinction measurements, weighting the intrinsic extinction, $x(\lambda) A_{i\,\mathrm{Sch}}$, by the unextincted flux weights derived from the Brackett-$\gamma$ map: 
\begin{equation} 
A(\lambda)_{\mathrm{integrated,simulated}}=-2.5\times \log \left( \frac { \sum_{i\in \mathrm{ISO}} 10^{-0.4\,x(\lambda)\,
A_{i\,\mathrm{Sch}}\,F_i(\,\lambda=\mathrm{Br}\,\gamma\,)_{\mathrm{unext}} } }
 { \sum_{\mathrm{i\in \mathrm{ISO}}} F_i(\,\lambda=\mathrm{Br}\,\gamma\,)_{\mathrm{unext} }  }
 \right) \label{eq:simulatedextinction}
\end{equation}
For every SINFONI line at wavelength $\lambda$ we find an $x(\lambda)$ which, when substituted in Equation \ref{eq:simulatedextinction}, results in our measured extinction $A(\lambda)_{\mathrm{integrated,measured}}$. The calculation is carried out over the same area (ISO) used in the extinction measurement for each line. The bias at a given wavelength $\lambda$ is then: \begin{equation} \Delta A(\lambda)= A(\lambda)_{\mathrm{integrated, true}}-A(\lambda)_{\mathrm{integrated,simulated}} 
\end{equation} 
This correction must be applied to $A(\lambda)_{\mathrm{integrated,measured}}$ 
in order to obtain the correct (extinction-law-conserving) extinction. 
The biggest correction occurs for the highest extinction (Paschen-$\beta$). Even for this line the correction is only 0.08 mag, smaller than the extinction error of 0.11 mag. 
 
% done in janalysis/rel.ext4.dpuser

\section{B: Interpolating the extinction curve}
\label{sec:int_ext}

\begin{figure}
\begin{center}
\includegraphics[width=0.415 \columnwidth,angle=-90]{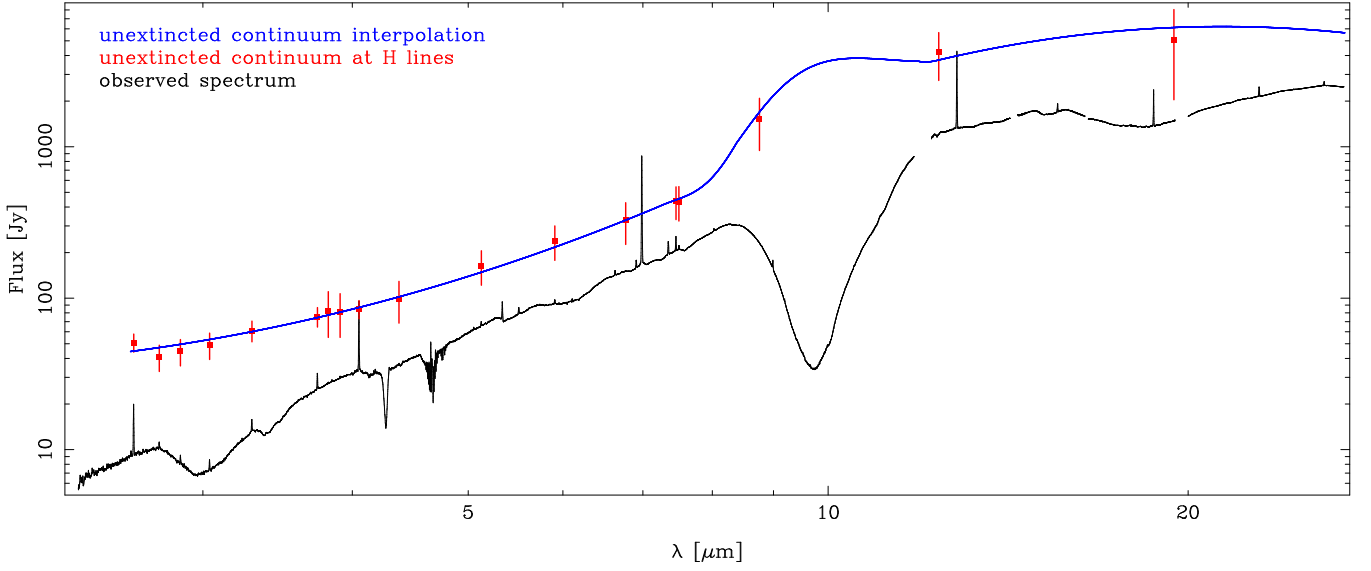}
\caption{Measured and unextinced MIR continuum towards the GC. The black line is the observed ISO-SWS spectrum. 
We unextinct the continuum around the hydrogen lines by
the hydrogen line extinctions there (red boxes). We interpolate these points to obtain the unextincted continuum (blue line). 
} 
\label{fig:cont_int}
\end{center}
\end{figure}

We use the continuum of the ISO-SWS spectrum in order to constrain the shape of the extinction curve in the MIR.
The unextincted continuum cannot be modeled with a single simple model such as a single black- or graybody. 
Instead, we correct the measured continuum close to the hydrogen lines by using the extinction of the lines.
We then interpolate the resulting points using simple models for sections of the spectrum, described in the paragraph below. 
After this, we obtain the extinction curve by dividing the 
interpolated extinction-corrected continuum by the observed continuum. By applying this method, we use the extinction of the hydrogen lines to obtain extinction values in regions where no hydrogen lines are available.

Between 2.62 and 7.6 $\mu$m we fit the extinction corrected continuum points with a second order polynomial in log-log space.
At longer wavelengths however the interpolation is less certain, due to the fact that there are fewer lines, larger 
extinction errors and silicate emission in the GC region \citep{Kemper_04}. 
For the continuum up to 12 $\mu$m we use the GC continuum of \citet{Kemper_04}, which contains silicate emission,
scaled such that it fits our two data points 
in this wavelength range.
A 240 K blackbody is a good approximation to the extincted continuum between 12 and 15 $\mu$m. In
this range there are no significant extinction features. We extrapolate this blackbody up to 26 $\mu$m, scaling it such that the unextincted continuum close to both lines in this range is fit by the blackbody, and use it as a model for the continuum between 12 and 26 $\mu$m.
We combine the three different continua with smooth transitions around 7 $\mu$m and 12 $\mu$m, see Figure~\ref{fig:cont_int}.

In a similar way, we make a smooth transition to the $\alpha=-2.11$ power-law shortward of 2.75 $\mu$m.
We continue the interpolation by a $\alpha=-2.02$ power-law shortward of 1.28 $\mu$m
We also correct unphysical jumps in the extinction curve, such as the one at 4.07 $\mu$m, due to calibration errors of the ISO-SWS data. 
For this we  adjust the extinction curve locally with linear interpolation. We interpolate the extinction curve we obtain, excluding spectral lines, 
with splines (Table~\ref{tab:int_ext}). The spectral resolution of this interpolation is not high enough to fully resolve all features, such as, e.g., the 
CO feature at 4.7 $\mu$m.

\begin{deluxetable}{lll} 
\tabletypesize{\scriptsize}
\tablecolumns{3}
\tablewidth{30pc}
\tablecaption{Interpolated infrared extinction curve \label{tab:int_ext}}
\tablehead{ $\lambda$[$\mu$m] & A(GC) & $\delta$A (GC) 
}
\startdata
1.00009 & 13.1075 & 0.3 \\
1.00984 & 12.8529 & 0.2925 \\
1.0197 & 12.6033 & 0.285 \\
1.02964 & 12.3586 & 0.2776 \\
1.03969 & 12.1186 & 0.2701 \\
1.04983 & 11.8832 & 0.2626 \\
1.06007 & 11.6525 & 0.2551 \\
1.07041 & 11.4262 & 0.2477 \\
1.08086 & 11.2043 & 0.2402 \\
1.0914 & 10.9867 & 0.2327 \\
\enddata
\tablecomments{
Our extinction curve. The second column is the average extinction towards the central 
14'' $\times$20'' of the GC.
 The errors includes all effects. Some effects matter only for comparison over big wavelength ranges.
Locally the error is smaller for most of ISO-SWS range (2.6 to 27 $\mu$m).
This curve can be used for $A(\lambda)$ in Formula~\ref{eq:extsim} for obtaining the extinction for other objects and filters. 
Scaled to other absolute extinctions 
it can be also  useful outside the Galactic Center.
The full table is available in the electronic edition.}
\end{deluxetable}

\section{C: Effective broadband filter extinctions}
\label{sec:filt_ext}

A complication in correcting for extinction in broadband flux measurements is that the extinction that should be 
associated with a given filter varies with both the intrinsic spectrum of the source and the strength of the extinction.

Since we use spectral lines in this paper to constrain the extinction law, the wavelengths to associate with the extinction measurements are 
 well-defined. Also the interpolated extinction  (Appendix~\ref{sec:int_ext}) has well defined wavelengths.  
Thus, we can explicitly calculate the extinction values for a given broadband filter, like \citet{Espinoza_09}, using the equation:

\begin{equation} A_{\mathrm{filter}}=-2.5\times \log \left( \frac {\int \lambda\,F_{\lambda}(\lambda) \, S(\lambda) \, R(\lambda)\, d\lambda} {\int \lambda\,F_{\lambda}(\lambda) \, S(\lambda)\, d\lambda} \right).
\label{eq:extsim}
\end{equation}
Here F$_{\lambda}(\lambda)$ is the intrinsic flux density of the source, which when multiplied with the wavelength, $\lambda$, is proportional to the number distribution of photons with wavelength. $S(\lambda)$ is the wavelength  dependent  throughput of telescope, instrument and atmosphere. We also define here $R(\lambda) = 10^{\left(-0.4\,A(\lambda)\right)}$ as a ``reddening factor'' to separate the extinction from the intrinsic flux. 

We use our interpolated extinction curve for $A(\lambda)$ (see Section~\ref{sec:int_ext} and Table~\ref{tab:int_ext}) in Formula~\ref{eq:extsim} to calculate the effective broadband filter extinction values for a range of commonly used infrared filters. The results are presented in Table \ref{tab:filt_ext}. As source spectrum we use primarily a blackbody of 9480 K. For stars of this temperature, there is no major difference between the real spectrum and the correct blackbody in the infrared. The use of such a star also has the advantage that a similar star is used for the definition of the Vega magnitude system.
 In addition, we also compute the effective filter extinctions for a bright Sgr~A* like spectrum \citep{Eisenhauer_05,Gillessen_06} which follows a power-law  with a slope $\beta=0.5$ 
 in $\nu L_{\nu}$ \citep{Hornstein_07,Katie_09} as example of a very red source. 
For $S(\lambda)$ we use the atmospheric transmission multiplied with the instrument transmission. 
For all Paranal filters we use the atran transmission models \citep{Lord_92} from Cerro Panchon for the atmosphere, with airmass 1 and 2.3 mm water vapor column from the Gemini web site\footnote{http://www.gemini.edu/?q=node/10789 \label{geminifootnote}
}.
For the VISIR filters we use the filter transmissions from the instrument web site\footnote{http://www.eso.org/sci/facilities/paranal/instruments/visir/inst/index.html}.
We obtain the VIRCAM filters from the instrument web site\footnote{http://www.eso.org/sci/facilities/paranal/instruments/vircam/inst/}.
For the NIRC2 filters, which we obtain from the instrument web site\footnote{http://www2.keck.hawaii.edu/inst/nirc2/filters.html}, we use a Mauna Kea atmosphere
of airmass 1.5 and 1.6 mm water vapor column from the Gemini web site (see footnote \ref{geminifootnote}).
For these instruments only in case of VIRCAM the wavelength dependent quantum efficiency and mirror reflectivity are available. We use them for our calculation.
 However, including QE and mirror reflectivity change the extinction  only by less than 0.07\% relatively. 
For the other instruments we assume that the throughput of the instrument is apart of the filter not
wavelength dependent within a filter. 
For 2MASS we use the full transmissions including the atmosphere from the project web site\footnote{http://www.ipac.caltech.edu/2mass/releases/second/doc/sec3$\mathunderscore$1b1.html\#s16}.
We obtain the NICMOS transmissions from the instrument web site\footnote{http://www.stsci.edu/hst/nicmos/design/filters} as in case of the IRAC 
transmissions\footnote{http://ssc.spitzer.caltech.edu/irac/calibrationfiles/spectralresponse}.

Although, strictly speaking, it is necessary to calculate the effective broadband extinction on a per spectrum basis, 
the differences in the effective extinction for different source SEDs are mostly relatively small, even for the high extinction of the GC. 
For example, the difference in effective extinction between a blue spectrum (Vega) and a red spectrum (Sgr~A) is only 0.056 mag in the H-band
 and 0.026 mag in the Ks-band. 

We test the influence of difference magnitudes of extinctions on broad band extinctions. 
The H-band extinction derived from A$_{Br\,\gamma}=2.62$ is 4.65 mag,
while the H-band extinction derived from A$_{Br\,\gamma}=0.5$ is 0.91 mag.
 Assuming linear scaling of the H-band extinction with the Brackett-$\gamma$ extinction the expected H-band extinction 
 (scaled up from A$_{Br\,\gamma}=0.5$) is $0.91 \mathrm{mag} \times 2.62/0.5=4.77$ mag. 
Thus, the deviation from linearity (the non-linearity) is 0.12 mag in this case.
In the Ks-band the non-linearity is, with 0.029 mag, much smaller for the same value of 
A$_{Br\,\gamma}$. We find that the non-linearity is of the order
of the measurement error for $A_{\mathrm{band}}\leq4$.

For most objects, for which the extinction is not significantly higher than in the GC, it is 
sufficient to use the extinction closest to the measured extinction in Table~\ref{tab:filt_ext}
renormalized to the correct absolute extinction. 
For very high extinctions or different filters it is necessary to calculate the filter extinctions from Formula~\ref{eq:extsim} using the interpolated
extinction curve given in Table~\ref{tab:int_ext}. 

\begin{deluxetable}{llllll}
\tabletypesize{\scriptsize}
\tablecolumns{6}
\tablewidth{0pc}
\tablecaption{Extinction of broadband filters for different Brackett-$\gamma$ extinctions \label{tab:filt_ext}}
\tablehead{Instrument & Filter & A (A$_{Br\,\gamma}=2.62$) &  A (A$_{Br\,\gamma}=0.5$) &  A (A$_{Br\,\gamma}=6$) &  A (A$_{Br\,\gamma}=2.40$)\\
                              & & 9480 K & 9480 K & 9480 K & $\beta=$0.5}
\startdata
NACO & J  & $ 8.16 \pm 0.12 $ & $1.62 \pm 0.03$  & $17.90 \pm 0.26$ & $7.42 \pm 0.11$ \\
NACO & H  & $ 4.65 \pm 0.11 $ & $0.91 \pm 0.02$  & $10.29 \pm 0.25$ & $4.21 \pm 0.10$\\
NACO & Ks  & $ 2.67 \pm 0.11 $ & $0.52 \pm 0.02$  & $6.02 \pm 0.26$ & $2.42 \pm 0.10$ \\
NACO & L'  & $ 1.20 \pm 0.14 $ & $0.23 \pm 0.03$  & $2.73 \pm 0.32$ & $1.09 \pm 0.13 $\\
NACO & M'  & $ 1.05 \pm 0.25 $ & $0.20 \pm 0.05$  & $2.37 \pm 0.57$ & $0.95 \pm 0.23$ \\
VISIR & PAH1 (8.6 $\mu$m)   & $ 1.73 \pm 0.54 $ & $0.34 \pm 0.10$  & $3.83 \pm 1.23$ & $1.60 \pm 0.50$ \\
VISIR & PAH2$\mathunderscore$2 (11.88 $\mu$m)   & $ 1.61 \pm 0.36 $ & $0.31 \pm 0.07$  & $3.66 \pm 0.83$ & $1.47 \pm 0.33$ \\
NICMOS & 110M & $ 10.42 \pm 0.19$ & $2.07 \pm 0.04$ & $ 22.86 \pm 0.40$ & $ 9.49 \pm 0.18$ \\
NICMOS & 145M & $6.00 \pm 0.11$ & $ 1.16 \pm 0.02$ & $13.44 \pm 0.25$ & $5.47 \pm 0.10 $ \\
NICMOS & 160W & $4.89 \pm 0.11$ & $ 0.97 \pm 0.02$ & $10.61 \pm 0.25$ & $4.40 \pm 0.10$ \\
NICMOS & 170M & $ 4.33 \pm 0.11 $ & $0.83 \pm 0.02 $ & $9.80 \pm 0.25 $ & $3.95 \pm 0.10 $ \\
NICMOS & 222M & $2.50 \pm 0.11 $ & $0.48 \pm 0.02 $ & $ 5.70 \pm 0.26 $ & $2.28 \pm 0.10$ \\
IRAC & Band 1 &  $1.47 \pm 0.14$ & $0.28 \pm 0.03 $ & $3.27 \pm 0.32$  & $1.31 \pm 0.13 $  \\
IRAC & Band 2 &  $1.06 \pm 0.22$ & $0.20 \pm 0.04 $ & $2.40 \pm 0.50$ & $0.97 \pm 0.20$  \\
IRAC & Band 3 &  $ 0.91 \pm 0.25$ & $ 0.17 \pm 0.05$ & $ 2.08\pm 0.57 $ & $0.83 \pm 0.23$ \\
IRAC & Band 4 &  $1.02 \pm 0.31 $ & $0.21 \pm 0.06 $ & $2.19 \pm 0.67$ & $1.01 \pm 0.30 $  \\
NIRC2 & H & $4.75 \pm 0.11$ & $0.93 \pm 0.02 $ & $10.56 \pm 0.25 $ & $4.31 \pm 0.10 $ \\
NIRC2 & K' & $2.73 \pm 0.11 $  & $ 0.53 \pm 0.02 $ &  $6.15 \pm 0.26 $ & $2.48 \pm 0.10 $  \\ 
NIRC2 & L' & $1.24 \pm 0.14 $ & $0.24 \pm 0.03 $ & $2.80 \pm 0.32 $ & $1.12 \pm 0.13 $\\
NIRC2 & Ms & $1.17 \pm 0.25 $ & $0.23 \pm 0.05 $ & $2.66 \pm 0.57 $ & $ 1.07 \pm 0.23 $ \\
VIRCAM & Y & $ 12.40 \pm 0.27 $ & $2.40 \pm 0.05 $ & $27.77 \pm 0.58 $ & $11.33 \pm 0.25 $\\
VIRCAM & J & $ 8.21 \pm 0.12 $ & $1.60 \pm 0.03 $ & $18.30 \pm 0.26 $ & $ 7.48 \pm 0.11 $\\
VIRCAM & H & $4.68 \pm 0.11 $ & $0.91 \pm 0.02 $ & $10.41 \pm 0.25 $ & $ 4.24 \pm 0.10 $\\
VIRCAM & Ks & $2.67 \pm 0.11$ & $0.51 \pm 0.02 $ & $6.04 \pm 0.26 $ & $ 2.42 \pm 0.10 $\\
2MASS & J & $ 8.26 \pm 0.12 $ & $1.64 \pm 0.03 $ & $18.14 \pm 0.26$ & $7.51 \pm 0.11 $\\
2MASS & H & $ 4.65 \pm 0.11 $ & $0.90 \pm 0.02 $ & $10.37 \pm 0.25$ & $4.22 \pm 0.10 $\\ 
2MASS & Ks & $ 2.58 \pm 0.11 $ & $0.50 \pm 0.02 $ & $5.84 \pm 0.26$ & $2.34 \pm 0.10 $ \\
\enddata
\tablecomments{We calculate filter extinctions for different Brackett-$\gamma$  (2.166 $\mu$m) extinctions.
 We use blackbodies and power-law source spectra. The error is the uncertainty due to the extinction error.
 The first column is the average extinction of stars in the central 14'' $\times$20'' of the GC. The last column is the extinction
 towards the power-law source Sgr~A* (i.e. using stars in the central r$\leq0.5$'' of the GC for scaling of A$_{Br\,\gamma}$ ).}
\end{deluxetable}

\section{D: Comparison with other methods for the determination of the NIR extinction slope}
\label{sec:alpha_oth}

We derive the NIR extinction slope $\alpha$ from absolute extinction values at known wavelength.
Another way commonly used to compute the near infrared extinction slope $\alpha$ is to use stellar colors, 
using an assumed wavelength to associate with the broadband filters (we hereafter call this the effective wavelength method). 
For example, the following equation is commonly used to compute the extinction slope $\alpha$ from JHK(s) colors \citep{Stead_09}: 
 \begin{equation}
\frac {E_{J-H}}{E_{H-Ks}}=\frac{\left(\frac{\lambda_J}{\lambda_H}\right)^{\alpha}-1 } {1-\left(\frac{\lambda_{Ks}}{\lambda_H}\right)^{\alpha}}
\label{eq:ejhk}
 \end{equation}
where $\lambda_J$ is some assumed wavelength of the $J$ filter, and so on for the other filters. 
Obviously, it is necessary to know the wavelength that can correctly be associated with the filters for the given source and extinction in order 
to derive $\alpha$. One possibility would be to use the isophotal wavelength ($\lambda_{\mathrm{iso}}$) \citep{Tokunaga_05}:
\begin{equation}
F_{\lambda}(\lambda_{\mathrm{iso}})=\frac{\int F_{\lambda}(\lambda)S(\lambda) d\lambda  } {\int S(\lambda) d\lambda }
\label{eq:isoph}
\end{equation}
This means that $\lambda_\mathrm{iso}$ is the wavelength at which the monochromatic flux $F_{\lambda}(\lambda_{\mathrm{iso}})$
equals the mean flux in the passband.

More commonly used for extinction purposes is the effective wavelength \citep{Tokunaga_05}:
\begin{equation} \lambda'_{\mathrm{eff\,unext}} =\frac{\int \lambda^2 \,F_{\lambda}(\lambda) \, S(\lambda)\, \,d\lambda} 
  {\int \lambda \,F_{\lambda}(\lambda) \, S(\lambda)\, \,d\lambda  }
\label{eq:effun}
\end{equation} 
The effective wavelength defined this way is the average wavelength of received photons,
 weighted by the number distribution of received photons at the detector, appropriate for photon counting detectors.
This formula is used in two variants: for the calculation of $\lambda'_{\mathrm{eff\,unext}}$ the source spectra are not extincted.

For the calculation of the other variant ($\lambda'_{\mathrm{eff\,ext}}$) the source spectra are extincted \citep{Stead_09,Schoedel_09b}:
\begin{equation} \lambda'_{\mathrm{eff\,ext}} =\frac{\int \lambda^2 \,F_{\lambda}(\lambda) \, S(\lambda)\, R(\lambda) \,d\lambda} 
 {\int \lambda \,F_{\lambda}(\lambda) \, S(\lambda)\, R(\lambda) \,d\lambda  }
\label{eq:effex}
\end{equation}

We test the accuracy of the methods concentrating on the two definitions of the effective wavelength. 
To do this we compare the explicitly calculated filter extinctions to those obtained with the effective wavelength method for an extincted, 
9480K Vega-like blackbody. As test filters we use the NACO JHKs filters plus atmosphere and our $\alpha=-2.11$ extinction law. 
Using Formula~\ref{eq:extsim} we obtain 
slightly higher extinctions for the explicit calculation than obtained by applying the extincted effective wavelength method (Formula~\ref{eq:ejhk}), 
with differences 
of $A_{Ks\,\mathrm{true}}-A_{Ks\,\lambda'\,\mathrm{eff\,ext}}=0.05$  for A$_{\mathrm{Br}\,\gamma}=2.5$ (for lower extinction the difference is much less 
 $A_{Ks\,\mathrm{true}}-A_{Ks\,\lambda'\,\mathrm{eff\,ext}}=0.013$  for A$_{\mathrm{Br}\,\gamma}=1$). Similarly we obtain 
  $A_{Ks\,\mathrm{true}}-A_{Ks\,\lambda'\,\mathrm{eff\,unext}}=-0.012$  for A$_{\mathrm{Br}\,\gamma}=2.5$ when we use
 the unextincted effective wavelengths.

\begin{figure}
\begin{center}
\includegraphics[width=0.4 \columnwidth,angle=-90]{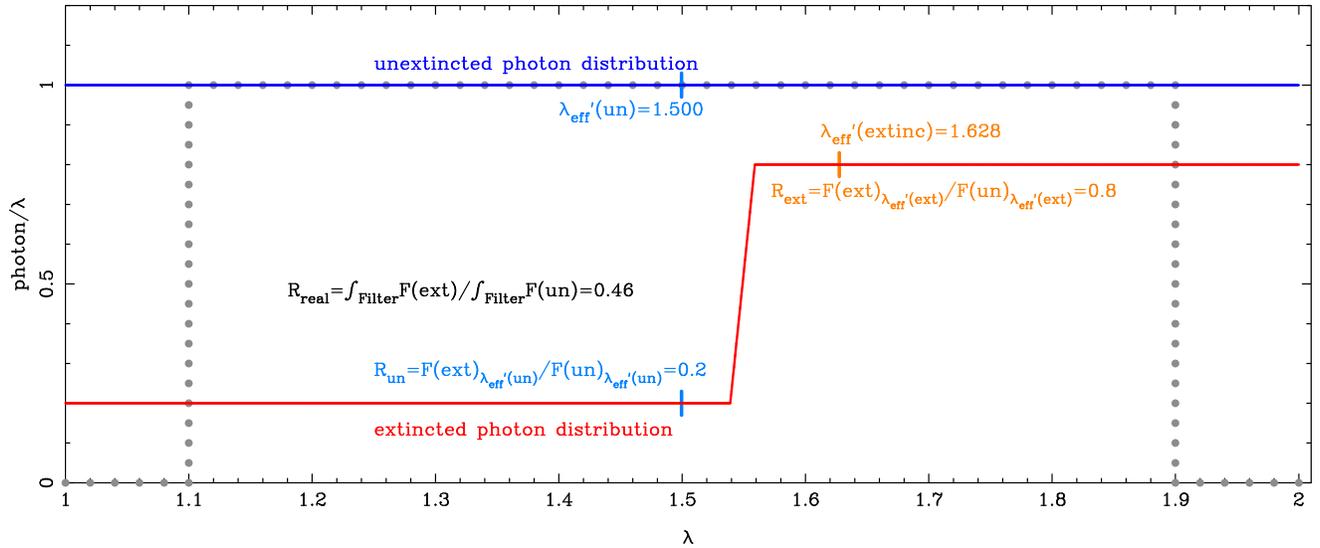}
\caption{Difference between the extinction at (extincted and unextincted) effective wavelengths, and the filter-integrated extinction.
A source (blue line, showing unextincted flux) is observed with a broad filter (gray dots). Via an (unrealistic) extinction, the source is extincted to the orange line. In this example the effective wavelength of the unextincted flux is in the part with high extinction, and the effective wavelength of the extincted flux is in the part with small extinction. The true reddening factor, however, is determined by the integrals over the filter range and has a value between the reddening factors 
at the two effective wavelengths. For more realistic extinction curves this effect is much smaller, but not always negligible.
} 
\label{fig:leff_exam}
\end{center}
\end{figure}

 The reason for these discrepancies is illustrated via an exaggerated example in Figure~\ref{fig:leff_exam}. To obtain the correct value of the true
 extinction it is necessary to consider the photon distribution, integrated over the filter, of both the unextincted and of the extincted source. Yet 
the two effective wavelength methods each consider only one of the above photon distributions. In the case of the unextincted source: because 
a larger proportion of the photons is transmitted through the extinction at longer wavelength than at $\lambda'_\mathrm{eff\,unext}$, the true
 filter extinction is smaller
than at $\lambda'_\mathrm{eff\,unext}$. The opposite is true at 
 $\lambda'_\mathrm{eff\,ext}$.

In summary both 'effective' wavelengths deviate from the true extinction, because these effective wavelengths are not the effective extinction
 wavelengths. As a result it is not correct to use these 'effective' wavelengths to calculate the extinction appropriate for broadband filters from 
our extinction curve. The true wavelength for extinction measurements $\lambda_{\mathrm{true}}$ is the wavelength at which the extincted object is extincted
by the same amount of extinction as in the explicit calculation  of Formula~\ref{eq:extsim}. $\lambda_{\mathrm{true}}$ depends on the optical system, 
object SED, strength of extinction and the shape of the extinction law.
However, since \citet{Stead_09,Schoedel_09b}, e.g., measured the broadband extinctions directly their extinctions
are correct at least for the specific source, filter and extinction strength combinations used in the determination.

\begin{figure}
\begin{center}
\includegraphics[width=0.50 \columnwidth,angle=-90]{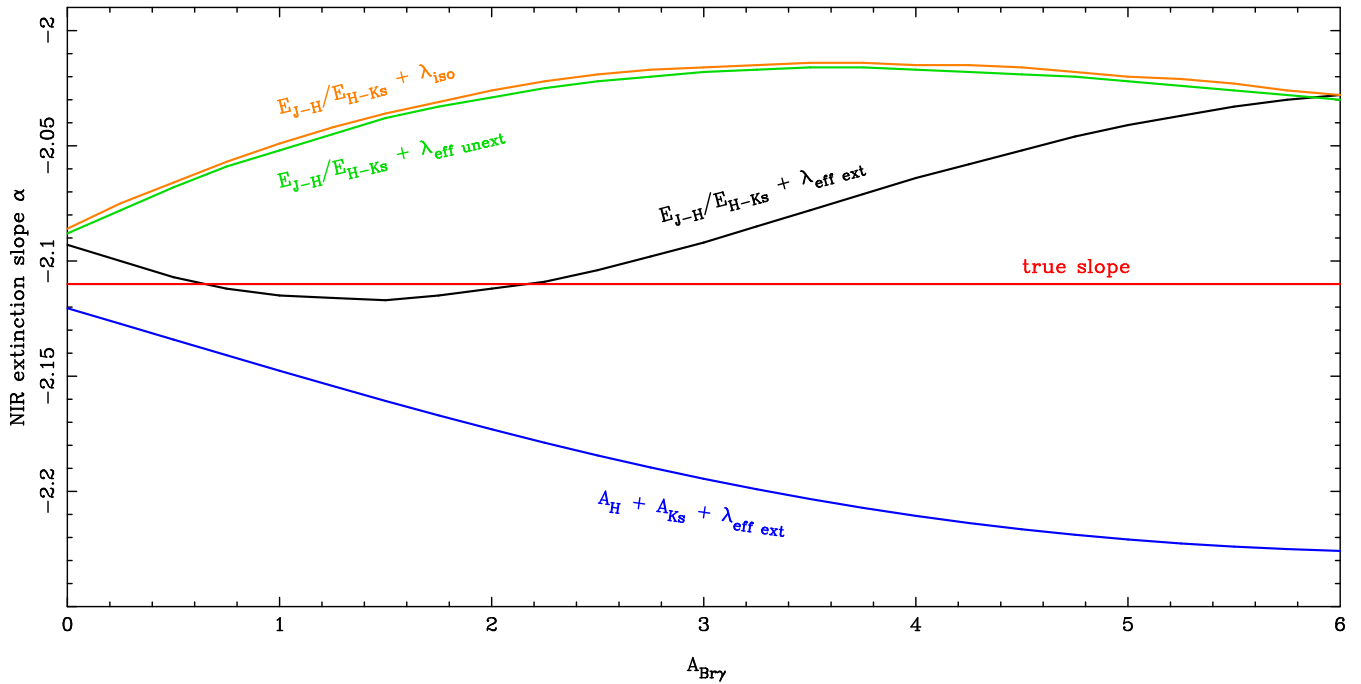}
\caption{Comparison of the true near infrared extinction slope $\alpha$ with the $\alpha$ obtained using common methods. We use the NACO JHKs filters with atmosphere and a 9480 K blackbody, simulating common measurements. For obtaining $\alpha$ from $E_{J-H}/E_{H-Ks}$ we use Formula~\ref{eq:ejhk}. The results are displayed as follows: (orange) the use of the isophotal wavelength in Formula~\ref{eq:ejhk}; (green) the use of the effective wavelength of the unextincted flux in  Formula~\ref{eq:ejhk}; (black)
 the use of the effective wavelength of the extincted flux in Formula~\ref{eq:ejhk}; and (blue) the computation of $\alpha$ from absolute A$_{H}$ and A$_{Ks}$, using the extincted effective wavelengths. The true slope is shown as the red line.
} 
\label{fig:alpha2}
\end{center}
\end{figure}

More important is the issue of effective wavelengths for the determination of $\alpha$ from measured broadband extinctions. 
To test this we calculate alpha, using either $\lambda_{\mathrm{iso}}$,  $\lambda'_{\mathrm{eff\,unext}}$ or $\lambda'_{\mathrm{eff\,ext}}$ in Formula~\ref{eq:ejhk} (method of relative extinction). We also calculate $\alpha$ from the absolute A$_{H}$ and A$_{Ks}$ using $\lambda'_{\mathrm{eff\,ext}}$, as is done in \citet{Schoedel_09b}.
We then compare the obtained values of $\alpha$ with the input $\alpha=-2.11$, see Figure~\ref{fig:alpha2}.

Since none of the methods obtains the true extinction values, it is not surprising that none of the methods obtains the true extinction slope $\alpha$. 
The best method for obtaining $\alpha$ is using $\lambda'_{\mathrm{eff\,ext}}$ and  Formula~\ref{eq:ejhk}: 
for extinctions A$_{\mathrm{Br}\,\gamma} \leq 4$ there is less than 0.046 deviation of $\alpha$ from the input $\alpha=-2.11$. 
The other methods can result in larger deviations of up to  
to 0.11 for the same range of extinction. % in $\alpha$.% for the input $\alpha=-2.11$ and A$_{\mathrm{Br}\,\gamma}\leq4$. 

The difference between the $\alpha$ obtained using  $\lambda'_{\mathrm{iso}}$ and $\lambda'_{\mathrm{eff\,unext}}$  is
only $\Delta\alpha \approx$0.01. In contrast \citet{Stead_09} obtain a difference of around $\Delta\alpha=$0.2.
 The reason for this could be the
slightly different photometric system, 
but perhaps more likely, the difference between a 9480 K blackbody and a K2III spectrum. Given this big difference in $\Delta\alpha$,
 it is unclear if using $\lambda'_{\mathrm{eff\,unext}}$ 
is also the best approximation to the true $\alpha$ for instruments and source spectra other than tested here.

\end{document}